\documentclass[twocolumn]{article}

\usepackage{arxiv}

\usepackage{amsmath, amssymb}
\usepackage[utf8]{inputenc} 
\usepackage[T1]{fontenc}    
\usepackage[colorlinks=true, allcolors=blue]{hyperref}       

\usepackage{url}            
\usepackage{booktabs}       
\usepackage{amsfonts}       
\usepackage{nicefrac}       
\usepackage{microtype}      
\usepackage{graphicx}
\usepackage[sort]{natbib}
\setcitestyle{comma,numbers,open={[},close={]}}
\usepackage{doi}
\usepackage{siunitx}
\usepackage{enumitem}
\setlist[enumerate,1]{leftmargin=!, itemsep=-1ex, partopsep=1ex,parsep=1ex}
\usepackage{stmaryrd}
\usepackage{caption}
\usepackage{subcaption}
\usepackage{booktabs}
\usepackage{tabularx}
\usepackage{upgreek}
\usepackage{bm}
\usepackage{adjustbox}

\newcommand{\ind}[1]{{}^{#1}\!}
\newcommand{\C}{\mathcal{C}}
\newcommand{\T}{\boldsymbol{\Uptheta}}
\newcommand{\TT}{\Tilde{\T}}
\newcommand{\TP}{\Tilde{\text{P}}}
\newcommand{\Tp}{\Tilde{p}}
\newcommand{\ve}[1]{\textbf{#1}}
\newcommand{\te}[1]{\boldsymbol{#1}}

\newcommand{\R}{\mathbb{R}}

\title{Inverse design of spinodoid structures using Bayesian optimization}
\author{
    Alexander Ra{\ss}loff\\
    Institute of Solid Mechanics\\
    TUD Dresden University of Technology\\
    01062 Dresden, Germany
    \And
    Paul Seibert\\
    Institute of Solid Mechanics\\
    TUD Dresden University of Technology\\
    01062 Dresden, Germany
    \And
    Karl A.\ Kalina\\
    Institute of Solid Mechanics\\
    TUD Dresden University of Technology\\
    01062 Dresden, Germany
    \And
    Markus K\"astner\thanks{Corresponding author, email: \texttt{markus.kaestner@tu-dresden.de}.}\\
    Institute of Solid Mechanics\\
    TUD Dresden University of Technology\\
    01062 Dresden, Germany
    }

\renewcommand{\shorttitle}{Inverse design of spinodoid structures using Bayesian optimization}
\hypersetup{
    pdftitle={Rassloff2024-InverseDesignSpinodoids-Preprint},
    pdfsubject={Preprint of original research article},
    pdfauthor={Rassloff},
    pdfkeywords={Inverse design, Bayesian optimization, Structure-property linkage, Architected materials, Spinodoid},
}

\begin{document}

\twocolumn[
    \begin{@twocolumnfalse}
    \maketitle
    
    \begin{abstract}
        Tailoring materials to achieve a desired behavior in specific applications is of significant scientific and industrial interest as design of materials is a key driver to innovation. Overcoming the rather slow and expertise-bound traditional forward approaches of trial and error, inverse design is attracting substantial attention. Targeting a property, the design model proposes a candidate structure with the desired property. This concept can be particularly well applied to the field of architected materials as their structures can be directly tuned.
        The bone-like spinodoid materials are a specific class of architected materials. They are of considerable interest thanks to their non-periodicity, smoothness, and low-dimensional statistical description. 
        Previous work successfully employed machine learning (ML) models for inverse design. The amount of data necessary for most ML approaches poses a severe obstacle for broader application, especially in the context of inelasticity.
        That is why we propose an inverse-design approach based on Bayesian optimization to operate in the small-data regime. Necessitating substantially less data, a small initial data set is iteratively augmented by in silico generated data until a structure with the targeted properties is found. The application to the inverse design of spinodoid structures of desired elastic properties demonstrates the framework's potential for paving the way for advance in inverse design.
    \end{abstract}
    \end{@twocolumnfalse}
]


\section{Introduction}\label{sec:introduction}
        Tailored materials are of widespread interest in both industry and academia. Being able to design the internal structure of a material to achieve a desired behavior is a key enabler for innovative applications. In this context, architected materials have attracted considerable attention. Their defining structure, i.e., their morphology on a mesoscopic scale, can be precisely controlled, for instance by additive manufacturing~\cite{schaefer2024, gebhardt2022a, guenther2022, sun2023}.
        In contrast, the heterogeneous microstructure of materials comprising, e.g., the texture and morphology of metal grains, can hardly be directly controlled. 

		Traditionally, expertise and a lot of experimentation is necessary to design structures with preferable properties. Furthermore, this \emph{forward design} approach of trial and error, i.e., choosing a structure, generating it, and determining its properties, can span multiple years. 
        This process can be significantly accelerated by in silico experiments, in which synthetic structures are generated~\cite{seibert2022b, henrich2023, eshlaghi2023} and numerically simulated to obtain their properties~\cite{aldakheel2023, eidel2023, henkes2021}. Extensive knowledge regarding the determination of effective properties and scale-transition has been developed in the context of multiscale modeling~\cite{kalina2023, schroeder2014a, miehe1999, schroeder2016}. Extracting and setting up quantitative linkages between an adequate description of the structure and the properties of interest further assist in this approach~\cite{meyer2022, heidenreich2023, fuhg2022}. Still, it is intensive in time and resources to find even acceptable choices in the usually vast design spaces. 
        For this reason, data-driven \emph{inverse} design approaches are extensively studied.
        
		In \emph{inverse design}, contrary to the forward concept, the targeted property serves as input. Two main categories can be distinguished: direct and indirect inverse design.
		In \emph{direct inverse design} property-structure linkages are employed to map directly from the property space to the design space of the structure. Therein, machine learning approaches are especially present. The reader is kindly referred to the reviews of Zheng~et~al.~\cite{zheng2023} and Lee~et~al.~\cite{lee2023} for a comprehensive introduction. Hereafter, a brief overview is given.

        An established approach is the utilization of both an inverse \emph{and} forward model, where the latter effectively supports the training of the former. Bastek~et~al.~\cite{bastek_inverting_2022} employed a supervised learning approach in a framework that combines stochastic and physics-guided neural networks for the design of truss-based architected materials, trained on 3\,000\,000 data points.
        The idea of a tandem MultiLayer Perceptron (MLP) is also pursued in the work of Kumar et al.~\cite{kumar2020} for successfully designing the stiffness tensor of architected materials based on a data set of 21\,300 structure-property pairs. Deng et~al.~\cite{deng2024} utilized the same model for an application in an experimental context. Another example is the work of Van’t Sant~et~al.~\cite{vantsant2023} in which growth-based cellular architected materials are inversely designed for desired stiffnesses based on 800\,000 structure-property pairs.
        Recently, also video denoising diffusion models have been employed for the inverse design of 2D cellular architected materials for specific stress-strain curves. Bastek~et~al.~\cite{bastek2023} used 53\,007 pairs of 2D images and nonlinear stress-strain response for the training of their model. Such a denoising algorithm is utilized by Vlassis and Sun~\cite{vlassis2023} in the context of fine-tuning nonlinear material properties of 2D topologies.

        Unsupervised machine learning methods have been applied to inverse design, too. Challapalli~et~al.~\cite{challapalli2021} employed Generative Adversarial Networks (GANs) to propose lattice cells of desired performance. Zheng~et~al.~\cite{zheng2023a} designed 3D cellular materials of targeted stiffness and porosity through a conditional GAN that is trained on 10\,000 data points. A convolutional neural network in combination with a modified Cycle-GAN is utilized by Tian~et~al.~\cite{tian2022} for the design of 2D structures with deformation-dependent Poisson's ratio based on 8\,000 molecular dynamics simulations.
        Also large transformer networks, that are commonly used in natural language processing, have been used for the generation of structures on the basis of text inputs and might be considered inverse design models~\cite{yang2021c}. The training of such models necessitates a huge amount of data, in case of the aforementioned work 400\,000\,000 image-text pairs.

        Once the model is established, most direct inverse design methods allow to obtain the desired structure in rather short time. However, the requirement of large training data sets poses a critical challenge. In contrast, most indirect inverse design approaches can operate on less initial data.

        \emph{Indirect inverse design} utilizes only forward structure-property linkages to predict the property of a given structure followed by an optimization or selection step. Examples comprise high-throughput screening, genetic algorithms, or Bayesian optimization.
		In high-throughput screening, large data bases are scanned for structures with properties close to the desired target. A common application can be found in alloy design~\cite{curtarolo2013}.
        Recently, Thalkolkaran~et~al.~\cite{thakolkaran2023} presented an optimization framework that utilizes a physics-enhanced machine learning model. It works directly on 321 experimental data points for designing a class of architected materials to fit a targeted stress-strain curve.
        Rixner and Koutsourelakis~\cite{rixner2022} employed an active learning strategy for the optimization of 2D random material microstructures in the small-data regime considering the process-structure-property linkages.
        A genetic algorithm is used by Wang~et~al.~\cite{wang2022} for designing shell-based materials for customized loading curves using a initial data set of 7\,000 structure-loading curve pairs. Deng~et~al.~\cite{deng2022} also utilized an evolution strategy in combination with an MLP trained on 30\,000 data points for finding architected materials of desired nonlinear mechanical responses.
        
		In Bayesian optimization, a structure-property model and an acquisition function are used to select points in the design space that should be investigated. The model needs to include a measure of uncertainty and is most commonly a Gaussian process regression. The acquisition function assigns each point in the design space a scalar value indicating how worth it is to investigate a structure at this specific point. This acquisition metric is based on the mean value and its uncertainty. It guides the optimization either towards potentially good structures (exploitation) or tries to minimize the uncertainty in the design space (exploration) or a combination of both. Gongora~et~al.~\cite{gongora2020} applied this strategy in an experimental context using 600 data points for designing barrel like structures for a certain toughness. Cauchy symmetric structures are designed through Bayesian optimization in the work of Sheikh~et~al.~\cite{sheikh2022}. Kusampudi~et~al.~\cite{kusampudi2023d} utilized a Bayesian optimization to inversely design the 2D microstructure of a dual-phase steel in a numerical study. 

        Bayesian optimization is a rather universal approach for exploring and exploiting structure-property linkages. The ability to operate on small to moderate data sets constitutes a major advantage. However, the optimization of a cost function and the iterative procedure are computationally demanding, especially in high dimensional design spaces.

		Regardless of the specific choice of the design approach, the topical challenge in setting up the forward or inverse structure-property linkages is the acquisition of enough data. Machine learning models need training data as can be seen from the aforementioned direct inverse design examples.
        Yet, the generation of such large data sets is expensive and hinders the broader application to real-world scenarios. In this context, iterative indirect approaches that necessitate less initial data are promising.

		  The adequate encoding of the structure poses a second significant challenge.
        The ability to describe the structural key characteristics in few meaningful numbers, the descriptors~\cite{torquato2002a, schmidt2023, reck2023}, is a prerequisite for any model. By this, the design space can be restricted to a reasonable number of dimensions.
        Additionally, the generation of a structure from these descriptors must be possible~\cite{seibert2022b, scheunemann2015, balzani2014}. While some machine learning models seem to directly link structures in terms of images to properties, they commonly comprise the encoding within their architecture.
        
		Kumar et al.~\cite{kumar2020} overcame the latter challenge for a specific class of architected materials, so-called \emph{spinodoids}. 
        The morphology of these potentially anisotropic structures can be described by few interpretable variables, resulting in a low-dimensional design space.
        This is contrary to more universal approaches, in which either machine learning is employed to get a low-dimensional but rather abstract latent space encoding~\cite{zhang2024, wang2023, lissner2023a} or high-dimensional descriptors are employed that require further dimensionality reduction~\cite{seibert2022b, seibert2022}.
		The bone-like spinodoid materials are a particularly interesting nature-inspired class of structure thanks to their smoothness as well as non-periodicity. They are successfully utilized in two-scale frameworks~\cite{zheng2021a}.
        Kumar et al.~\cite{kumar2020} pursued a direct inverse design approach for predicting the design parameters given a desired stiffness tensor. As mentioned before, the tandem MLP model is trained on a large set of 21\,300 structure-property pairs.

        The present article aims at overcoming the issue of necessitating large amounts of data for inverse structure design by employing Bayesian optimization. At the example of spinodoids, an iterative approach is presented.
        Starting with a small data set of few structures of known descriptor and properties, structure-property linkages are modeled in the whole descriptor space and promising new structures are proposed in an active learning approach. Their properties are derived from numerical simulations. These candidate structures are added to the data set. This loop is run until the design goal, herein, the acquisition of a spinodoid structure of desired properties, is reached.

        This framework and its necessary steps are described in \autoref{sec:Method}. Applying this inverse design concept to three examples of increasing complexity, the results are presented in \autoref{sec:Result}. The article ends with a discussion in \autoref{sec:Discussion}.

        In the following, vectors and arrays are indicated by bold, upright letters ($\ve{x},\:\bm{\upxi}$), their coordinates by non-bolt, upright letters ($\text{x}_i,\:\upxi_i$), and other scalar quantities by italic letters ($n,\:\beta$).
        In the present work, the morphology of spinodoid materials is considered only, that is hereafter referred to as structure.
	
	\section{Methods}\label{sec:Method}
        \subsection{General workflow}\label{subsec:Concept}
            This contribution's objective of inversely designing spinodoid structures is pursued in an indirect, iterative approach to ensure applicability to the small-data regime. That is particularly important as data, especially if experimentally obtained, is in general expensive to acquire. 
            Augmenting a small initial data set within the design process in an actively guided manner facilitates a significant reduction of the data needed.
               
\begin{figure*}[t]
    				\centering
    				\includegraphics[width=\textwidth]{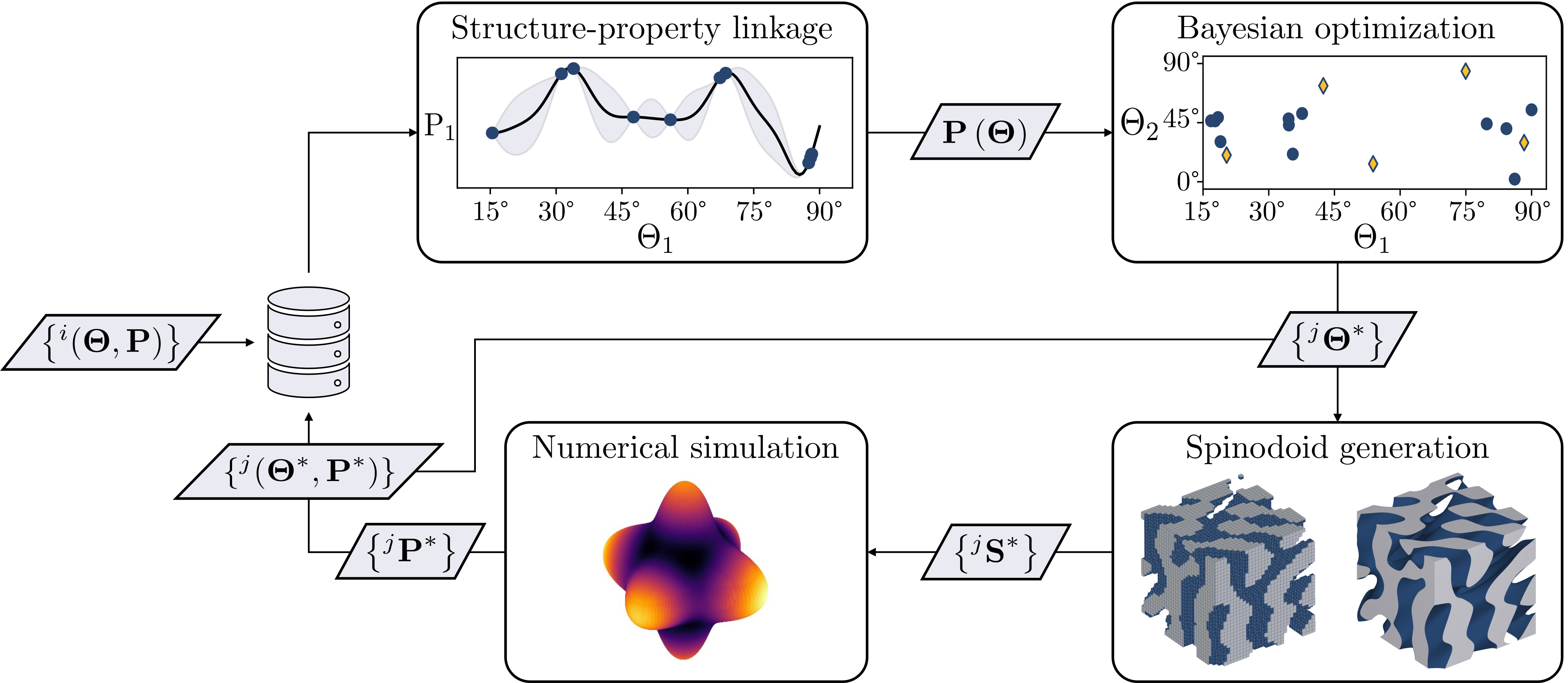}
    				\caption{Active learning loop for the inverse design of spinodoid structures. Starting with an initial set of $n_\text{init}$ descriptor-property pairs $\mathcal{D}=\left\{\ind{i}\left(\T,\textbf{P}\right)\right\}_{i=1}^{n_\text{init}}$ Gaussian process regression is utilized to model the structure-property linkage $\textbf{P}(\T)$. Secondly, Bayesian optimization is employed to propose a set of $n_\text{cand}$ descriptors $\left\{\ind{j}\T^*\right\}_{j=1}^{n_\text{cand}}$ (indicated by the diamond) of new structures to be investigated. Using them, spinodoid structures $\left\{\ind{j}\textbf{S}^*\right\}_{j=1}^{n_\text{cand}}$ are generated as voxelized simulation domains. In a last step, the properties of the corresponding structures $\left\{\ind{j}\textbf{P}^*\right\}_{j=1}^{n_\text{cand}}$ are derived from numerical simulations. The so obtained set of in silico generated structure-property pairs $\mathcal{D}^*=\left\{\ind{j}\left(\T^*,\textbf{P}^*\right)\right\}_{j=1}^{n_\text{cand}}$ is added to the data set $\mathcal{D}\leftarrow\mathcal{D}\cup\mathcal{D}^*$ and the next iteration starts. For the sake of simplicity, the index range is not indicated in the figure.}\label{fig:AugLoop}
            \end{figure*}

            \autoref{fig:AugLoop} illustrates the idea of such an active learning loop.
            The starting point is a data set of $n_\text{init}$ descriptor-property pairs $\mathcal{D}=\left\{\ind{i}\left(\T,\textbf{P}\right)\right\}_{i=1}^{n_\text{init}}$. The descriptor $\T \left(\ve{S}\right)$ characterizes a structure $\ve{S}$ in a statistical and translation-invariant manner. If two structures are similar then so are the corresponding descriptors. The property array $\ve{P}\left(\ve{S}\right)$ comprises all properties of interest for the design of a specific structure. The indicated dependence on $\ve S$ is omitted in the following.
            
            In a first step, the structure-property linkage\footnote{The structure-property linkage maps a descriptor to the property of the corresponding structure.} $\textbf{P}(\T)$ is derived. Here, a Gaussian process regression, see \autoref{subsec:GPR}, is utilized to yield an estimate of the properties at each location in the descriptor space alongside a measure of uncertainty. 
            This is the required input to the actual optimization.
    
            The optimization step, see \autoref{subsec:BO}, requires the definition of a quantity to be optimized, that is here referred to as cost $\mathcal{C}\left(\ve{P}\right)$. Depending on the optimization goal, which could be either an exploration of the descriptor space, i.e., a reduction of uncertainty, or an exploitation, i.e., the objective to find an optimal structure, or a mix of both, the so-called acquisition function $\alpha\left(\mathcal{C}\right)$ is chosen. Maximizing it in the descriptor space yields a set of $n_\text{cand}$ candidate descriptors $\left\{\ind{j}\T^*\right\}_{j=1}^{n_\text{cand}}$ that, based on the current structure-property linkage, might potentially achieve the desired properties. They are selected either in regions of low uncertainty and assumed matching values or in regions of high uncertainty, i.e., a trade-off between exploitation and exploration.
    
            For the numerical simulation, a set of spinodoid structures $\left\{\ind{j}\textbf{S}^*\right\}_{j=1}^{n_\text{cand}}$ is now generated from the proposed descriptors as explained in~\autoref{subsec:Spino}. In the subsequent step, numerical simulations are performed to obtain the properties of each candidate $\left\{\ind{j}\textbf{P}^*\right\}_{j=1}^{n_\text{cand}}$.
    
            In a last step, the acquired set of $n_\text{cand}$ new descriptor-property pairs $\left\{\ind{j}\left(\T,\textbf{P}\right)\right\}_{j=1}^{n_\text{cand}}$ is added to the data set
            \begin{equation*}
                \mathcal{D} = \left\{\ind{i}\left(\T,\textbf{P}\right)\right\}_{i=1}^{n_\text{init}} \cup \left(\bigcup_{k=1}^{i_\text{iter}} \ind{k}\left\{\ind{j}\left(\T,\textbf{P}\right)\right\}_{j=1}^{n_\text{cand}} \right) \:,
            \end{equation*}
            where $i_\text{iter}$ denotes the index of the current iteration.
            This augmented data set is the starting point for the next iteration, starting with deriving updated structure-property linkages et cetera. This procedure is repeated until a certain criterion, e.g., the number of iterations or a specific cost value, is reached.
    
            The following subsections describe the details of the method introduced. In~\autoref{subsec:Spino} the generation of spinodoid structures is outlined. Details on the Gaussian process regression are given in~\autoref{subsec:GPR}, followed by~\autoref{subsec:BO} describing the Bayesian optimization itself.    
            For the numerical simulations DAMASK's~\cite{roters2019} efficient spectral solver is employed. More information is given in~\autoref{app:DAMASK}.

        \subsection{Spinodoid structures}\label{subsec:Spino}
            The bone-like spinodal structures have smooth, non\--inter\-secting surfaces and are of great interest for architected materials as they can be seamlessly tailored for specific properties~\cite{vidyasagar2018, kumar2020, senhora2022}. These structures are naturally created in spinodal decomposition processes during phase separation. Solving the time-dependent Cahn-Hilliard equation to get the phase-field solution is impractical for design purposes as it is rather computationally demanding. Following the idea of Cahn~\cite{cahn1961}, Kumar~et~al.~\cite{kumar2020} proposed an efficient approach to model and generate spinodal-like structures, so-called spinodoids. This approach is recapitulated hereafter in brief.

            Cahn~\cite{cahn1961} showed that the concentration phase-field $\varphi(\ve{x})$ of one of two phases at a spatial position ${\ve{x}\in\mathbb{R}^3}$ in the early phases of spinodal decomposition can be approximated by a Gaussian random field
            \begin{equation}
                \varphi(\ve{x}) \approx \sqrt{\dfrac{2}{n_\text{wave}}} \sum_{i=1}^{n_\text{wave}} \cos{\left( \beta \ve{n}_i \cdot \ve{x} + \gamma_i \right)} \:,
            \end{equation}
            i.e., a superposition of $n_\text{wave}$ standing waves of the same wavenumber $\beta>0$, if $n_\text{wave}$ is sufficiently large. The direction and phase angle are denoted by ${\ve{n}\sim\mathcal{U}_{\mathcal{S}^2}}$ and ${\gamma\sim\mathcal{U}_{[0, 2\pi)}}$, respectively, where $\mathcal{U}$ indicates a uniform distribution and ${\mathcal{S}^2:=\left\{ \ve{k} \in \mathbb{R}^3 \mid \left| \ve{k} \right|=1 \right\}}$ is the surface of the 3D unit sphere.

\begin{figure*}[t]
				\begin{subfigure}[b]{110pt}
					\centering
					\includegraphics[width=100pt]{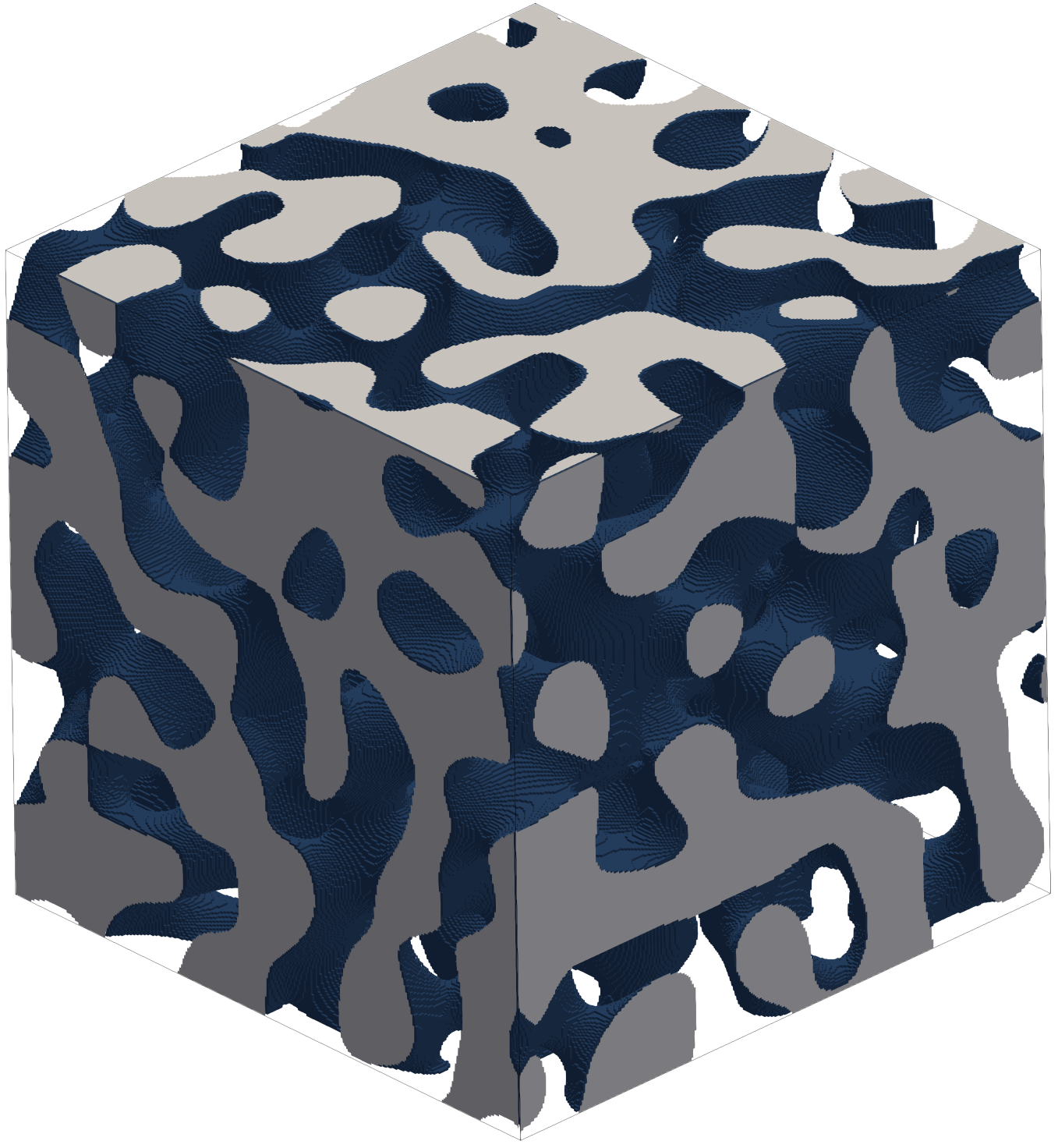}
					\caption{$\bm{\uptheta}^\intercal=\left(\SI{90}{\degree},\, \SI{0}{\degree},\, \SI{0}{\degree}\right)$}\label{fig:SpinodoidIsotropic}
				\end{subfigure}
				\hfill
				\begin{subfigure}[b]{110pt}
					\centering
					\includegraphics[width=100pt]{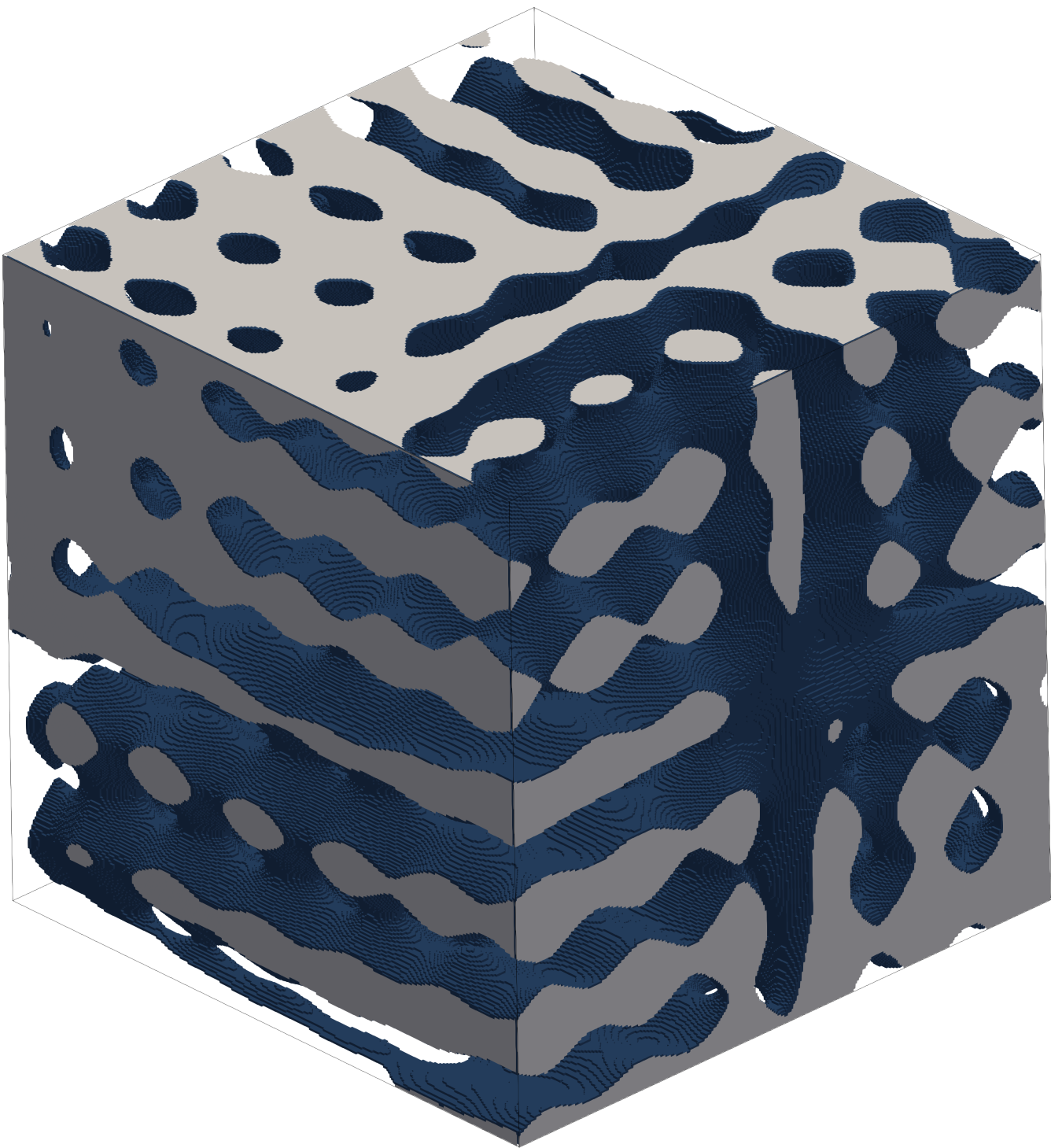}
					\caption{$\bm{\uptheta}^\intercal=\left(\SI{15}{\degree},\, \SI{15}{\degree},\, \SI{15}{\degree}\right)$}\label{fig:SpinodoidCubic}
				\end{subfigure}
				\hfill
				\begin{subfigure}[b]{110pt}
					\centering
					\includegraphics[width=100pt]{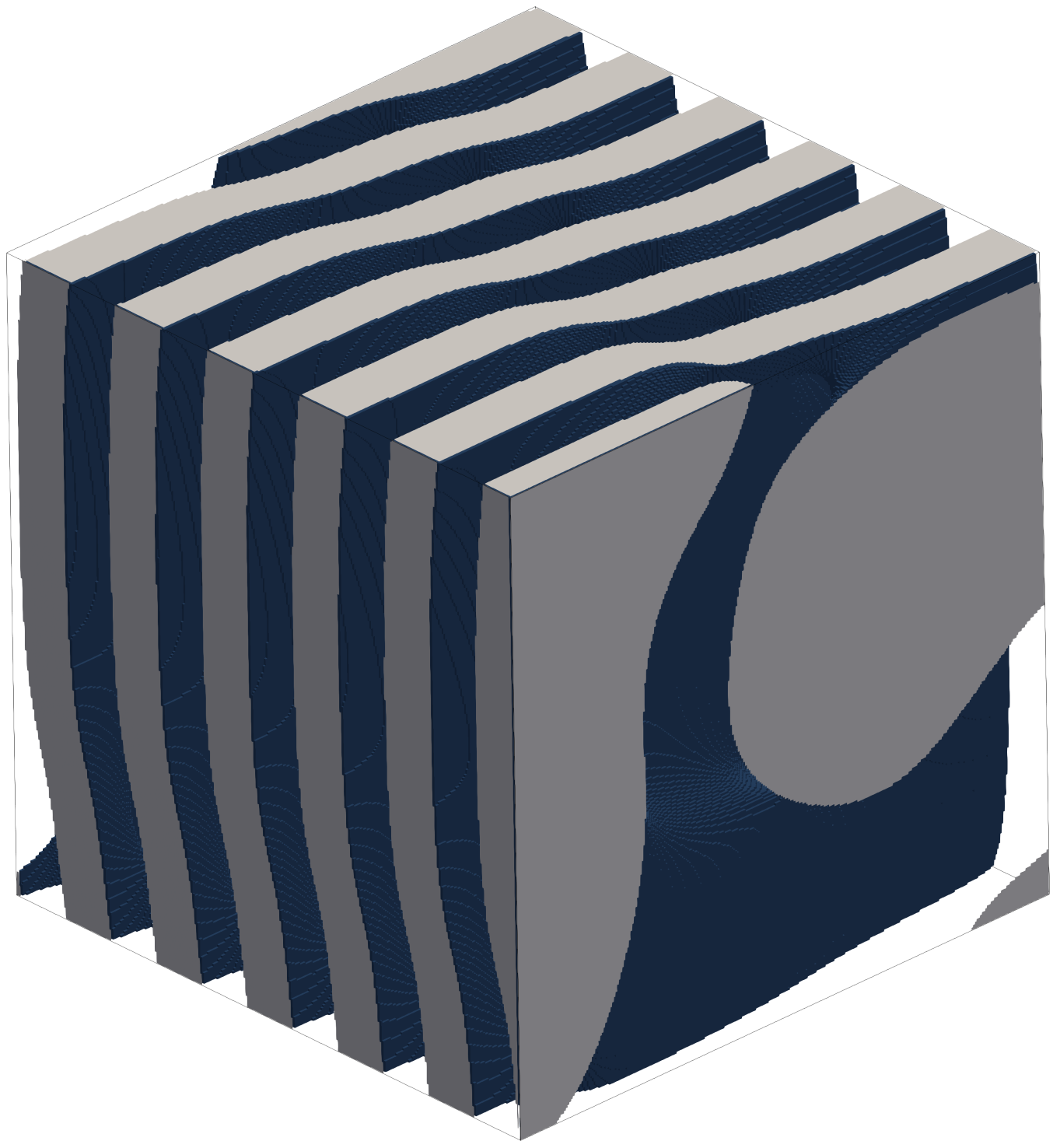}
					\caption{$\bm{\uptheta}^\intercal=\left(\SI{15}{\degree},\, \SI{0}{\degree},\, \SI{0}{\degree}\right)$}\label{fig:SpinodoidLamellar}
				\end{subfigure}
				\hfill
				\begin{subfigure}[b]{110pt}
					\centering
					\includegraphics[width=100pt]{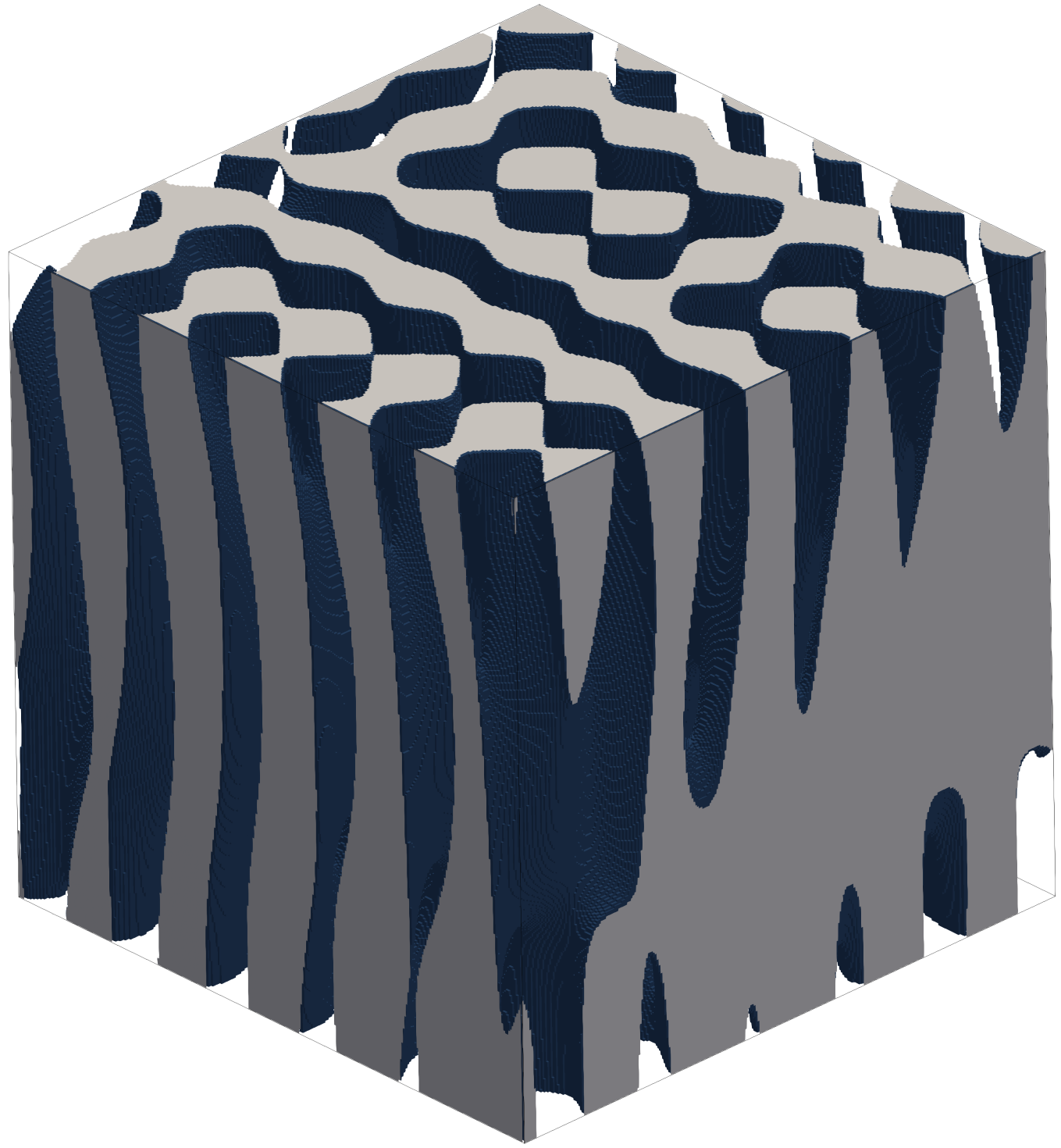}
					\caption{$\bm{\uptheta}^\intercal=\left(\SI{15}{\degree},\, \SI{15}{\degree},\, \SI{0}{\degree}\right)$}\label{fig:SpinodoidColumnar}
				\end{subfigure}
				\hfill
                \begin{subfigure}{45pt}
					\centering
					\includegraphics[height=1.4cm]{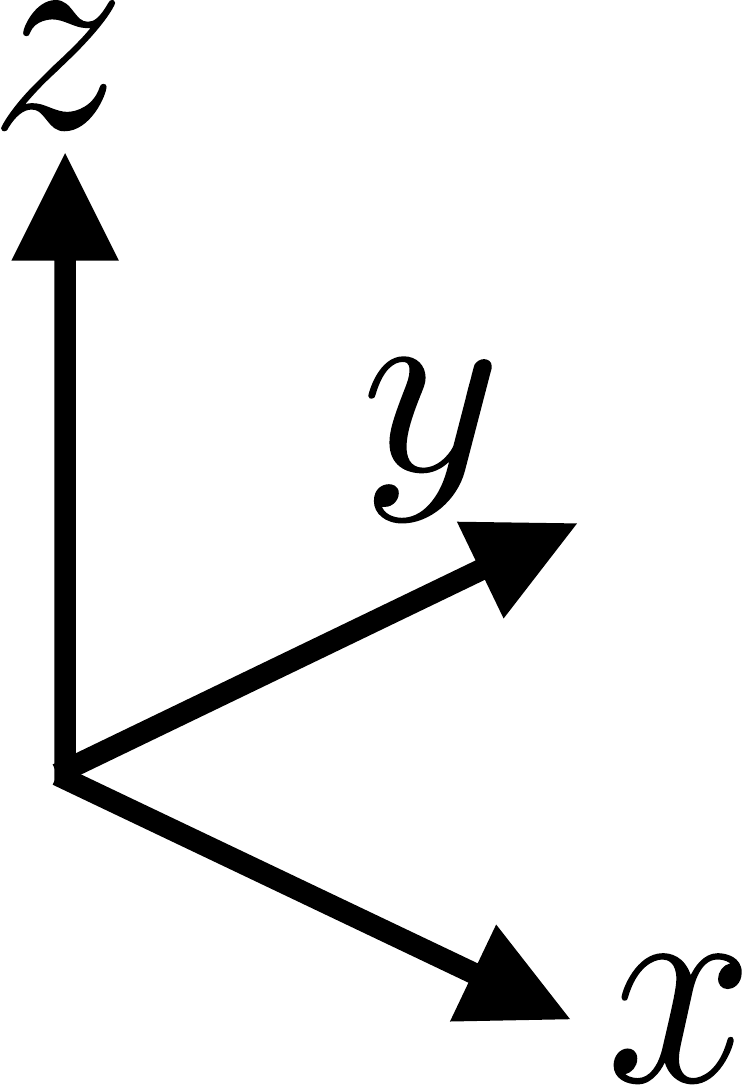}
					\vspace{20pt}
				\end{subfigure}
							
				\begin{subfigure}[b]{110pt}
					\centering
					\includegraphics[height=3.25cm]{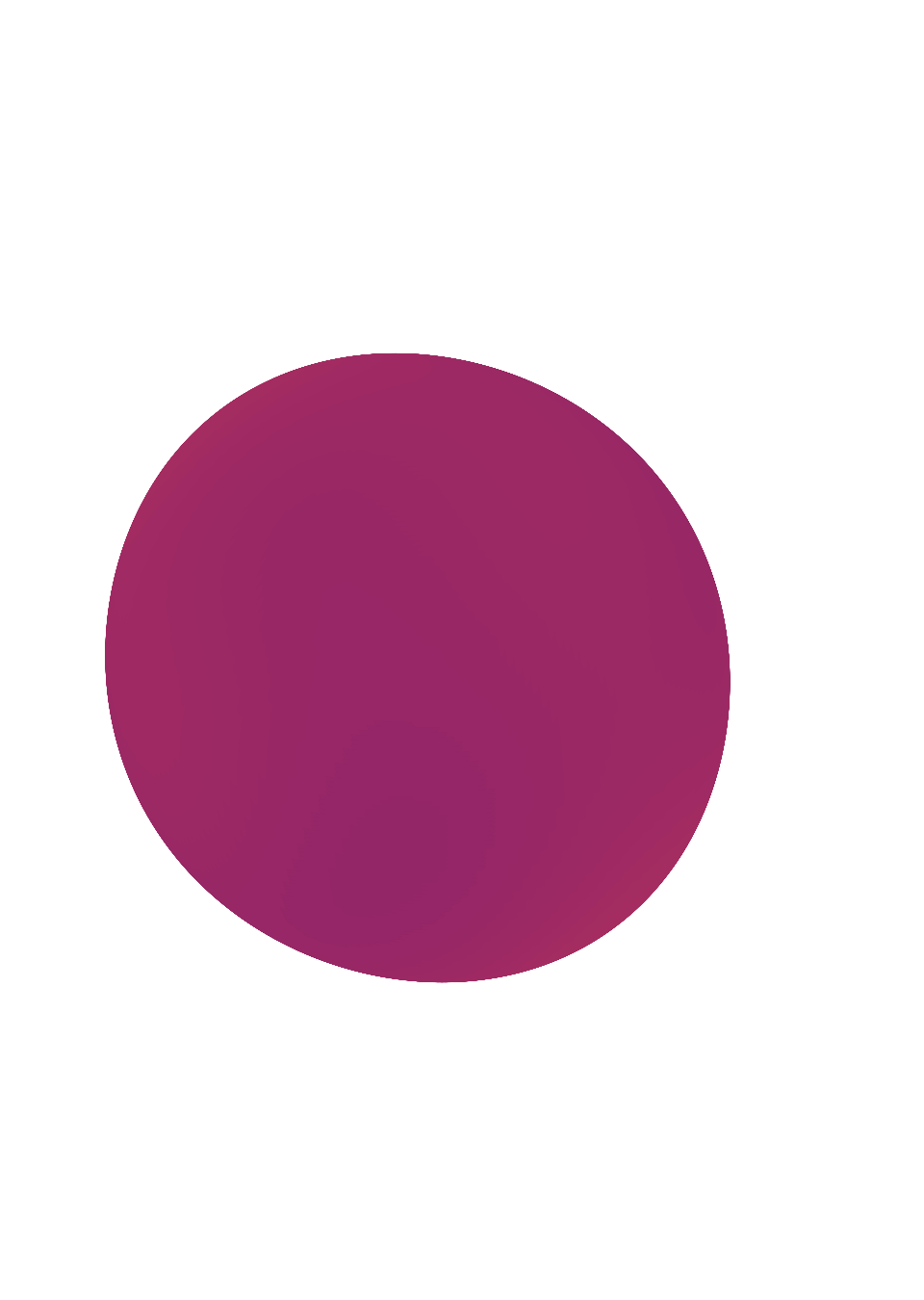}
					\caption{Isotropic}\label{fig:SpinodoidIsotropicES}
				\end{subfigure}
				\hfill
				\begin{subfigure}[b]{110pt}
					\centering
					\includegraphics[height=3.25cm]{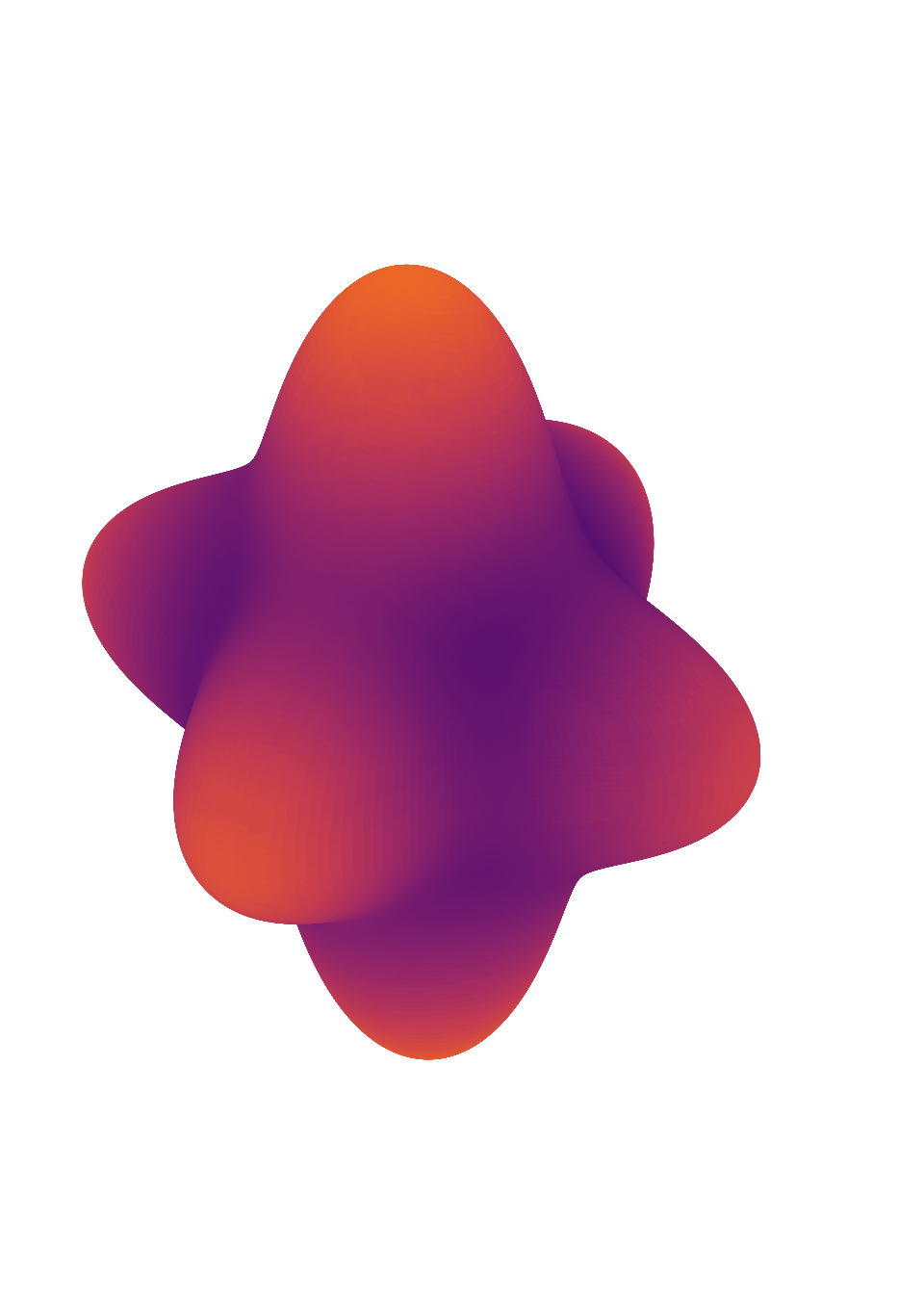}
					\caption{Cubic}\label{fig:SpinodoidCubicES}
				\end{subfigure}
				\hfill
				\begin{subfigure}[b]{110pt}
					\centering
					\includegraphics[height=3.25cm]{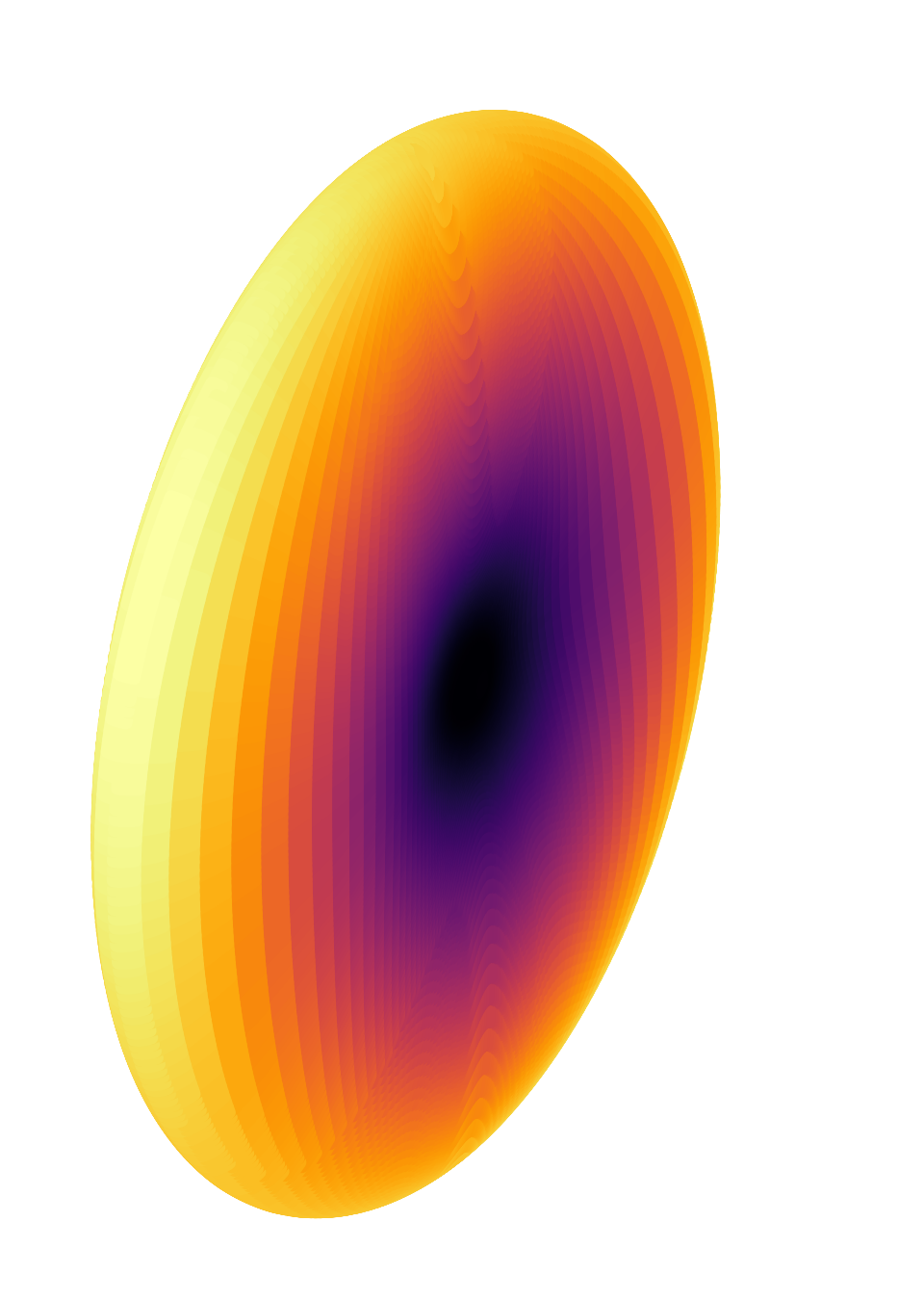}
					\caption{Lamellar}\label{fig:SpinodoidLamellarES}
				\end{subfigure}
				\hfill
				\begin{subfigure}[b]{110pt}
					\centering
					\includegraphics[height=3.25cm]{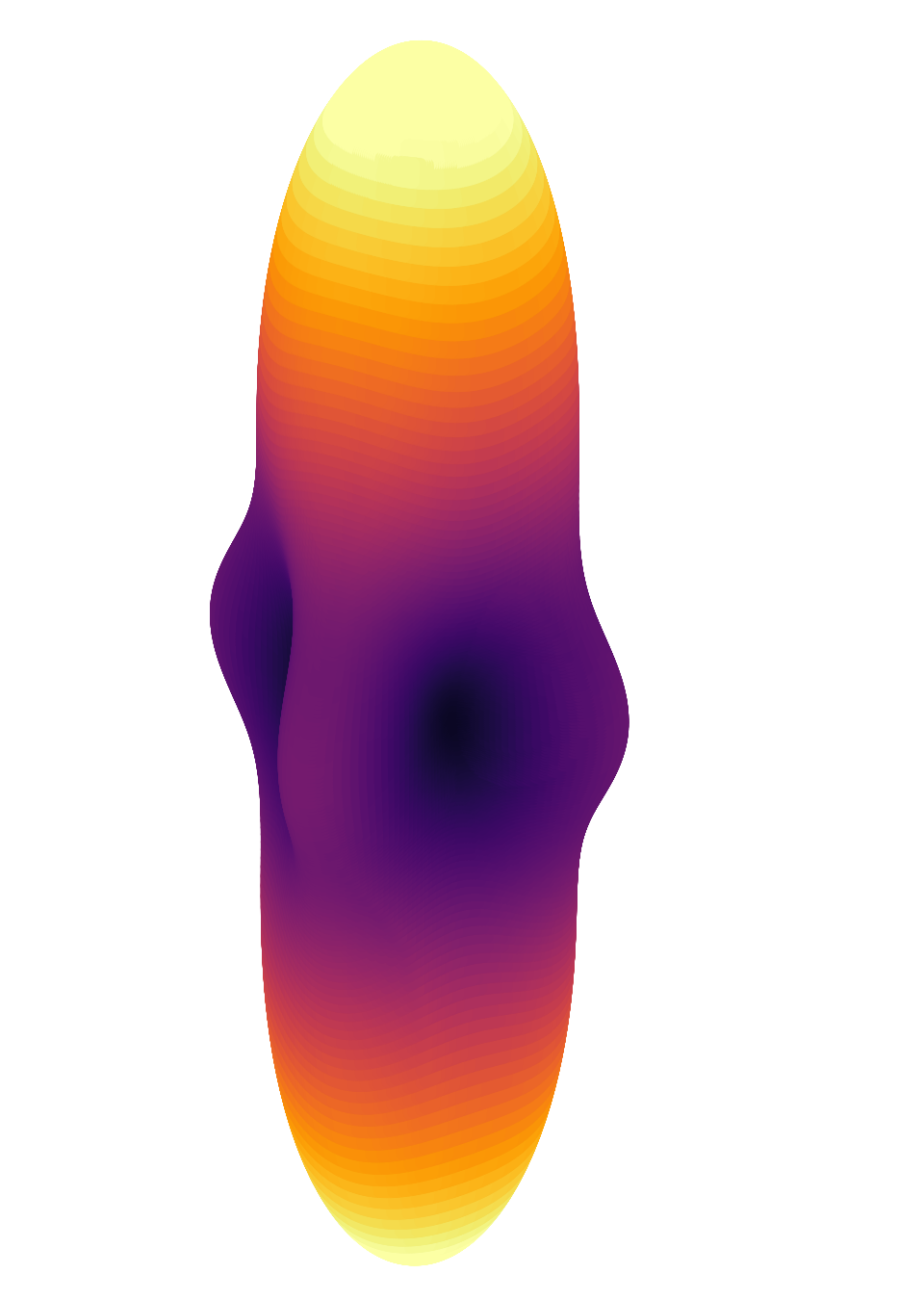}
					\caption{Columnar}\label{fig:SpinodoidColumnarES}
				\end{subfigure}
				\hfill
				\begin{subfigure}{45pt}
					\centering
					\includegraphics[height=3.00cm]{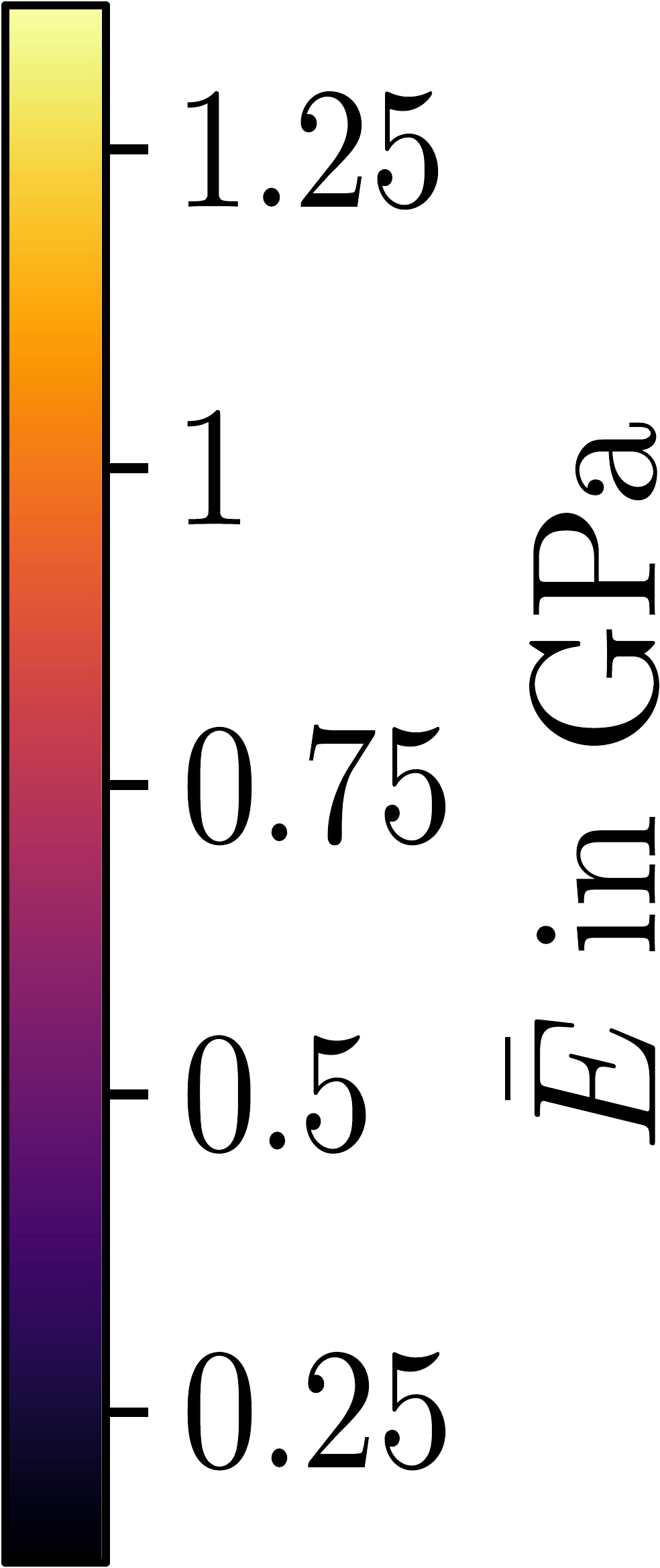}
					\vspace{20pt}
				\end{subfigure}
    
				\caption{Characteristic examples of spinodoid structures. The morphological angles $\bm\uptheta$ are given in the subcaptions (a)-(d). In (e)-(h) the elastic surfaces of a corresponding representative exemplar, i.e., the directional Young's moduli are given, visualizing the anisotropy class.}\label{fig:Spinodoids}
			\end{figure*}

            Kumar et al.~\cite{kumar2020} extended this isotropic Gaussian random field approach to anisotropy. By restricting the wave directions
            \begin{multline}
                \ve{n}_i \sim \mathcal{U}_{\mathcal{\tilde{S}}^2} \,,\; \mathcal{\tilde{S}}^2 := \bigl\{ \ve{k}\in S^2 \mid \left(\left| \ve{k}\cdot\ \hat{\ve{e}}_1 \right| > \cos{\uptheta_1} \right) \\ 
                \lor \left(\left| \ve{k}\cdot\ \hat{\ve{e}}_2 \right| > \cos{\uptheta_2} \right) 
                \lor \left(\left| \ve{k}\cdot\ \hat{\ve{e}}_3 \right| > \cos{\uptheta_3} \right) \bigl\} \:,
            \end{multline}
            to lie within cones of a certain angle $\uptheta_i$ around the respective Cartesian basis vector $\hat{\ve{e}}_i$, anisotropic structures can be generated as shown in~\autoref{fig:Spinodoids}.
            Choosing an appropriate threshold concentration\footnote{The Gauss error function is defined as $\text{erf}(z):=\dfrac{2}{\sqrt{\pi}} \int^z_0 \text{e}^{-t^2} \text{d}t$.} $\varphi_\text{thresh}:=\sqrt{2}\text{erf}^{-1}\left(2v_\text{f}-1\right)$ to meet the desired  volume fraction $v_\text{f}$ of the solid phase, the binary spinodoid structure $\ve{S}$ is obtained as 
            \begin{equation}
                \ve{S}(\ve{x}) := 
                \begin{cases}
                    1 & \text{if} \qquad \varphi(\ve{x})\leq \varphi_\text{thresh} \\
                    0 & \text{else}
                \end{cases} \quad .
            \end{equation}

            Using this approach, spinodoid structures with a large morphological variety can be generated. Keeping the volume element length $l_\text{VE}$, the resolution of the equidistant spatial discretization, i.e., the number of voxels $n_\text{voxel}$, the wavenumber $\beta$ and the number of waves $n_\text{wave}$ constant, the spinodoids are statistically fully described by their three morphological angles
            \begin{equation*}\label{eq:theta}
                \uptheta_i \in \mathcal{S}_\uptheta \quad \text{with} \quad \: i\in \left\{ 1,2,3 \right\} \:,
            \end{equation*}
            the volume fraction
            \begin{equation*}\label{eq:vf}
                v_\text{f} \in \mathcal{S}_{v_\text{f}} \: ,
            \end{equation*}
            and three rotational angles  
            \begin{equation*}\label{eq:phi}
                \upphi_i \in \mathcal{S}^i_\upphi \quad \text{with} \quad \: i\in \left\{ 1,2,3 \right\} \:,
            \end{equation*}
            describing a subsequent intrinsic rotation around the $z$-, $y$- and $x$-axis. The spaces of these quantities
            \begin{gather}
                \mathcal{S}_\uptheta := \{0\}\cup\left\{\vartheta\in\mathbb{R} \,\bigg|\, \frac{\pi}{12}\leq\vartheta\leq\frac{\pi}{2}\right\} \:,\label{eq:thetaSpace}\\
                \mathcal{S}_{v_\text{f}} :=\{ x \in \mathbb{R} \mid 0.3 \leq x\leq 0.8\}\:, \quad \text{and}\label{eq:VfSpace}\\
                \mathcal{S}^i_\upphi := 
                \begin{cases}
                    \left\{\vartheta\in\mathbb{R}\mid 0 \leq \vartheta \leq 2\pi\right\} & \text{if} \quad i\in \{1,3\} \\
                    \left\{\vartheta\in\mathbb{R}\mid 0 \leq \vartheta \leq \pi\right\} & \text{if} \quad i\in\{2\}
                \end{cases} \label{eq:phiSpace}\:,
            \end{gather}
            are defined such that connected domains are ensured and such that a rotation in each 3D spatial direction is possible. The state $\bm\uptheta=\bm{0}$, where all morphological angles are equal to zero, is not admissible.
            
            There are four borderline types of spinodoid morphologies that are controlled by the morphological angles $\bm{\uptheta}$:
            \begin{enumerate}[label=(\textit{\roman*})]
                \item Columnar spinodoids as depicted in~\autoref{fig:SpinodoidColumnar} with one stiff direction, see \autoref{fig:SpinodoidColumnarES}, for which a single morphological angle is $\SI{15}{\degree}$ and the remaining are equal to zero.
                \item Lamellar spinodoids as depicted in~\autoref{fig:SpinodoidLamellar} with two stiff directions, see \autoref{fig:SpinodoidLamellarES}, for which two angles are $\SI{15}{\degree}$ and the remaining is equal to zero.
                \item Cubic spinodoids as depicted in~\autoref{fig:SpinodoidCubic} with three stiff directions, see \autoref{fig:SpinodoidCubicES}, for which all angles are $\SI{15}{\degree}$.
                \item Isotropic spinodoids as depicted in~\autoref{fig:SpinodoidIsotropic} with equally stiff directions, see \autoref{fig:SpinodoidIsotropicES}, for which at least one angle is $\SI{90}{\degree}$.
            \end{enumerate}
            
            Summarizing the descriptors in one quantity, the total descriptor of a spinodoid
            \begin{multline}\label{eq:ThetaSpace}
                \T:=\left(\bm{\uptheta}^\intercal \; v_\text{f} \; \bm{\upphi}^\intercal \right) \in \mathcal{S}_\Uptheta\:, \\
                \mathcal{S}_\Uptheta := \mathcal{S}_\uptheta \times \mathcal{S}_\uptheta \times \mathcal{S}_\uptheta \times \mathcal{S}_{v_\text{f}} \times \mathcal{S}^1_\upphi \times \mathcal{S}^2_\upphi \times \mathcal{S}^3_\upphi
            \end{multline}
            is defined. This serves as input to the structure-property linkages, as described in the next subsection.

		\subsection{Gaussian process regression}\label{subsec:GPR}
            A structure-property linkage that not only maps the descriptor $\T$ to the expected value of a single property $\text{P}(\T)$ but also to a measure of uncertainty, more specifically to a distribution over property values ${p=p\left( \text{P} \right)}$, is a prerequisite for employing Bayesian optimization.
            The standard choice is a Gaussian process. It is defined as a stochastic process, i.e., a sequence of random variables, that has normally distributed finite dimensional marginal distributions. The reader is kindly referred to, e.g.,~\cite{quadrianto2010}, for a detailed introduction to Gaussian processes. Hereafter, a brief summary is given.
            
            A Gaussian process is a non-parametric, probabilistic model that is defined by a mean function $\mu(\T)$ and a covariance or kernel function $k(\T, \T')$, from which the mean array 
            \begin{equation*}
                \text{M}_i := \left( \mu\left(\ind{i}\T\right),\, \dots\,, \mu\left(\ind{N}\T\right) \right), \quad i \in \{1,\dots,n_\text{k}\}
            \end{equation*}
             and covariance matrix 
             \begin{equation*}
                \text{K}_{ij} := k(\ind{i}\T, \ind{j}\T), \quad i,j \in \{1,\dots,n_\text{k}\}  
             \end{equation*}
             are computed for all known $n_\text{k}$ descriptor-property pairs in the data set. The notation
             \begin{equation*}
                 f_\mathcal{GP} (\bullet) \sim \mathcal{GP} \left( \mu(\bullet), k(\bullet,\bullet)\right)
             \end{equation*}
            is commonly used to indicate that the function distribution $f_\mathcal{GP}$ is drawn from a Gaussian process $\mathcal{GP}$.

            The descriptors of the input data set are normalized to the interval $[0, 1]$ for each coordinate and the properties are standardized to have zero mean and a unit variance for each coordinate to form the $n_\text{k}$ known data points
            \begin{equation*}
                \Tilde{\mathcal{D}}=\left\{\left(\ind{i}\Tilde{\T}, \ind{i}\Tilde{\text{P}}\right)\right\}_{i=1}^{n_\text{k}} \:.
            \end{equation*}

            Gaussian process regression can be utilized to predict the distribution of the property 
            \begin{equation}
                \Tp^\star \mid \mathcal{\Tilde{D}}, \TT^\star \sim \mathcal{N}\left( \ve{M}^\star, \ve{K}^\star \right)
            \end{equation}
            at any location $\TT^\star$ in the descriptor space based on the data set $\Tilde{\mathcal{D}}$ by sampling from a normal distribution $\mathcal{N}$ with mean vector $\ve{M}^\star$ and covariance matrix $\ve{K}^\star$. Further details are given in~\autoref{app:GP}, where also $\ve{M}^\star$ and $\ve{K}^\star$ are defined in~\autoref{eq:GP_pred_mean} and~\autoref{eq:GP_pred_cov}.

            For the present work, the established Matérn kernel function
            \begin{equation}\label{eq:Matern}
              k_{\text{Matern}}(\T, \T') := \frac{2^{1 - \nu}}{\Gamma(\nu)}
              \left( \sqrt{2 \nu} d \right)^{\nu} K_\nu \left( \sqrt{2 \nu} d \right)
            \end{equation}
            is chosen, where $d := (\mathbf{\T} - \mathbf{\T'})^\top \bm{\uplambda}^{-2} (\mathbf{\T} - \mathbf{\T'})$ is the distance between $\T$ and $\T'$ scaled by the lengthscale parameter $\bm{\uplambda}$, ${\nu \in \left\{0.5, 1.5, 2.5 \right\}}$ is a smoothness parameter, and $K_\nu$ is the modified Bessel function\footnote{$K_\nu(x):=\int_0^\infty \mathrm{e}^{-x\cosh{t}}\cosh(\nu t)\,\text{d}t$}, and $\Gamma$ is the gamma function\footnote{$\Gamma(\nu):=\int_0^\infty t^{\nu-1}\mathrm{e}^{-t}\,\text{d}t\:,\nu\in\left\{x\in\mathbb{R}|x>0\right\}$}.
            To allow for a better fitting, a parameter $s_\text{K}$ is introduced to scale the covariance matrix
            \begin{equation*}
                \text{K}_{ij} := s_\text{K}k_\text{Matern}(\T_i, \T_j) \: .
            \end{equation*}
            These hyperparameters $\bm{\uplambda}$ and $s_\text{K}$ are inferred from the data by maximizing the logarithmic marginal likelihood.
        
		\subsection{Bayesian optimization}\label{subsec:BO}
	       	The goal of the optimization step is to find points in the descriptor space which should be added to an initial data set, such that these points correspond to good properties or help decreasing the overall uncertainty of property predictions. 

\begin{figure}[t]
				\centering
				\includegraphics[width=1\columnwidth]{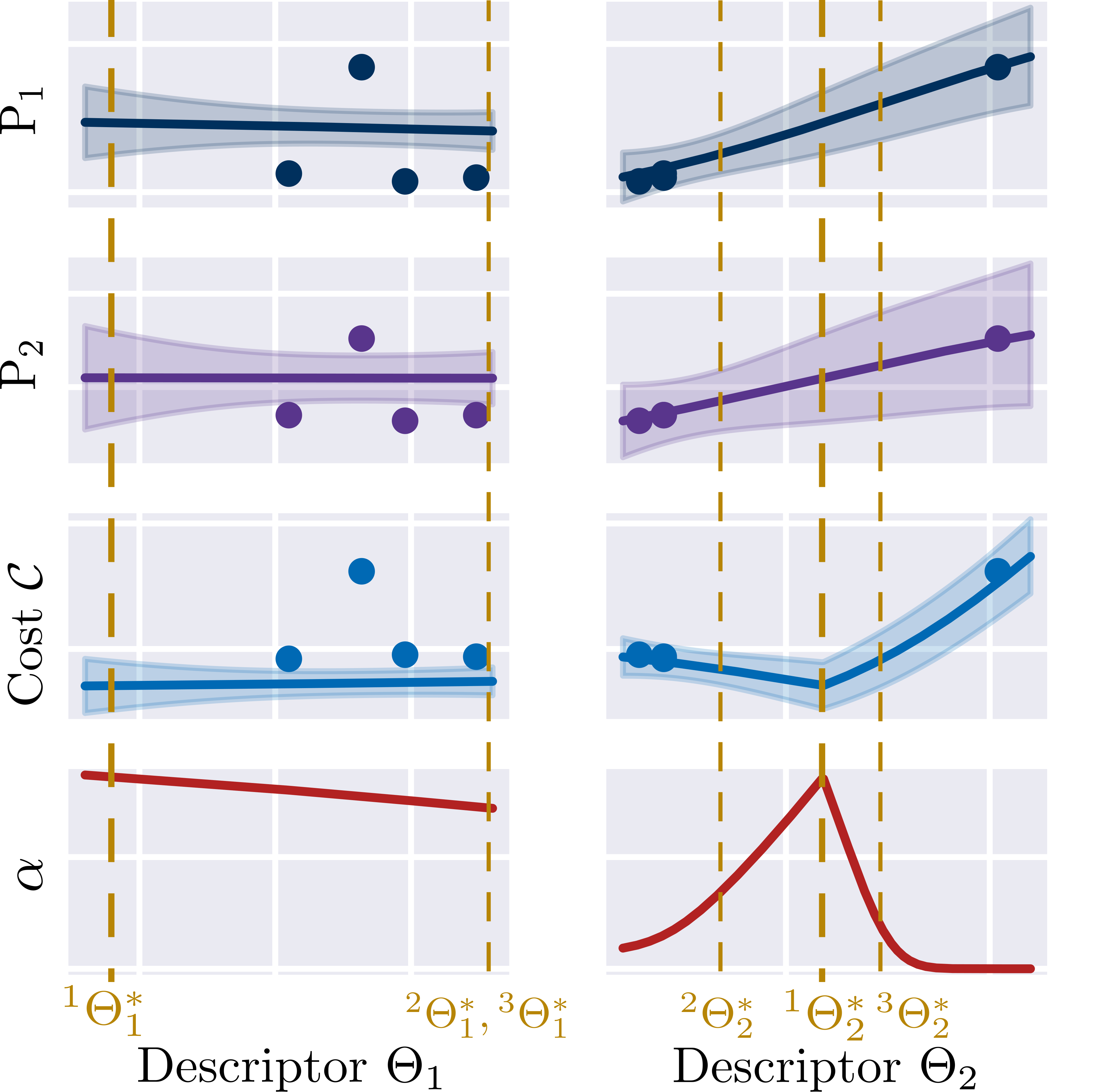}
				\caption{Gaussian process for two examplary properties $\text{P}_1$ and $\text{P}_2$, distribution of the cost $\C\left(\text{P}_1, \text{P}_2\right)$ and acquisition value $\alpha\left(\C\right)$ in a 2D descriptor space. New candidate descriptors $\ind{i}\te\Theta^*$ are selected based on the acquisition value.}\label{fig:GP+BO}
			\end{figure}

            The optimization is conducted in three main steps, as illustrated in~\autoref{fig:GP+BO}.
            \begin{enumerate}[label=(\textit{\roman*})]
                \item The structure-property linkage is set up. Using Gaussian process regression, the distribution of each property $\text{P}_i$ in the descriptor space is modeled.
                \item The distribution of the cost $\C(\ve{P})$ is computed from the property distributions.
                \item The acquisition function $\alpha(\C)$ is maximized to propose suitable descriptors of candidate structures $\ind{i}\T^*$.
            \end{enumerate}

            In this work, the cost 
            \begin{multline}
                \label{eq:Cost}
                \C\left(\ve{P}\right) := - \sum_{i=1}^{n_\text{m}} w_i \dfrac{\text{P}_i}{\text{P}_i^\text{ref}} \\
                + \sum_{i=n_\text{m}+1}^N w_i \left[\exp{ \left( c \biggl\langle\dfrac{\text{P}_i}{\text{P}^\text{thresh}_i} - 1 \biggl\rangle \right)} -1\right]
            \end{multline}
            is formulated rather flexibly, where $w_i$ denote scalar weights, $\text{P}_i^\text{ref}$ and $\text{P}_i^\text{thresh}$ denote reference and threshold values, respectively, $c=2$ is a constant factor, $n_\text{m}$ denotes the number of main properties that are to be maximized, $N$ denotes the total number of properties, and $\langle\bullet\rangle$ denote Macaulay brackets\footnote{Also known as ramp function or in machine learning as rectified linear unit (ReLU).}. As the optimization is formulated as a minimization problem, a low cost is aimed at. This can be achieved by increasing the first summation term, i.e., by maximizing the $n_\text{m}$ main properties, and by minimizing the second penalizing summation term in keeping the properties below its threshold value.
            While~\autoref{eq:Cost} defines the design goal in finding structures of possible high or low properties, the formulation could be modified to aim at a specific desired property value $\text{P}^\text{des}$, e.g., by penalizing the distance to it.
            
            Instead of computing the cost $\C$ for all points in the data set and applying the Gaussian process regression on it directly, i.e., to model
            \begin{equation*}
                f_{\mathcal{GP}}: \mathcal S_{\Uptheta} \to \R, \quad \T \mapsto \C(\ve{P}) \:,
            \end{equation*}
            here, the properties are modeled
            \begin{equation*}
                \ve{f}_{\mathcal{GP}}:  \mathcal S_{\Uptheta} \to \R^N, \quad\T \mapsto \ve{P}
            \end{equation*}
            and then a cost distribution $\C\left( \ve{f}_{\mathcal{GP}} \left( \T \right) \right)$ is derived from them. Otherwise, the cost function would have to be inferred from the data, complicating the modeling in most cases and necessitating more data. 
            The cost distribution's mean
            \begin{equation*}
                \mu\left(\C \left(\T\right) \right) = \C\left( \mu\left(\text{f}_{\mathcal{GP},i} \left(\T\right) \right),\, \cdots,\,\mu\left(\text{f}_{\mathcal{GP},N}\left(\T\right) \right)  \right)
            \end{equation*}
            is computed as the cost of the properties' means. The covariance matrix of the cost
            \begin{equation*}
                \ve{K}_{\C} := \sum_{i=1}^{N} w_i^2\ve{K}_{\text{P}_i}
            \end{equation*}
            is here defined as the weighted sum of the properties' covariance matrices.

            The acquisition function $\alpha$ guides the optimization as candidates 
            \begin{equation*}
                \T^* := \underset{{\T \in \mathcal{S}_{\Uptheta}}}{\arg\max}\, \alpha\left(\C\left( \ve{f}_{\mathcal{GP}} \left( \T \right) \right)\right)
            \end{equation*}
            are chosen by maximizing the acquisition value. The expected improvement function
            \begin{equation}\label{eq:acq_ana}
                \alpha(\T) := \mathbb{E}\left[ \langle\C^*-\C\rangle  \mid \C \sim \mathcal{N}\left( \mu(\T), \sigma^2(\T) \right) \right]
            \end{equation}
            is selected, where $\mu$ is the mean value and $\upsigma$ the standard deviation. $\C^*$ denotes the lowest cost in the data set.
			
			For a faster design process, the proposition of multiple candidates per iteration is desired. For this purpose, so-called batched acquisition functions can be used as readily available in the Python package \textit{botorch}~\cite{balandat2020a}. The main idea is not to solve the maximization problem analytically but by a sampling strategy.
            A Monte-Carlo approach is utilized to approximate~\autoref{eq:acq_ana} by
            \begin{equation}\label{eq:acq_MC}
                \alpha(\T) \approx \dfrac{1}{N} \sum_{i=1}^N \langle \C^*-\C_i \rangle\,, \quad \C_i \sim \mathcal{N}\left( \mu(\T), \sigma^2(\T) \right)
            \end{equation}
            through sampling $N$ cost values from its distribution at $\T$. 
            To allow for an optimization of a batch of $q$ descriptors 
            \begin{equation*}
                \underline{\T} := \left( \T_1, \dots, \T_q \right)\:,
            \end{equation*}
            \autoref{eq:acq_MC} is extended to
            \begin{equation}\label{eq:qEI}
                \alpha_\text{b}(\underline{\T}) \approx \dfrac{1}{N} \sum_{i=1}^N \max_{j=1,\dots,q}{  \langle\C^* - \underline{\C}_{ij}\rangle}\:,
            \end{equation}
            where $\underline{\C}_i$ is sampled from the joint posterior distribution.

	\section{Results}\label{sec:Result}
        The inverse design procedure proposed in~\autoref{subsec:Concept} is applied to three examples of increasing dimensions of the design space. Starting from three different data sets as given in~\autoref{tab:DataSets}, the design goal is to maximize the effective Young's modulus $\bar{E}$ in certain directions.
        \begin{table*}
            \centering
            \caption{Descriptor settings for the data sets of the present examples. $\mathcal{U}$ denotes the uniform distribution.}
            \fontsize{7.5}{8}\selectfont
            \begin{tabularx}{\textwidth}{c|c|ccc|c|ccc}
                \toprule
                \normalsize i & \normalsize $n_\text{init}$ & \normalsize $\uptheta_1$ & \normalsize $\uptheta_2$ & \normalsize $\uptheta_3$ & \normalsize $v_\text{f}$ & \normalsize $\upphi_1$ & \normalsize $\upphi_2$ & \normalsize $\upphi_3$ \\
                \midrule
                I   & 4               & $\sim\mathcal{U}_{[\SI{30}{\degree},\SI{90}{\degree}]}$ & $\SI{0}{\degree}$             & $\SI{0}{\degree}$             & $\sim\mathcal{U}_{[0.3,0.45]\cup [0.65,0.8]}$ & $\SI{0}{\degree}$ & $\SI{0}{\degree}$ & $\SI{0}{\degree}$ \\
                II  & 16              & $\sim\mathcal{U}_{[\SI{30}{\degree},\SI{90}{\degree}]}$ & $\sim\mathcal{U}_{[\SI{30}{\degree},\SI{90}{\degree}]}$ & $\sim\mathcal{U}_{[\SI{30}{\degree},\SI{90}{\degree}]}$ & $\sim\mathcal{U}_{[0.3,0.45]\cup [0.65,0.8]}$ & $\SI{0}{\degree}$ & $\SI{0}{\degree}$ & $\SI{0}{\degree}$ \\
                III & 16              & $\SI{0}{\degree}$             & $\sim\mathcal{U}_{[\SI{30}{\degree},\SI{90}{\degree}]}$ & $\sim\mathcal{U}_{[\SI{30}{\degree},\SI{90}{\degree}]}$ & $\sim\mathcal{U}_{[0.3,0.45]\cup [0.65,0.8]}$ & $\sim\mathcal{U}_{[\SI{0}{\degree},\SI{360}{\degree})}$ & $\sim\mathcal{U}_{[\SI{0}{\degree},\SI{75}{\degree}]\cup[\SI{115}{\degree},\SI{180}{\degree})}$ & $\sim\mathcal{U}_{[\SI{0}{\degree},\SI{360}{\degree})}$ \\
                \bottomrule
            \end{tabularx} 
            \label{tab:DataSets}
        \end{table*}

        The section starts by explaining details on the implementation, followed by three design examples.

		\subsection{Implementational details}\label{subsec:Implementation}
            The methods explained in~\autoref{sec:Method} are implemented in an autonomous High Performance Computing (HPC) framework as the iterative inverse design approach is computationally demanding. The loop itself is controlled by a bash script that is run on a HPC cluster using the Batch-System SLURM (Simple Linux Utility for Resource Management).

            The 3D spinodoid structures are generated using the open-source \textit{MATLAB}~\cite{matlab2022b} toolbox \textit{GIBBON}~\cite{moerman2018}. The wave\-number is set to ${\beta=15\pi}$ and the number of waves to\linebreak${n_\text{wave}=1\,000}$. Following the results of a study on a suitable size of the volume elements as explained in~\autoref{app:Convergence} and a trade-off with performance, the volume element length is set to $l_\text{VE}=1$ at a spatial discretization into $n_\text{voxel}=64^3$ voxels.  

            The numerical simulations for obtaining effective Young's moduli $\bar{E}$ are conducted utilizing \textit{DAMASK}~\cite{roters2019} as described in~\autoref{app:DAMASK}. The material parameters are given in~\autoref{tab:material_parameters}. They are chosen to mimic polyvinyl chloride (PVC) for the solid part of the spinodoids.

            The optimization step is implemented in Python. The efficient \textit{PyTorch}-based~\cite{paszke2019} package \textit{botorch}~\cite{balandat2020a} is used as it allows for the usage of GPUs. Running on an NVIDIA A100, a single optimization step of reading and pre-processing the data, setting-up the Gaussian process model, and doing the Bayesian optimization takes approximately $\SI{2.5}{\minute}$ for a set of 400 points\footnote{The initial data sets are much smaller ($n_\text{init}\leq16$). The additional data points are generated in silico during the iterative design process.} (6D descriptor, 2D property). The smoothness parameter of the Matérn kernel function~(\autoref{eq:Matern}) is set to $\nu=2.5$. The batched expected improvement acquisition function~(\autoref{eq:qEI}) is approximated by $N=10\,240$ Monte-Carlo samples.
            The design loop is run for a fixed number of iterations.
            
            \autoref{tab:Resources} lists an approximation of the necessary resources required for the design loop of the present examples. 
            \begin{table}[t]
                \centering
                \caption{Approximate resources for the steps of the inverse design procedure for the present examples.}
                
                \small
                \begin{tabular}{l|lrr}
                     \toprule
                     Step& Hardware & Time & Resources \\
                     \midrule
                     Optimization& 1x NVIDIA A100 & $\SI{2.5}{\minute}$ & $\SI{2.5}{GPUmin}$ \\
                     Generation& 1x Intel Xeon 8470 & $\SI{2}{\minute}$ & $\SI{2}{CPUmin}$ \\
                     Simulation& 8x Intel Xeon 8470 & $\SI{3}{\minute}$ & $\SI{24}{CPUmin}$ \\
                     \bottomrule
                \end{tabular}
                \label{tab:Resources}
            \end{table}
            Assuming $40$ iterations, the proposition of $5$ candidates per iteration and $3$ simulations to obtain three properties per candidate, the resources $R$ accumulate to
            \begin{align*}
                R_\text{GPU} &= 40\cdot\SI{2.5}{GPUmin} = \SI{1.7}{GPUh} \\
                R_\text{CPU} &= 40\cdot 5\cdot \left( \SI{2}{} + 3\cdot \SI{24}{} \right) \si{CPUmin} = \SI{247}{CPUh} \: .
            \end{align*}

		\subsection{Inverse design in a 2D design space}\label{subsec:2D}
			The goal of the first example is to design a spinodoid structure such that it possesses the maximum possible effective Young's modulus in $z$-direction $\bar{E}_z$ while targeting a volume fraction below $v_\text{f}^\text{tresh}=0.55$. This is formulated in the cost function as
            \begin{equation}\label{eq:CostEz}
                \C := -\dfrac{\bar{E}_z}{\SI{2}{\giga\pascal}} + 2\left(\exp{\left( 2 \biggl\langle\dfrac{v_\text{f}}{0.55}-1\biggl\rangle \right)} -1\right)\:.
            \end{equation}
   
            The space
            \begin{equation*}
                \mathcal{S}_{\Theta^\text{2D}} := \mathcal{S}'_{\uptheta} \times \mathcal{S}_{v_\text{f}}
            \end{equation*}
            of the reduced 2D descriptor
            \begin{equation*}
                \T^\text{2D} := \left(\uptheta_1 \; v_\text{f} \right) \in \mathcal{S}_{\Theta^\text{2D}}
            \end{equation*}
            is limited to the first morphological angle $\uptheta_1$ and the volume fraction $v_\text{f}$, where
            \begin{equation}\label{eq:thetaSpacePrime}
                \mathcal{S}'_{\uptheta} := \left\{\vartheta\in\mathbb{R}\mid\frac{\pi}{12}\leq\vartheta\leq\frac{\pi}{2}\right\}
            \end{equation}
            and $\mathcal{S}_{v_\text{f}}$ is defined according to~\autoref{eq:VfSpace}. Limiting the angle $\uptheta_i$ to the interval as defined in~\autoref{eq:thetaSpacePrime} facilitates the optimization problem as this space is continuous in contrast to the total possible space for $\uptheta_i$ as defined in~\autoref{eq:thetaSpace} and the inadmissible state of $\bm\uptheta=\bm{0}$ is avoided. 
            
            The remaining descriptors are set to the values as given in~\autoref{tab:DataSets}. It can been seen that the initial data set consists of only four descriptor-property pairs. The free descriptors $\uptheta_1$ and $v_\text{f}$ are randomly sampled from a uniform distribution. The distributions are restricted to not contain the expected outcome of the inverse design process, that is a lamellar structure ($\uptheta_1=\SI{15}{\degree}$) of $v_\text{f}=0.55$ volume fraction.

\begin{figure*}[t]
				\label{fig:ResComp}
				\begin{subfigure}[b]{150pt}
					\centering
					\includegraphics[height=3.5cm]{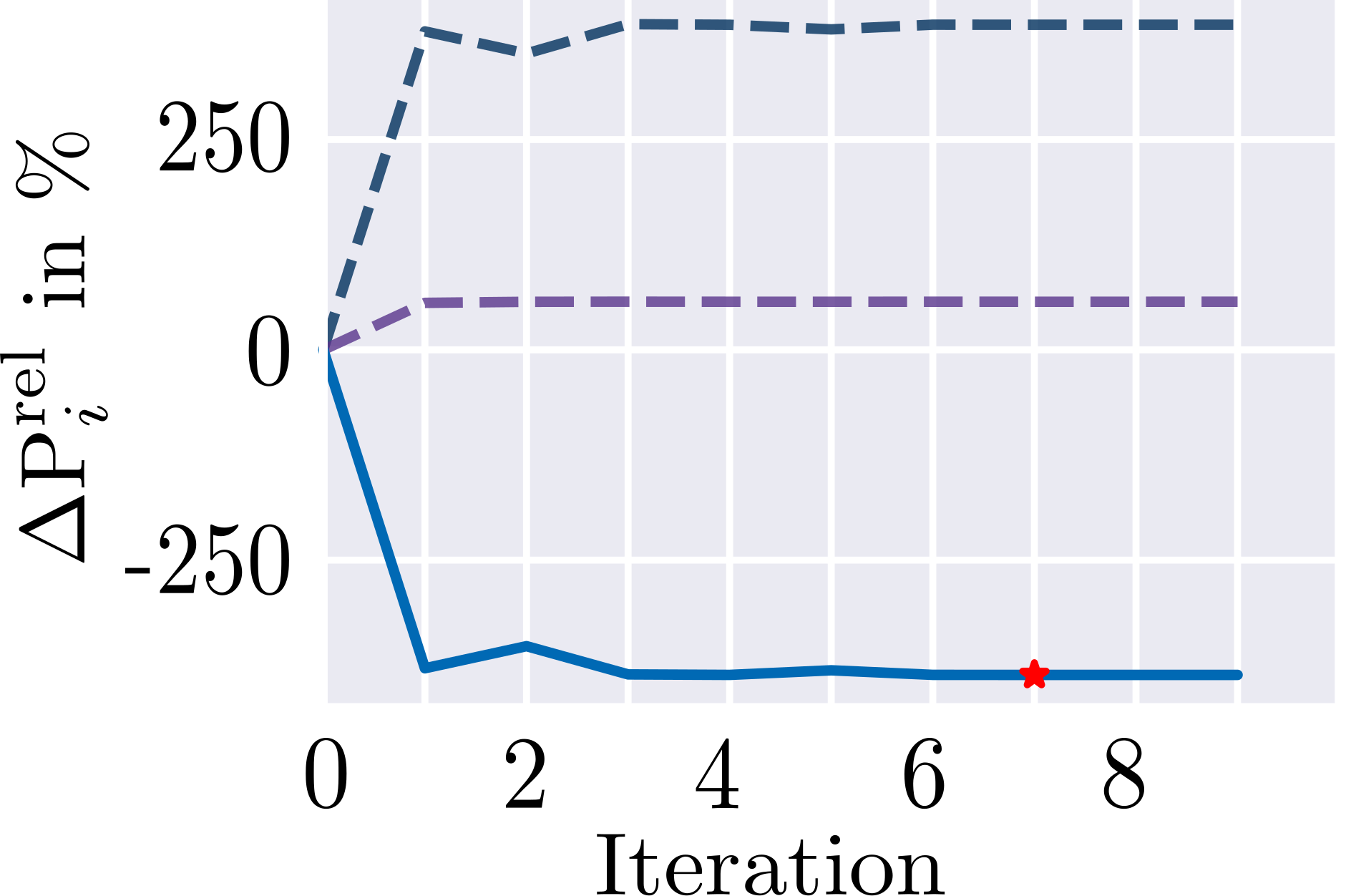}
					\caption{2D example (\autoref{tab:DataSets} I)}\label{fig:ResCompEvo2D}
				\end{subfigure}
				\hfill
				\begin{subfigure}[b]{140pt}
					\centering
					\includegraphics[height=3.5cm]{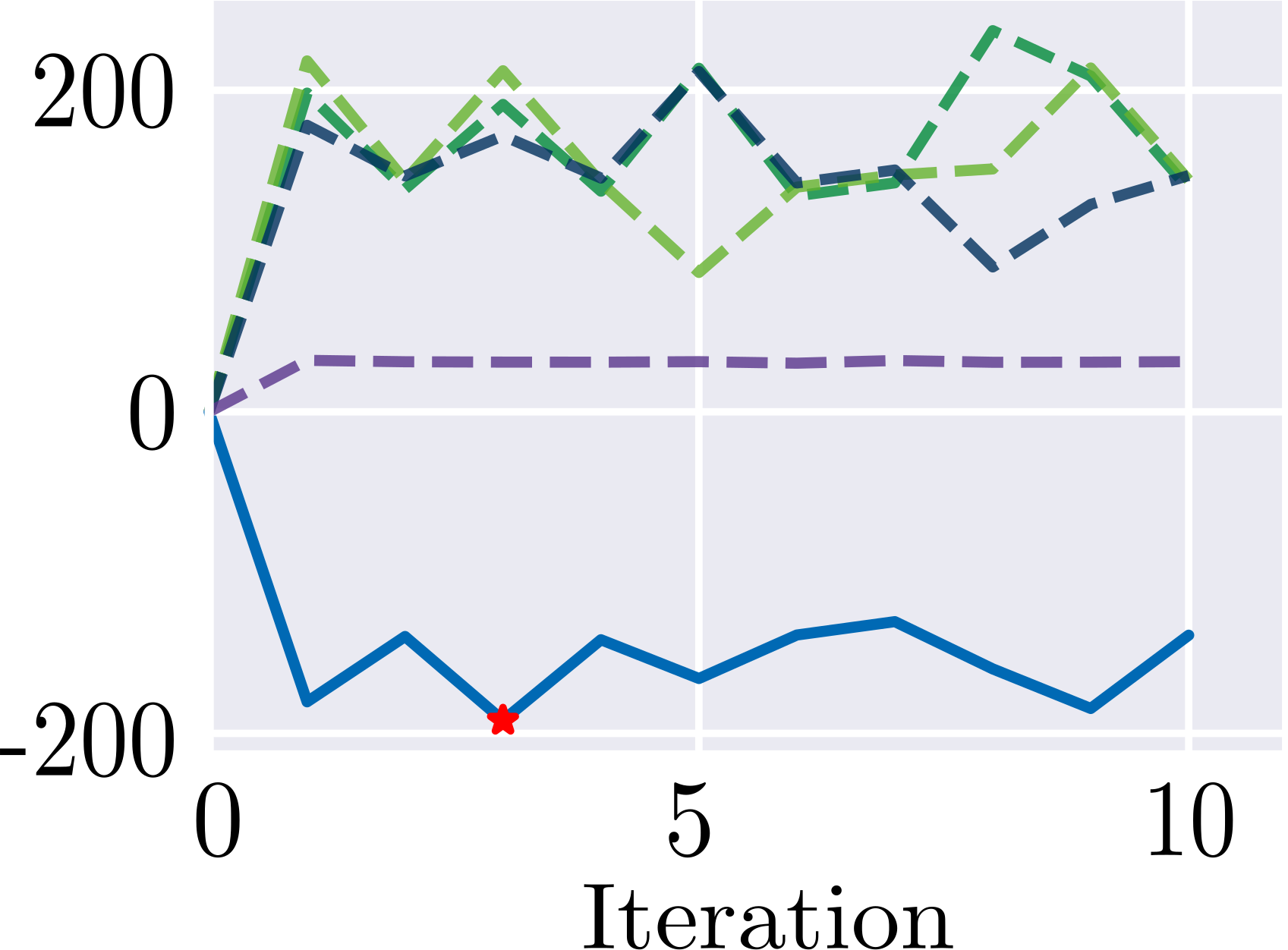}
					\caption{4D example (\autoref{tab:DataSets} II)}\label{fig:ResCompEvo4D}
				\end{subfigure}
				\hfill
				\begin{subfigure}[b]{140pt}
					\centering
					\includegraphics[height=3.5cm]{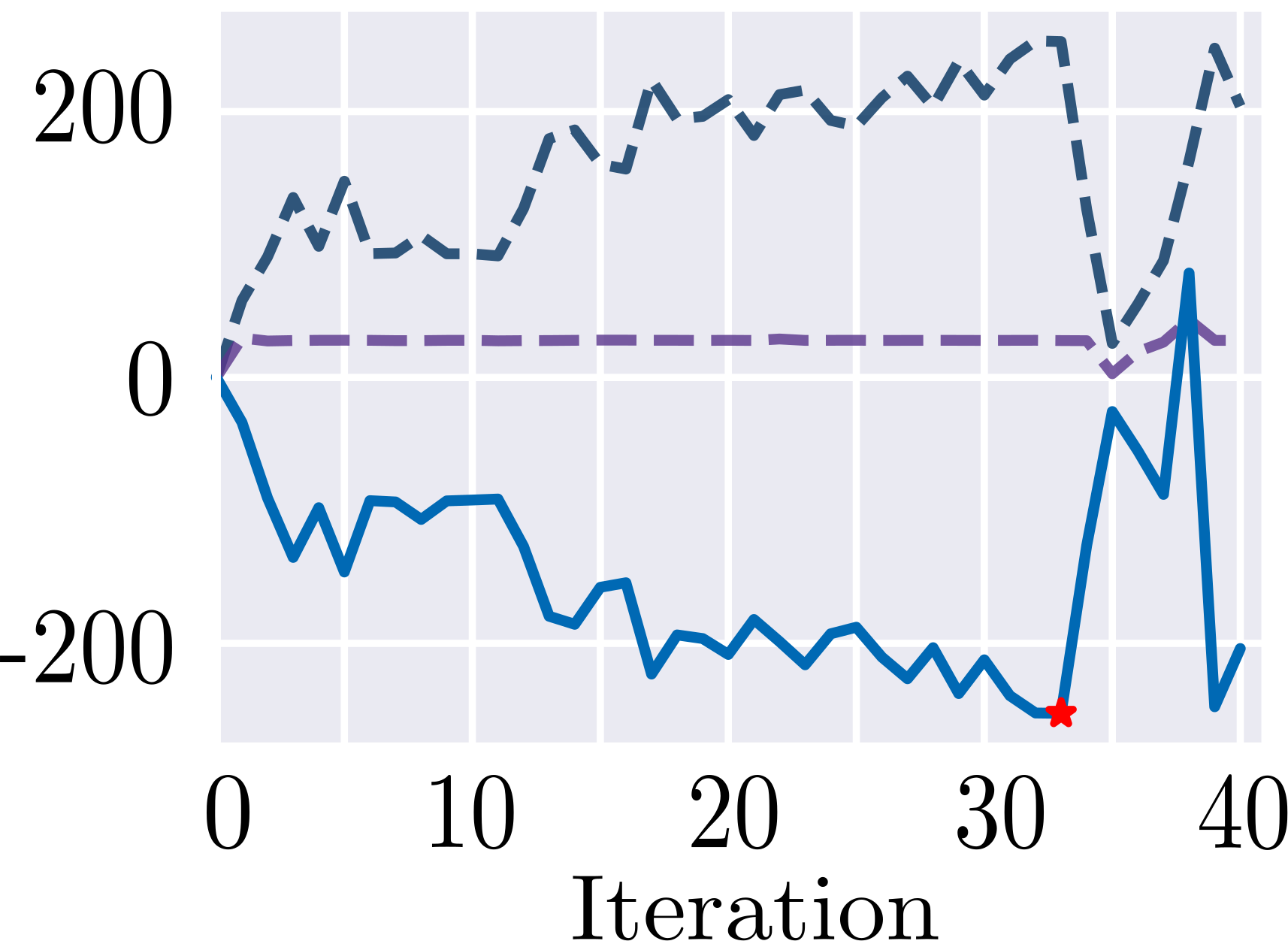}
					\caption{6D example (\autoref{tab:DataSets} III)}\label{fig:ResCompEvo6D}
				\end{subfigure}
				\hfill
				\begin{subfigure}{45pt}
					\centering
					\includegraphics[height=3.5cm]{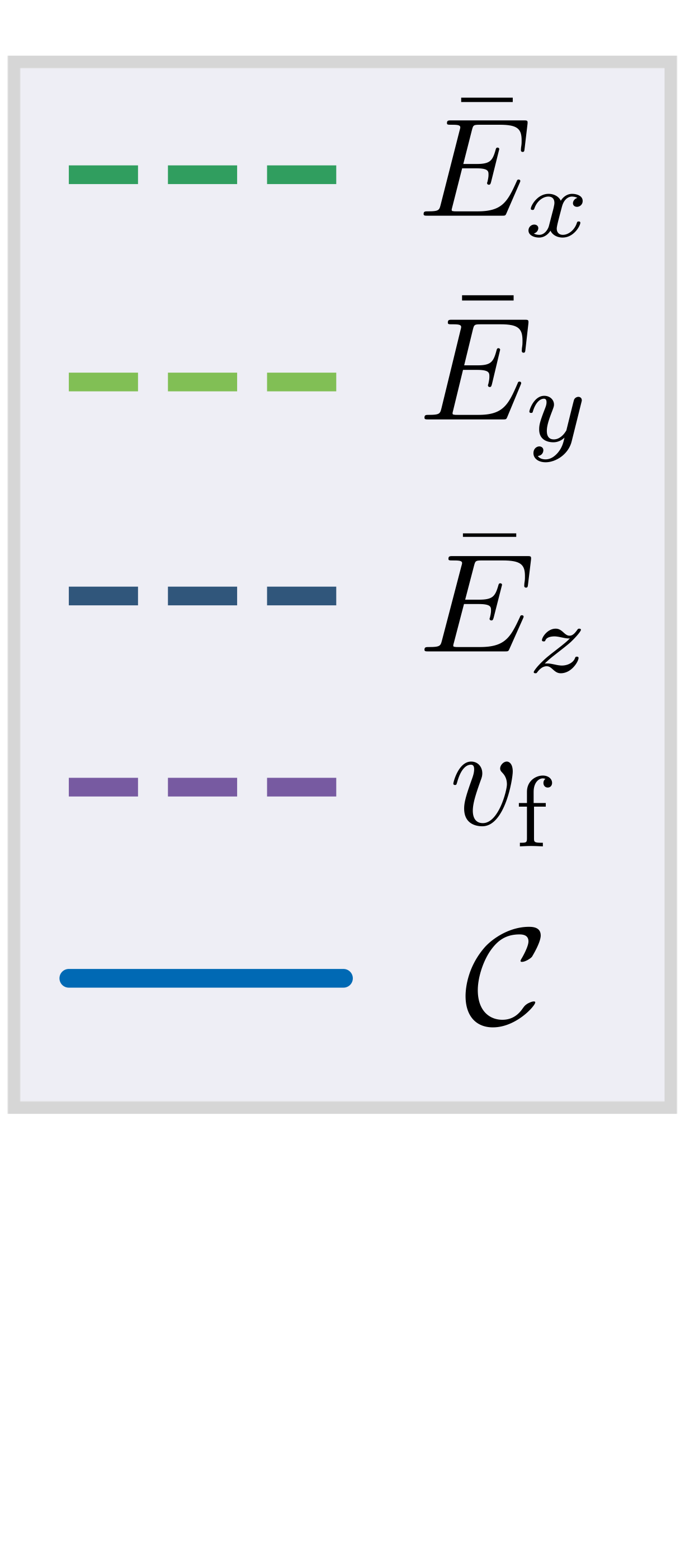}
					\vspace{8pt}
				\end{subfigure}
			
				\begin{subfigure}[b]{150pt}
					\centering
					\includegraphics[height=3.25cm]{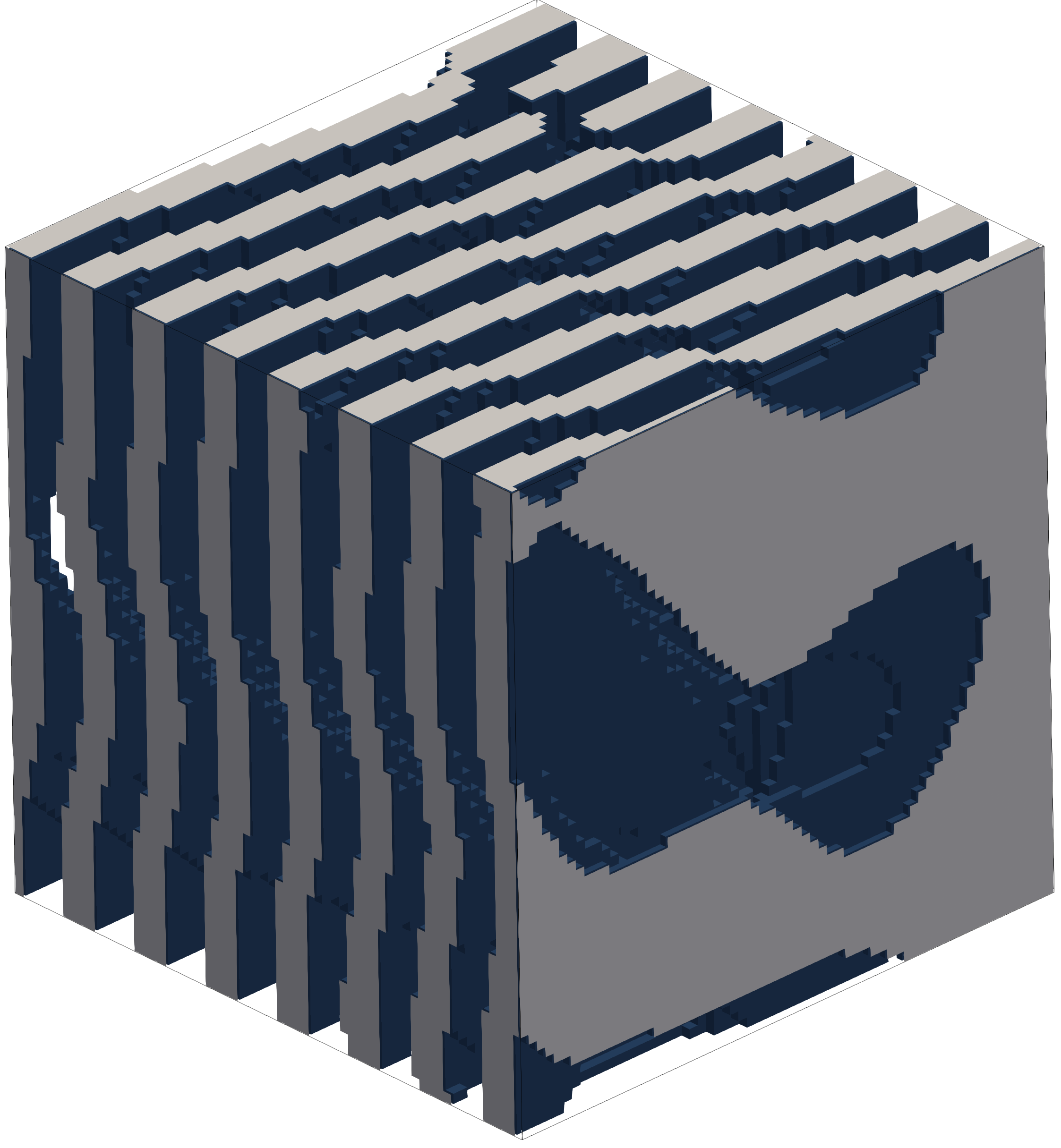}
					\caption{2D example (\autoref{tab:DataSets} I)}\label{fig:ResCompSVE2D}
				\end{subfigure}
				\hfill
				\begin{subfigure}[b]{140pt}
					\centering
					\includegraphics[height=3.25cm]{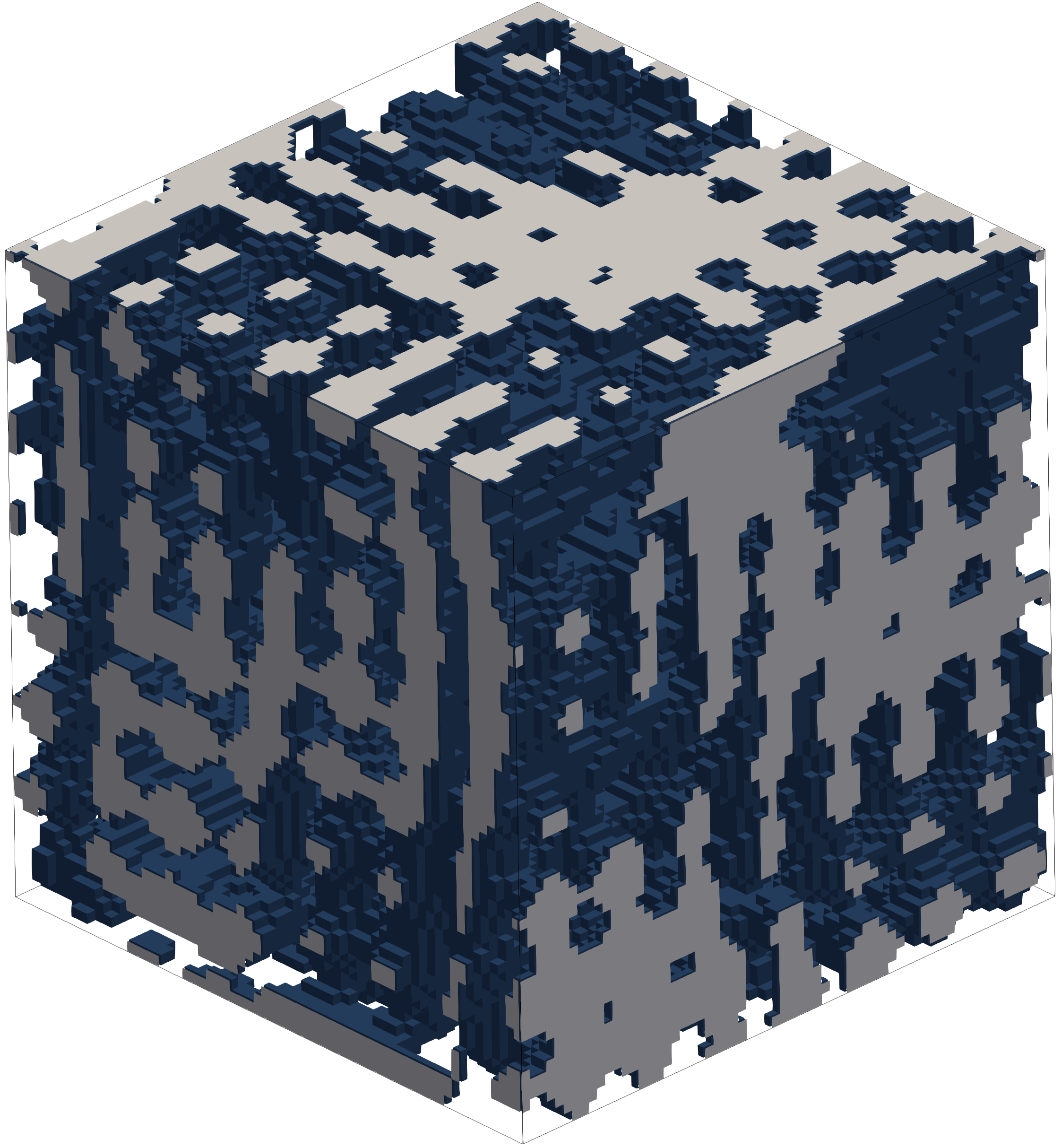}
					\caption{4D example (\autoref{tab:DataSets} II)}\label{fig:ResCompSVE4D}
				\end{subfigure}
				\hfill
				\begin{subfigure}[b]{140pt}
					\centering
					\includegraphics[height=3.25cm]{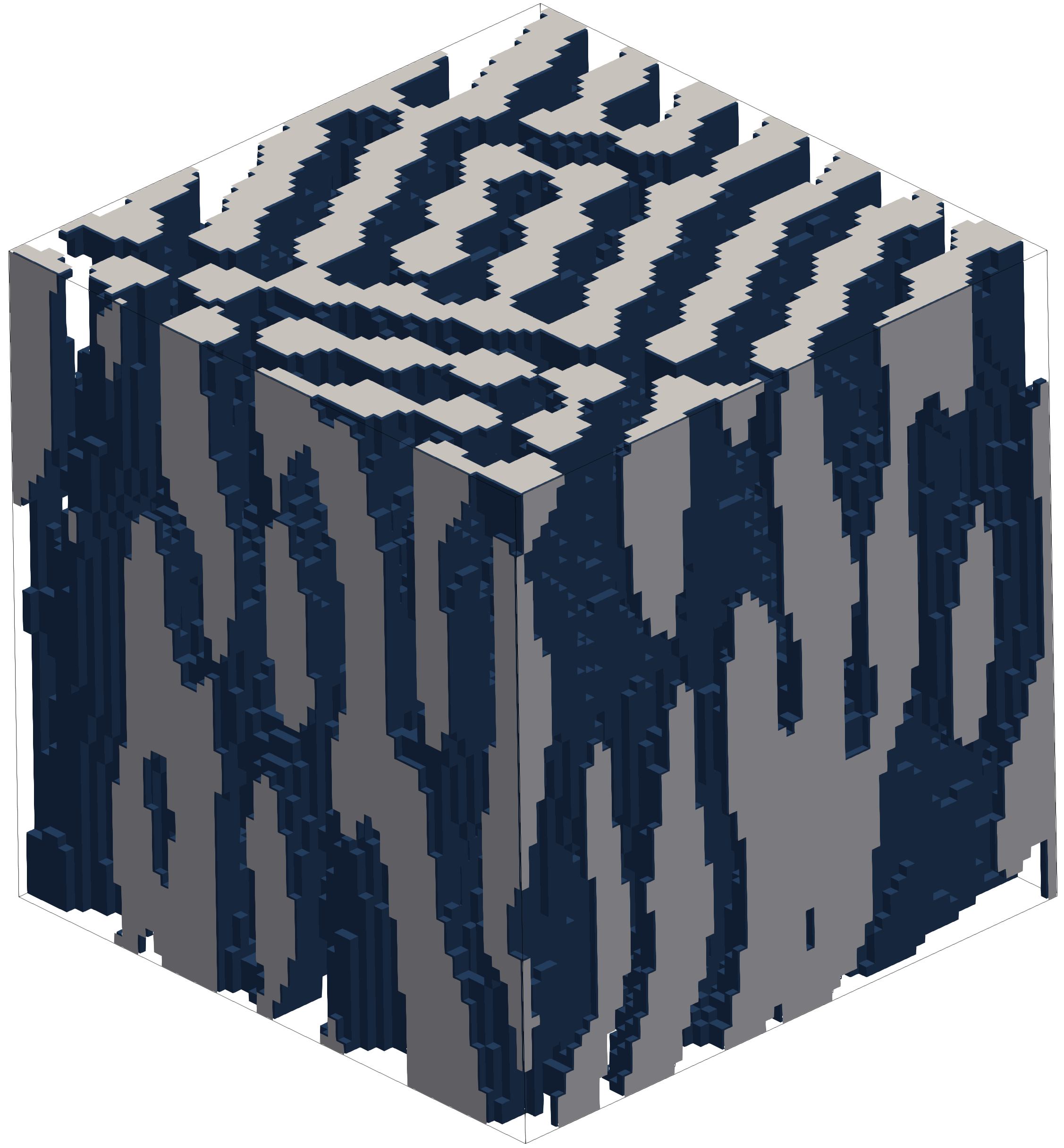}
					\caption{6D example (\autoref{tab:DataSets} III)}\label{fig:ResCompSVE6D}
				\end{subfigure}
				\hfill
				\begin{subfigure}{45pt}
					\centering
					\includegraphics[height=1.4cm]{figures/coordinate_system.pdf}
					\vspace{20pt}
				\end{subfigure}
				\caption{Evolution of the relative deviation of the properties $\Delta\text{P}^\text{rel}$ (\autoref{eq:RelProp}) of the structure of the lowest cost $\C$ referred to the initial structure of lowest cost over iterations for the corresponding example (a)-(c). The best spinodoid structures found are shown in (d)-(f) for the corresponding example.}
			\end{figure*}

			\autoref{fig:ResCompEvo2D} shows the evolution of the properties including the cost over iterations of the augmentation loop.
            For each iteration, the values of the structure with the lowest cost are plotted. The quantities are given as relative deviation
            \begin{equation}\label{eq:RelProp}
                \Delta \text{P}^\text{rel}_i := \dfrac{\text{P}_i - \text{P}_i^\text{init*}}{\text{P}_i^\text{init*}}
            \end{equation}
            from the initial data set's best structure's property $\text{P}^\text{init*}$.
			As can be seen from the line graph, already after the first iteration, a structure of significantly less cost is found. The overall best structure is found after seven iterations, indicated by the red star. The spinodoid is depicted in~\autoref{fig:ResCompSVE2D}. It has the expected descriptor
            \begin{equation*}
                \T^\text{2D*} := \left(\uptheta_1^* \: v^*_\text{f} \right) \quad
                \text{with}\quad \uptheta_1^* = \SI{15.0}{\degree},\, v^*_\text{f} = 0.55\:.
            \end{equation*}
            The Young's modulus $\bar{E}_z$ increases significantly, while the volume fraction $v_\text{f}$ converges to its threshold value. 

\begin{figure}[t]
				\centering
				\includegraphics[width=\columnwidth]{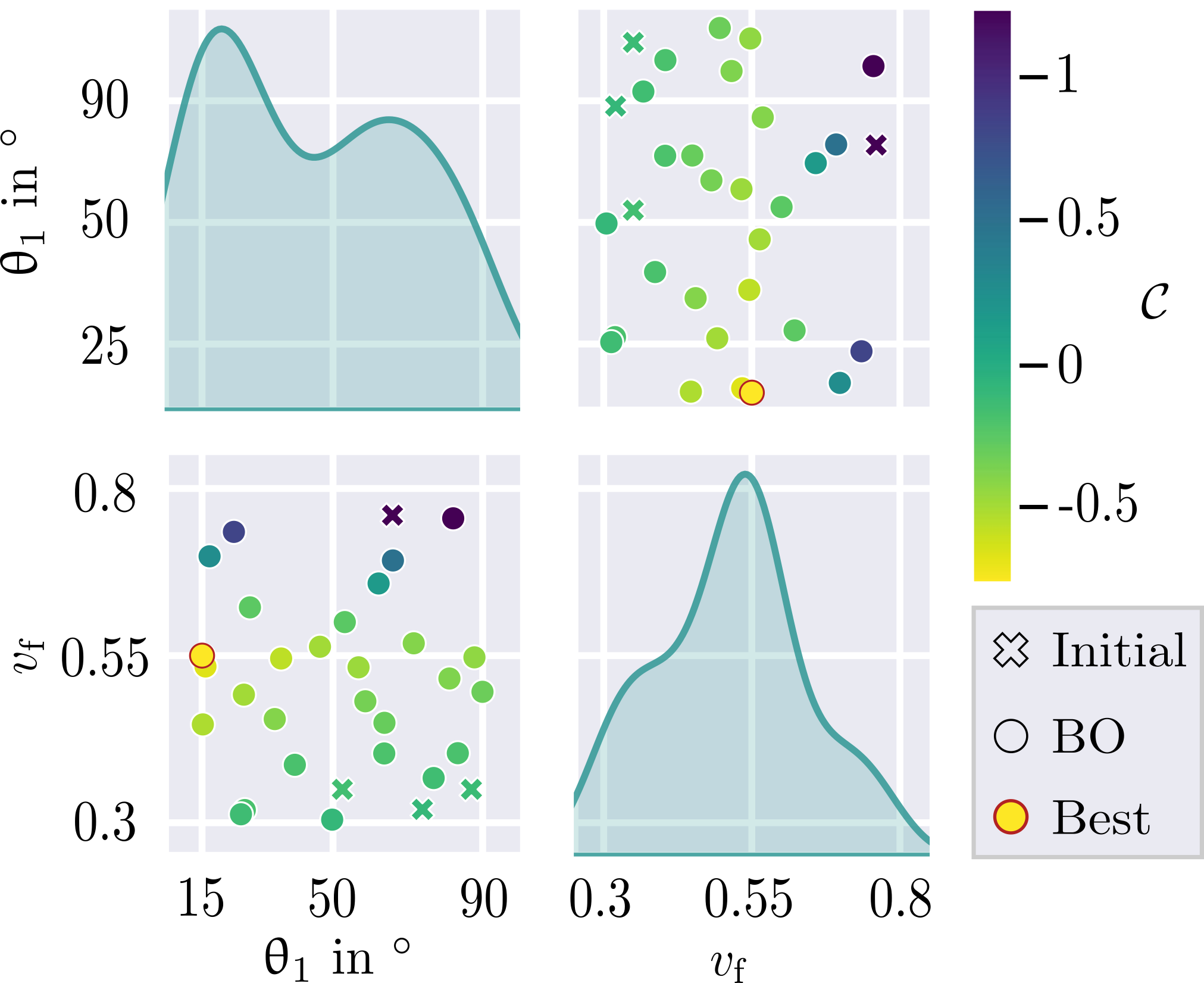}
				\caption{Correlation plot of both dimensions of the 2D design space of data set I (\autoref{tab:DataSets}) after nine iterations. The data points added during the Bayesian optimization (BO) are indicated as circles. All points are color-coded by their cost $\C$ as defined in~\autoref{eq:CostEz}.}\label{fig:2D_pairgrid}
			\end{figure}

            \autoref{fig:2D_pairgrid} illustrates the distribution of the data points in the descriptor space after nine iterations. The diagonal plots show the relative distribution of a descriptor coordinate as kernel density estimate.
            It can be seen that most of the data points added during the design process tend to the optimal respective value of $\uptheta^*_1=\SI{15}{\degree}$ and $v^*_\text{f}=0.55$. The off-diagonal scatter plots visualize potential cross correlations between two descriptors coordinates. Descriptors from the initial data set are indicated by crosses while the points added during the Bayesian optimization are represented by circles. The markers' color encode the cost $\C$ ranging from yellow for a desired low cost to purple for high cost. The red circle represents the best structure found.
            
            Three conclusions can be drawn from the data:
            \begin{enumerate}[label=(\textit{\roman*})]
                \item The best structure is found at $\uptheta_1=\SI{15}{\degree}$ and $v_\text{f}=0.55$ within 9 iterations, exactly meeting the expected value.
                \item A smaller angle $\uptheta_1$ correlates with lower cost at a constant volume fraction.
                \item The cost increases more sharply for increasing volume fractions above $v_\text{f}^\text{thresh}=0.55$ compared to the increase for decreasing volume fractions below this value.
            \end{enumerate}

            Considering the formulation of the cost function from \autoref{eq:CostEz}, the results are plausible and underline the potential of the present design approach. The morphology and volume fraction of the solid material influence the effective Young's modulus. In general, increasing the volume fraction increases the stiffness. This explains the decreasing cost for increasing volume fractions up to the threshold value $v_\text{f}^\text{thresh}$. The following decrease is explained by the penalization in the cost function. 
            Continuous structural features in a certain direction stiffen the material in this particular direction. That is the reason for low cost at small $\uptheta_1$ as this angle controls the transition from an isotropic morphology for large values towards a lamellar one for small values.

		\subsection{Inverse design in a 4D design space}\label{subsec:4D}
            The second example aims at a structure of high effective Young's moduli in all three Cartesian directions $x$, $y$, and $z$. Again, a volume fraction of $v_\text{f}^\text{tresh}=0.55$ should not be exceeded. Hence, the cost function
            \begin{multline}\label{eq:CostEall}
                \C := -\dfrac{\bar{E}_x}{\SI{2}{\giga\pascal}}-\dfrac{\bar{E}_y}{\SI{2}{\giga\pascal}}-\dfrac{\bar{E}_z}{\SI{2}{\giga\pascal}}\\
                + 2\left(\exp{\left( 2 \biggl\langle\dfrac{v_\text{f}}{0.55}-1\biggl\rangle \right)} -1\right)
            \end{multline}
            is formulated.

            The reduced descriptor for this example is limited to four coordinates
            \begin{multline}
                \T^\text{4D} := \left(\uptheta_1\; \uptheta_2\; \uptheta_3\; v_\text{f}\right) \in \mathcal{S}_{\Theta^\text{4D}}\:, \\
                \mathcal{S}_{\Theta^\text{4D}}:=\mathcal{S}'_{\uptheta}\times\mathcal{S}'_{\uptheta}\times\mathcal{S}'_{\uptheta}\times \mathcal{S}_{v_\text{f}} \:,
            \end{multline}
            with the spaces defined in \autoref{eq:thetaSpacePrime} and \autoref{eq:VfSpace}. Considering all coordinates of the morphological angle array $\bm{\uptheta}$ allows for the largest possible variations of the morphology.
            The remaining descriptors are set to constant values as given in~\autoref{tab:DataSets}. The initial data set consists of 16 descriptor-property pairs.

            As can be seen from the line graph in~\autoref{fig:ResCompEvo4D}, already after three iterations the best structure is found. Shown are the relative properties as defined in~\autoref{eq:RelProp}. The best structure of each iteration is of nearly $v_\text{f}=0.55$ volume fraction, while there are some variations in the directional Young's moduli. For the best structure at iteration 3 all three moduli are rather high, forming the best trade-off between all directions. In contrast, at iteration 5, only $\bar{E}_x$ and $\bar{E}_z$ are relatively high, thereby forcing $\bar{E}_y$ to a lower value.
            \autoref{fig:ResCompSVE4D} shows the optimally designed spinodoid with the descriptor
            \begin{multline*}
                \T^\text{4D*} := \left(\uptheta_1^*\; \uptheta_2^*\; \uptheta_3^*\; v^*_\text{f} \right)\\
                \quad \text{with} \quad \uptheta_1^* = \SI{15.0}{\degree},\, \uptheta_2^* = \SI{15.0}{\degree},\, \uptheta_3^* = \SI{15.0}{\degree},\, v^*_\text{f} = 0.55\:.
            \end{multline*}
            It has pronounced continuous features in the three spatial directions. They provide the desired cubic anisotropy.

\begin{figure*}[t]
				\centering
				\includegraphics[width=0.9\textwidth]{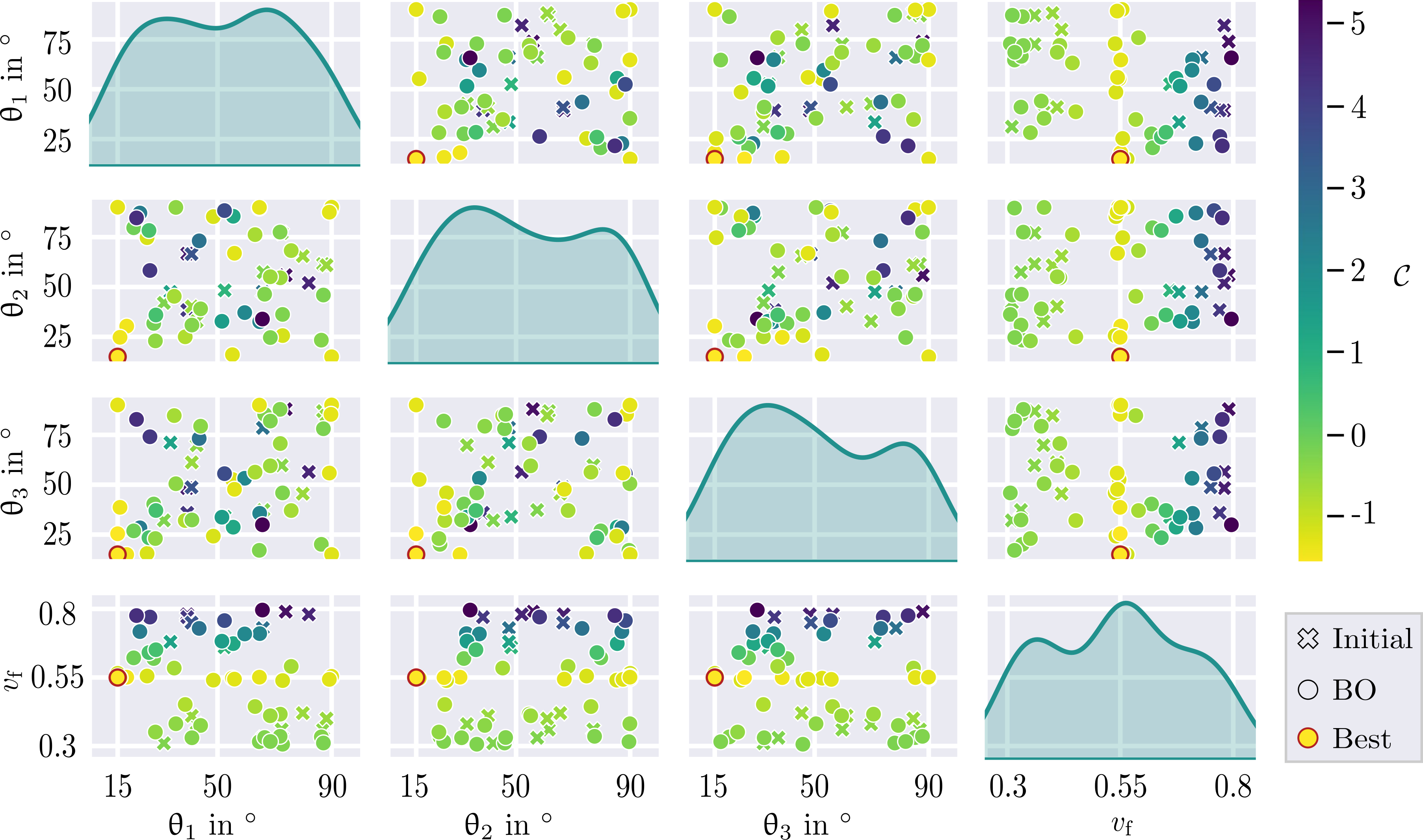}
				\caption{Correlation plot of all dimensions of the 4D design space of data set II (\autoref{tab:DataSets}) after 10 iterations. The data points added during the Bayesian optimization (BO) are indicated as circles. All points are color-coded by their cost $\C$ as defined in~\autoref{eq:CostEall}.}\label{fig:4D_pairgrid}
			\end{figure*}

            The occupation of the 4D descriptor space after ten iterations is visualized in~\autoref{fig:4D_pairgrid}.
            The distribution plots on the diagonal show no pronounced concentrations to specific values for the angles $\uptheta_i$ and only a slight peak at the threshold value for the volume fraction $v_\text{f}$.
            Considering the cost encoded in the colors, it is apparent that all structures of low cost (yellow) can be found at a volume fraction near the threshold value as expected. As explained for the 2D example in the preceding~\autoref{subsec:2D}, smaller volume fractions correspond to less low costs than higher ones as the latter are penalized in the cost function formulated in~\autoref{eq:CostEall}.
            
            Focusing on the correlation between volume fraction $v_\text{f}$ and the angles $\uptheta_i$, it can be concluded that the smallest possible value of $\uptheta_i=\SI{15}{\degree}$ is optimal as it is of lowest cost. The optimal structure, indicated by the red circle, is at exactly this location in the descriptor space.
            
            Another finding can be retrieved by examining the cross correlation plots of two angles: For both extreme values $\uptheta_i \in \left\{\SI{15}{\degree}, \SI{90}{\degree}\right\}$ significantly lower costs are observed than within this interval. This can be explained by the formation of isotropic spinodoid structures if one of the angles is at ${\uptheta_i=\SI{90}{\degree}}$. It is plausible that these isotropic structures are of low cost as they provide a decent trade-off between the stiffnesses in the desired three directions. However, the cubical structure at $\uptheta^*_i=\SI{15}{\degree}$ is of even lower cost as the material can be distributed such that the stiffness is increased in only three specific directions compared to all possible directions in the isotropic case.

            Designing spinodoid structures such that multiple properties are considered can be achieved by applying the proposed algorithm. The initial data set comprises 16 structures that are to be described in a 4D space, underlining the capability of the concept to operate on few data in spaces of multiple dimensions. To illustrate this further, another example in 6D is given in the following~\autoref{subsec:6D}.

		\subsection{Inverse design in a 6D design space}\label{subsec:6D}
		      A third inverse design process is exemplary conducted to examine the capability of the present design approach to handle a larger design space with few data and to handle ambiguities.
            For this reason, the reduced descriptor
            \begin{multline}
                \T^\text{6D} := \left(\uptheta_2\; \uptheta_3\; v_\text{f}\; \upphi_1\; \upphi_2\; \upphi_3 \right) \in \mathcal{S}_{\Theta^\text{6D}}\:, \\
                \mathcal{S}_{\Theta^\text{6D}}:=\mathcal{S}'_{\uptheta}\times\mathcal{S}'_{\uptheta}\times \mathcal{S}_{v_\text{f}} \times \mathcal{S}^1_\upphi \times \mathcal{S}^2_\upphi \times \mathcal{S}^3_\upphi \:,
            \end{multline}
            is constructed to comprise six coordinates. The spaces are defined in~\autoref{eq:thetaSpacePrime}, \autoref{eq:VfSpace}, and \autoref{eq:phiSpace}. 
            The descriptor now includes the rotational angles $\bm{\upphi}$, allowing for any possible rotation.
            
            For structures of specific symmetries, the representation of a rotation by 
            \begin{equation*}
               \bm{\upphi} \in \mathcal{S}^1_\upphi \times \mathcal{S}^2_\upphi \times \mathcal{S}^3_\upphi
            \end{equation*}
            is ambiguous, thus, posing a challenge to solving the design task.
            For incorporating the need of a rotation, the example is designed such that a high effective Young's modulus in $z$-direction $\bar{E}_z$ is aspired but the choice of the morphological angles $\uptheta_2$ and $\uptheta_3$ without any rotation does not allow for the formation of columnar features in $z$-direction. 
            To that end, $\uptheta_1=\SI{0}{\degree}$ is chosen. Now, the best possible morphology for a maximum unidirectional stiffening is a columnar structure at ${\uptheta_2=\uptheta_3=\SI{15}{\degree}}$. However, with this configuration and ${\bm{\upphi}=\ve{0}}$, the columns are aligned in $x$-direction. Consequently, a rotation such that the rotated $x$-axis aligns with the (unrotated) $z$-axis is a prerequisite for columnar features in $z$. Intuitively, a rotation of
            \begin{equation}\label{eq:DS6_rotation}
                \bm{\upphi} = \begin{pmatrix}
                    \SI{0}{\degree}\\
                    \SI{90}{\degree}\\
                    \SI{0}{\degree}
                \end{pmatrix}
            \end{equation}
            could be chosen, i.e., a rotation around the $y$-axis.
            The same formulation of the cost function as in~\autoref{eq:CostEz} is chosen. The number of initial descriptor-property pairs is restricted to 16 only as given in~\autoref{tab:DataSets}.

            \autoref{fig:ResCompEvo6D} supports the assumption that an optimization in 6D is more challenging as 33 iterations are necessary to find the best solution within a maximum of 40 iterations. In most iterations one of the five added structures is of lower cost then the previous ones. From iteration 2 to 12 the cost is stagnating as the Bayesian optimization does not propose better candidates. This indicates some kind of training or exploration interval in which more data is added to the data set such that the structure-property model is improved.
            After iteration 33 the cost increases sharply, hinting a possible exploration of different subspaces further away from the current optimum.

            To gain more insight on the evolution of the design process, intermediate results are summarized in~\autoref{tab:6D}.
            \begin{table*}[p]
                \centering
                \caption{Intermediate results of the design in 6D of data set III (\autoref{tab:DataSets}) for selected iterations $i$. Given are the descriptors $\bm{\uptheta}$, $v_\text{f}$, and $\bm{\upphi}$ as well as a visualization of the stiffness tensor $\mathbb{C}$ in terms of the elastic surface, the effective directional Young's modulus in $z$-direction $\bar{E}_z$, and the cost $\C$.}
                \label{tab:6D}
                \begin{tabular}{c|c|ccc|ccc}
                     $i$ & $\ind{i}\ve{S}$ & $\ind{i}\bm{\uptheta}$ & $\ind{i}v_\text{f}$ & $\ind{i}\bm{\upphi}$ & $\ind{i}\mathbb{C}$ & $\ind{i}\bar{E}_z$ & $\ind{i}\C$ \\
                     \toprule
                     0& \begin{tabular}{c}\includegraphics[height=69.4pt]{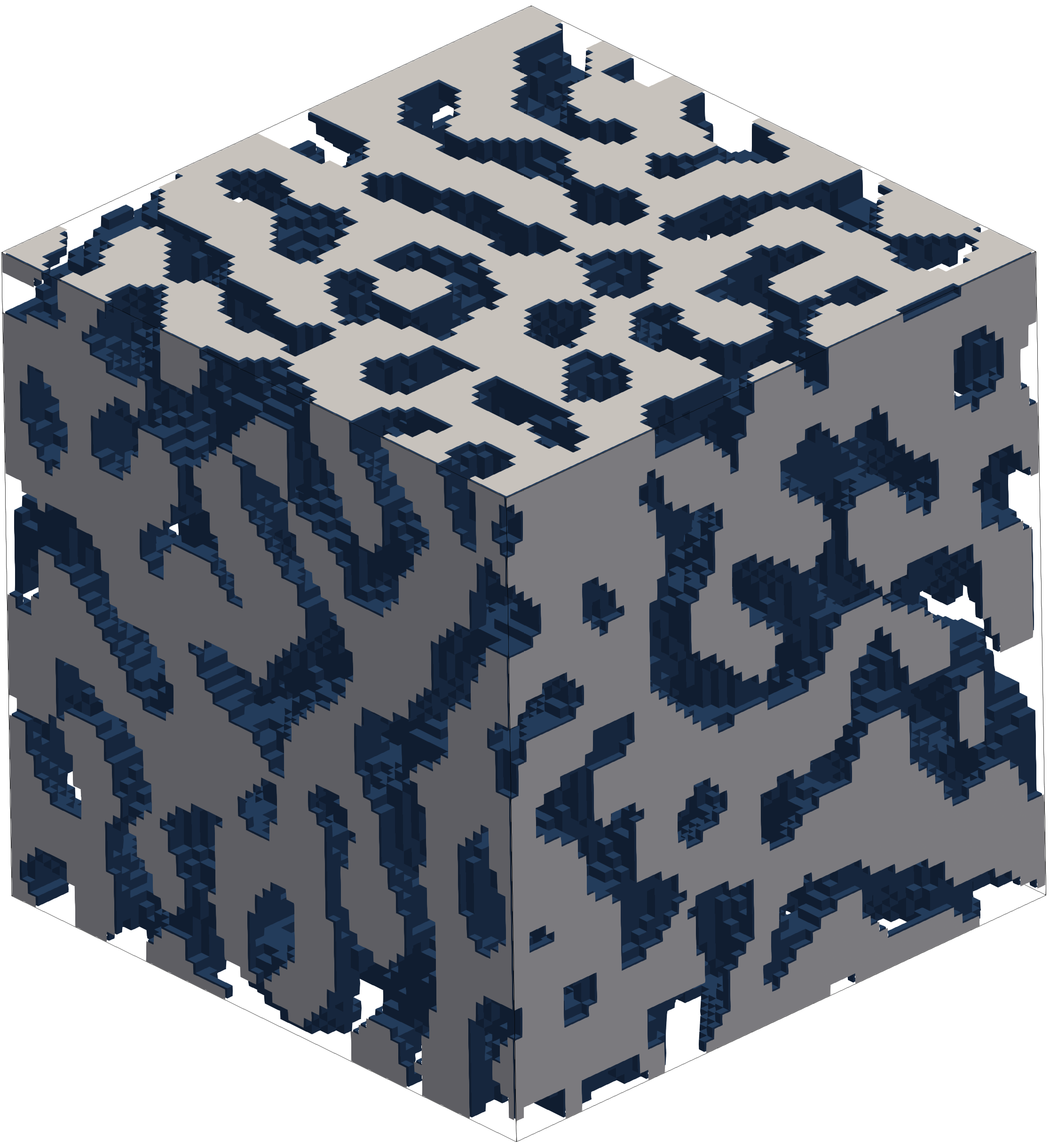} \includegraphics[height=30pt]{figures/coordinate_system.pdf}\end{tabular} & $\begin{pmatrix} \SI{0.0}{\degree} \\ \SI{66.1}{\degree} \\ \SI{30.3}{\degree} \end{pmatrix}$ & $0.43$ & $\begin{pmatrix} \SI{186.8}{\degree} \\ \SI{129.9}{\degree} \\ \SI{315.9}{\degree} \end{pmatrix}$ & \begin{tabular}{c} \includegraphics[height=30pt]{figures/coordinate_system.pdf}\includegraphics[height=69.4pt]{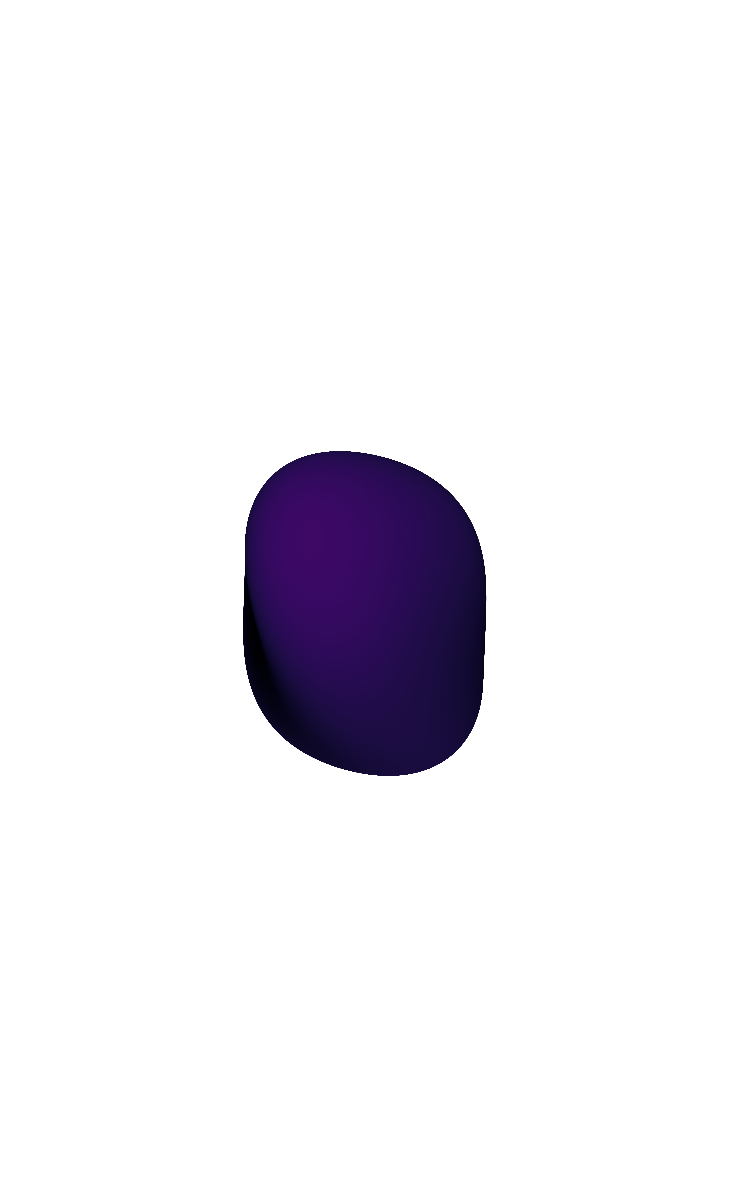}\end{tabular} \begin{tabular}{c}\includegraphics[height=55pt]{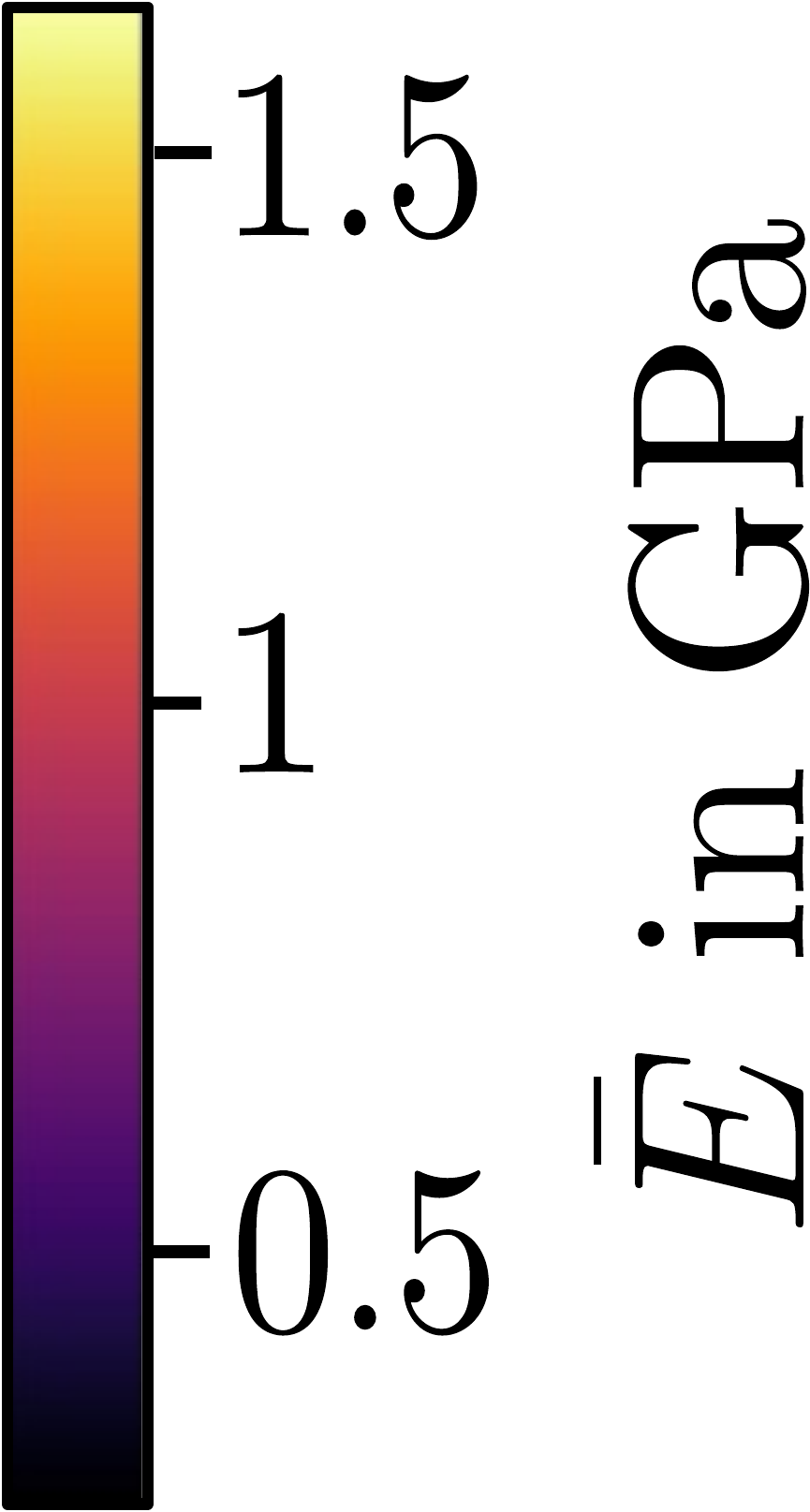}\end{tabular}  & $\SI{0.46}{\giga\pascal}$ & $-0.23$ \\
                     \midrule
                     5& \begin{tabular}{c}\includegraphics[height=69.4pt]{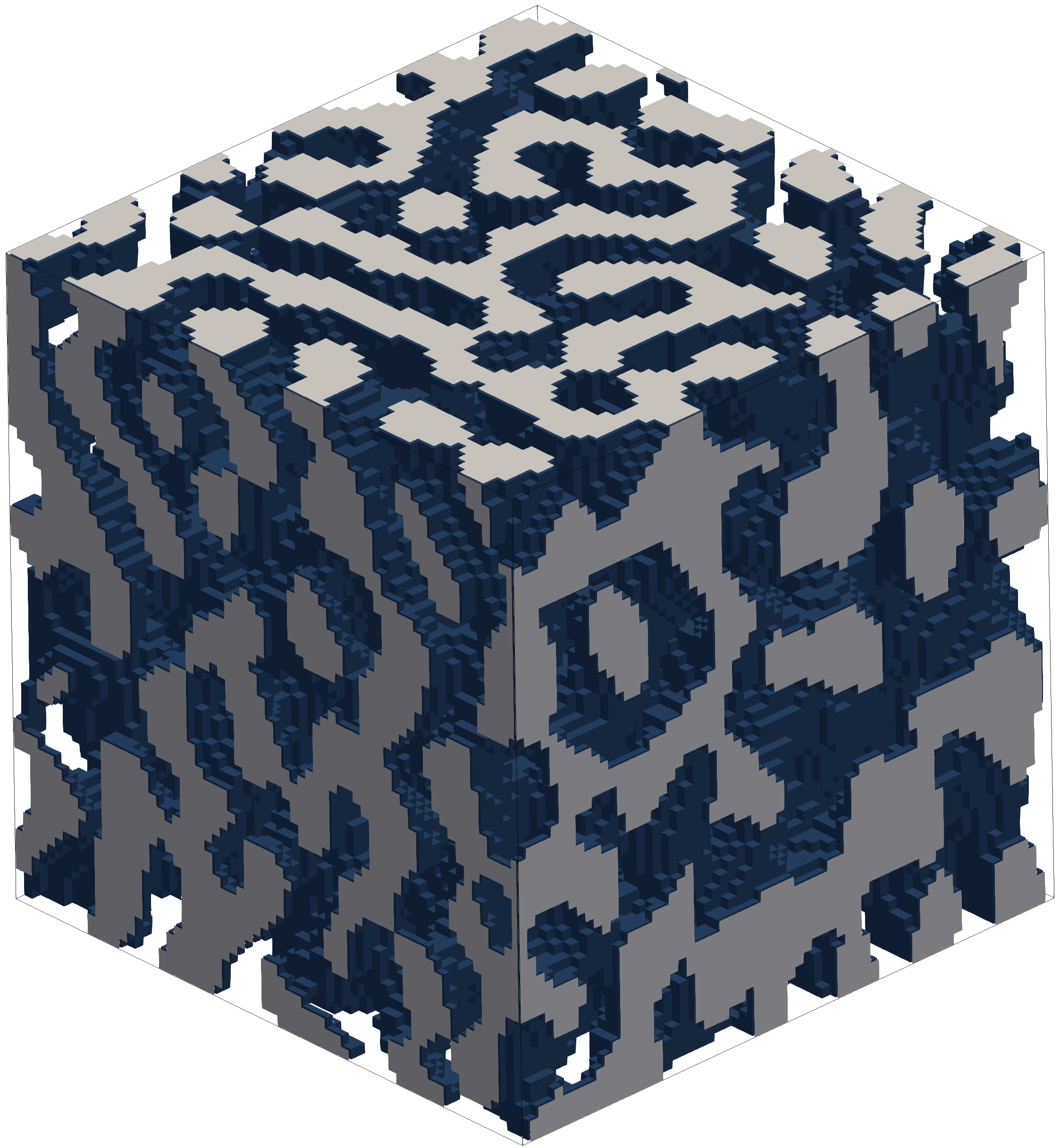} \includegraphics[height=30pt]{figures/coordinate_system.pdf}\end{tabular} & $\begin{pmatrix} \SI{0.0}{\degree} \\ \SI{50.8}{\degree} \\ \SI{45.9}{\degree} \end{pmatrix}$ & $0.55$ & $\begin{pmatrix} \SI{109.6}{\degree} \\ \SI{97.2}{\degree} \\ \SI{216.9}{\degree} \end{pmatrix}$ & \begin{tabular}{c}\includegraphics[height=30pt]{figures/coordinate_system.pdf}\includegraphics[height=69.4pt]{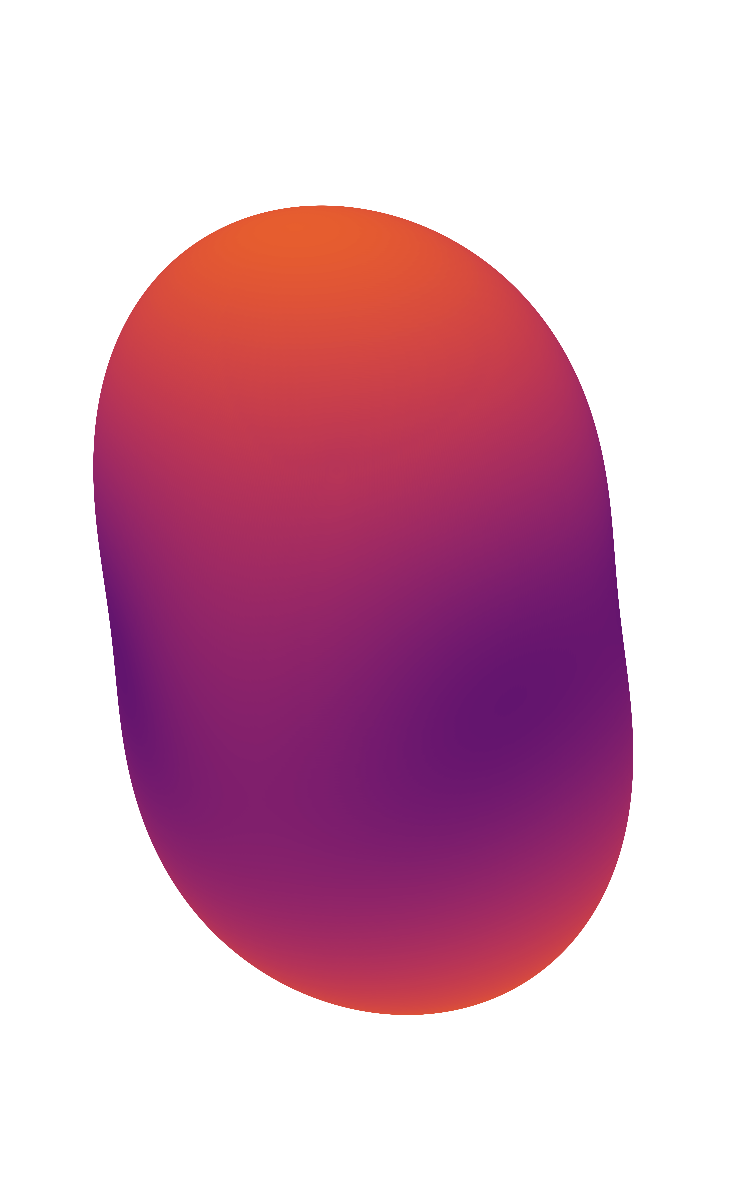}\end{tabular}\begin{tabular}{c}\includegraphics[height=55pt]{figures/DS16_legend2.pdf}\end{tabular} & $\SI{1.13}{\giga\pascal}$ & $-0.56$\\
                     10& \begin{tabular}{c}\includegraphics[height=69.4pt]{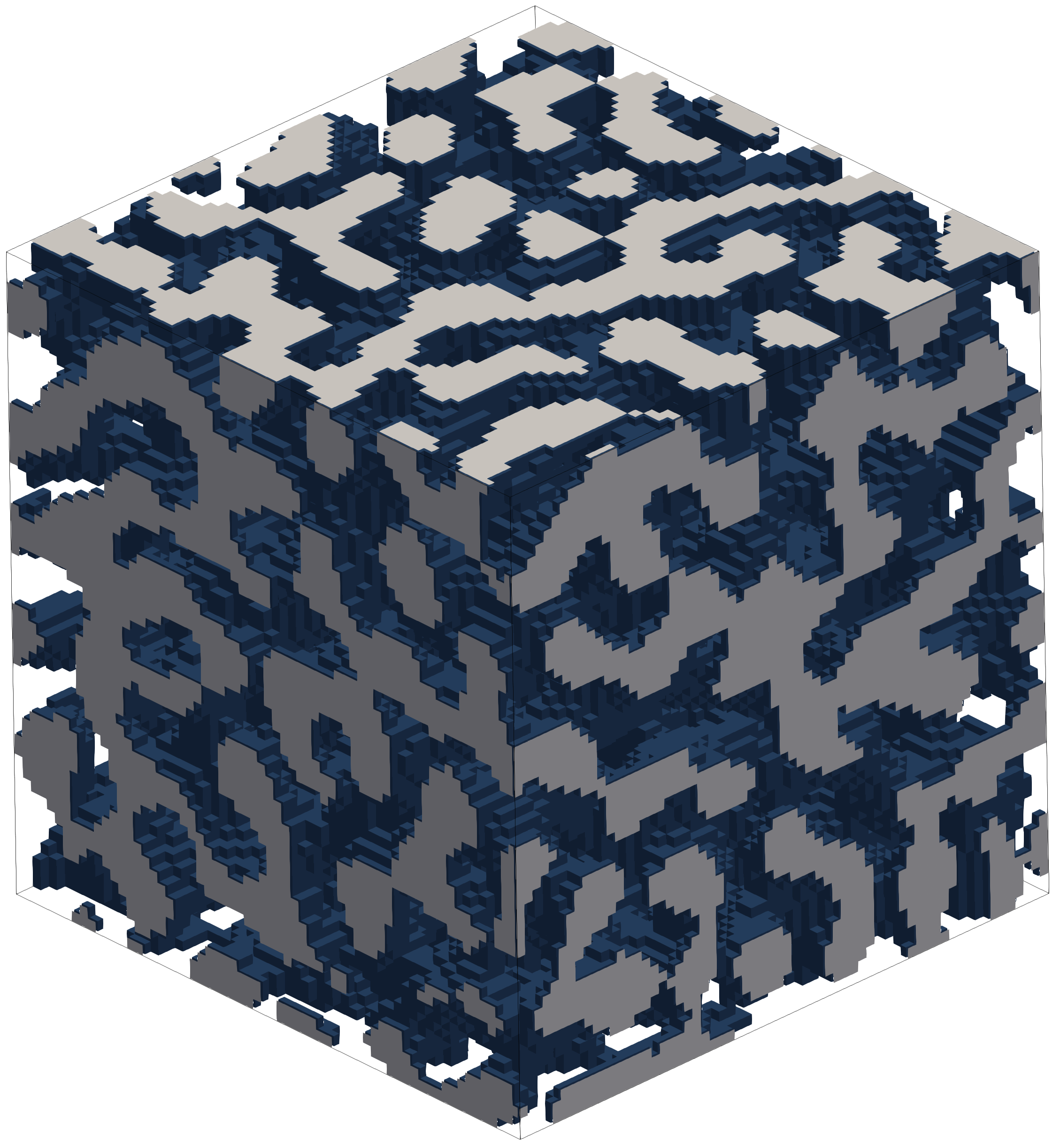} \includegraphics[height=30pt]{figures/coordinate_system.pdf}\end{tabular} & $\begin{pmatrix} \SI{0.0}{\degree} \\ \SI{88.8}{\degree} \\ \SI{49.9}{\degree} \end{pmatrix}$ & 0.55 & $\begin{pmatrix} \SI{225.3}{\degree} \\ \SI{48.4}{\degree} \\ \SI{279.6}{\degree} \end{pmatrix}$ & \begin{tabular}{c}\includegraphics[height=30pt]{figures/coordinate_system.pdf}\includegraphics[height=69.4pt]{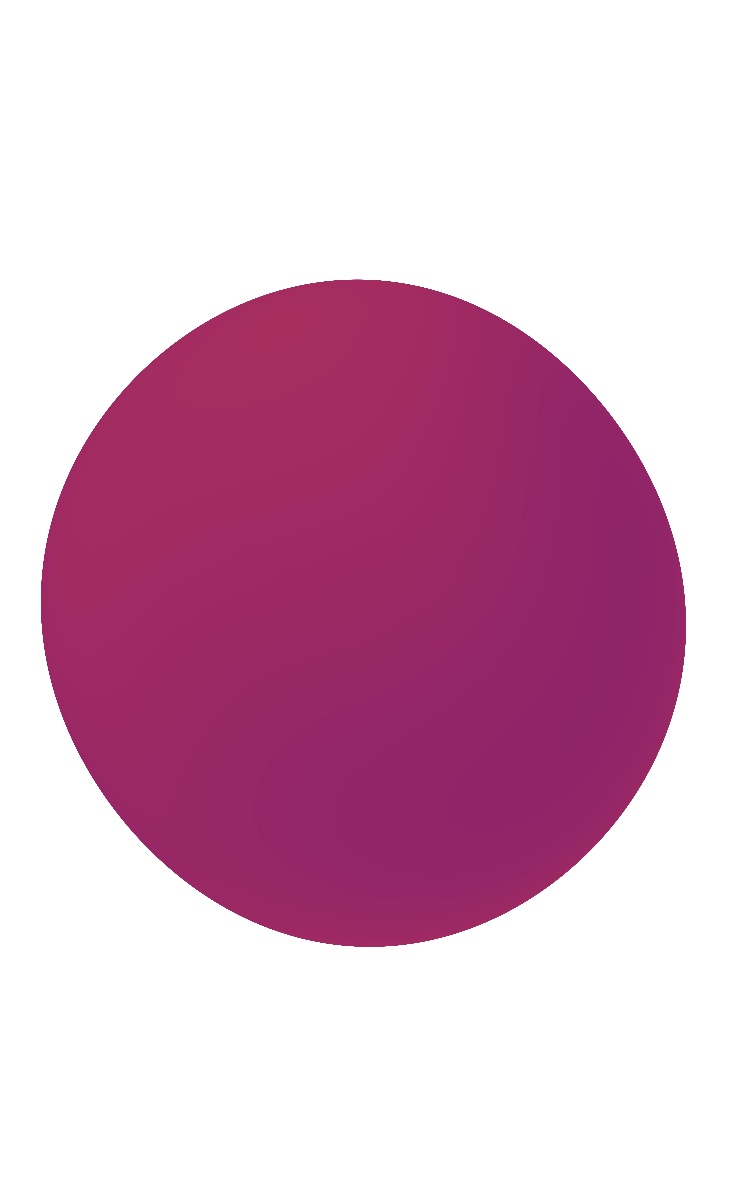}\end{tabular}\begin{tabular}{c}\includegraphics[height=55pt]{figures/DS16_legend2.pdf}\end{tabular} & $\SI{0.88}{\giga\pascal}$ & $-0.44$\\
                     15& \begin{tabular}{c}\includegraphics[height=69.4pt]{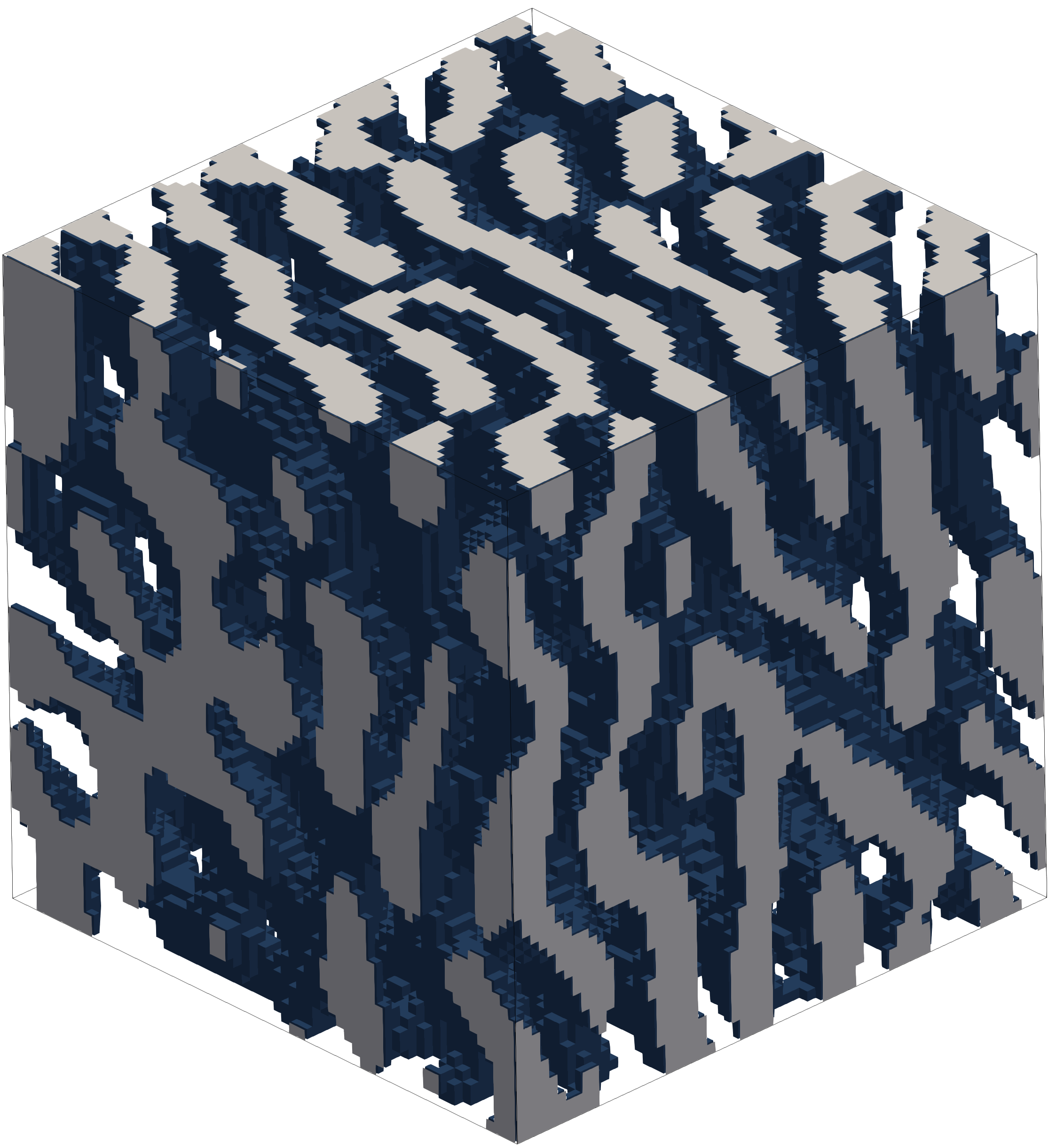} \includegraphics[height=30pt]{figures/coordinate_system.pdf}\end{tabular} & $\begin{pmatrix} \SI{0.0}{\degree} \\ \SI{37.1}{\degree} \\ \SI{15.0}{\degree} \end{pmatrix}$ & 0.55 & $\begin{pmatrix} \SI{137.5}{\degree} \\ \SI{75.7}{\degree} \\ \SI{107.8}{\degree} \end{pmatrix}$ & \begin{tabular}{c}\includegraphics[height=30pt]{figures/coordinate_system.pdf}\includegraphics[height=69.4pt]{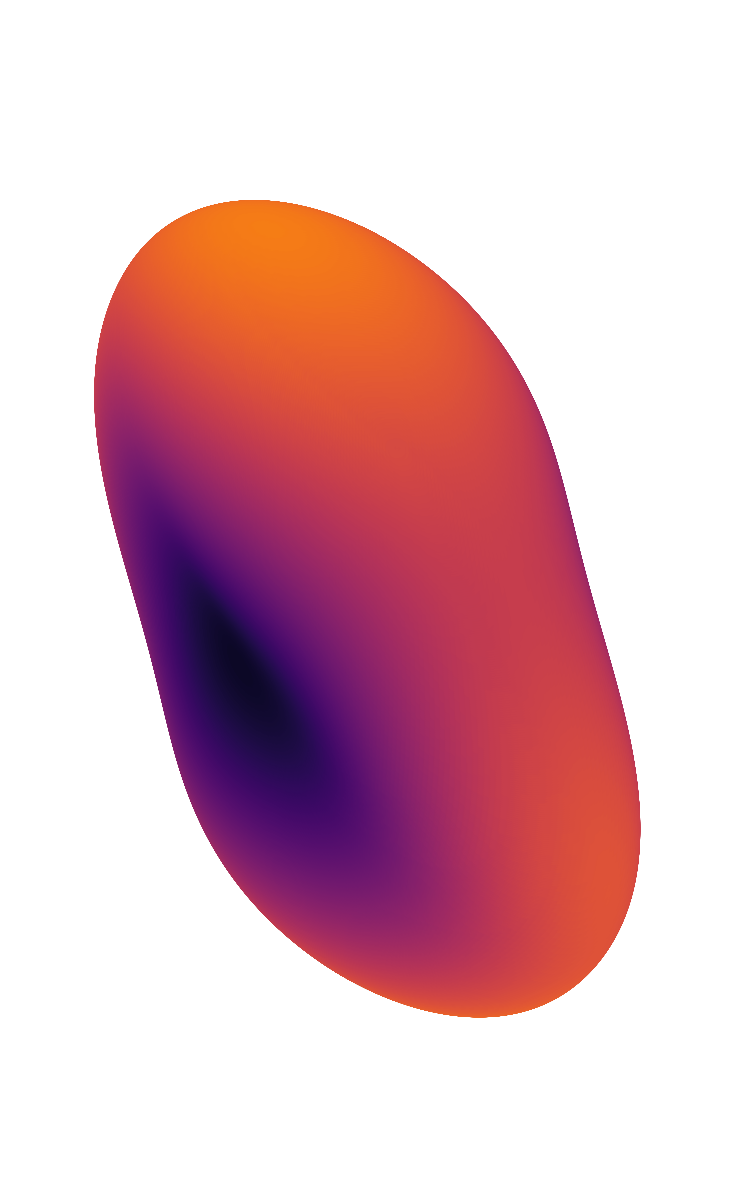}\end{tabular}\begin{tabular}{c}\includegraphics[height=55pt]{figures/DS16_legend2.pdf}\end{tabular} & $\SI{1.19}{\giga\pascal}$ & $-0.59$ \\
                     20& \begin{tabular}{c}\includegraphics[height=69.4pt]{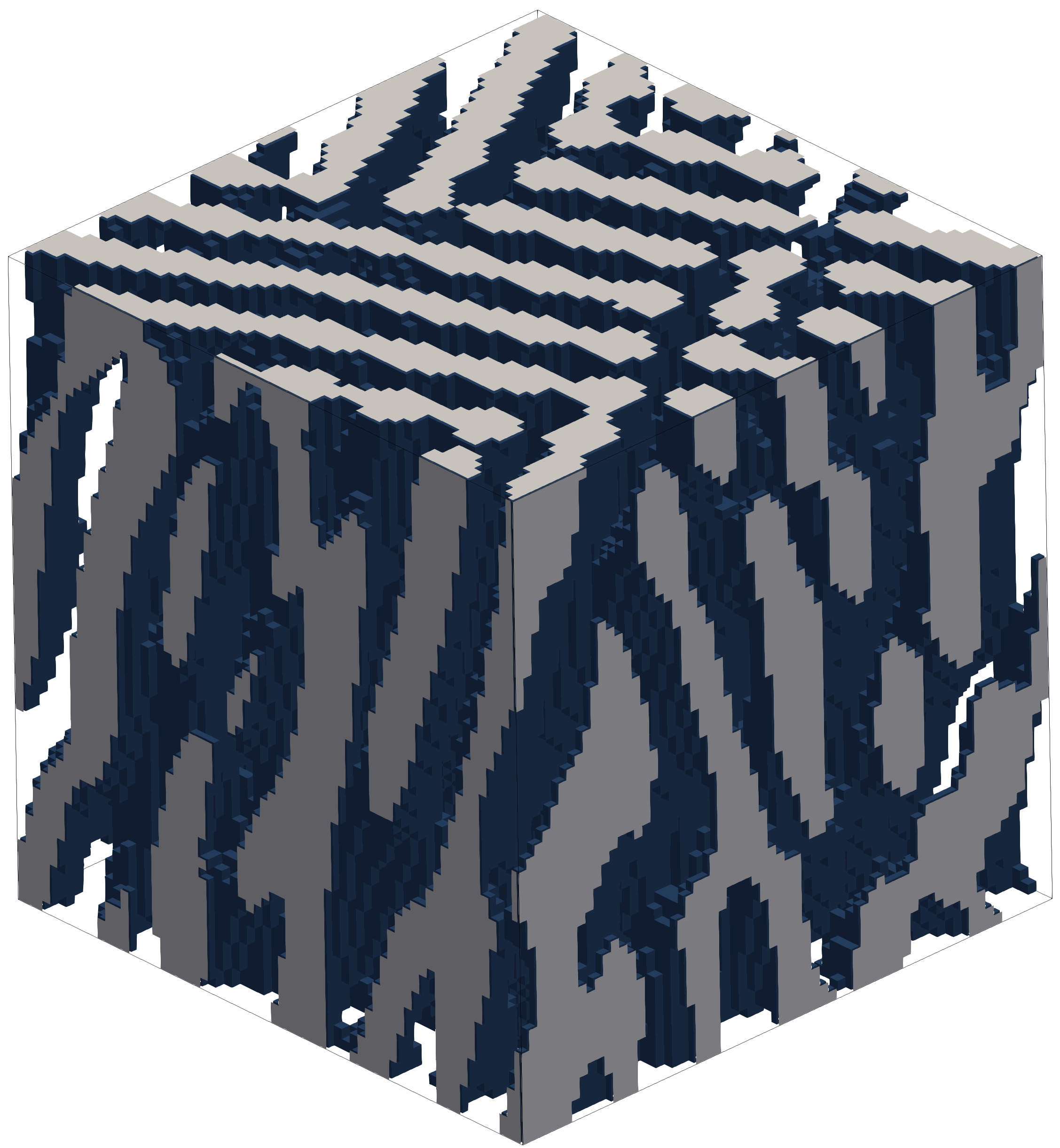} \includegraphics[height=30pt]{figures/coordinate_system.pdf}\end{tabular} & $\begin{pmatrix} \SI{0.0}{\degree} \\ \SI{16.9}{\degree} \\ \SI{16.3}{\degree} \end{pmatrix}$ & 0.55 & $\begin{pmatrix} \SI{167.0}{\degree} \\ \SI{104.4}{\degree} \\ \SI{10.7}{\degree} \end{pmatrix}$ & \begin{tabular}{c}\includegraphics[height=30pt]{figures/coordinate_system.pdf}\includegraphics[height=69.4pt]{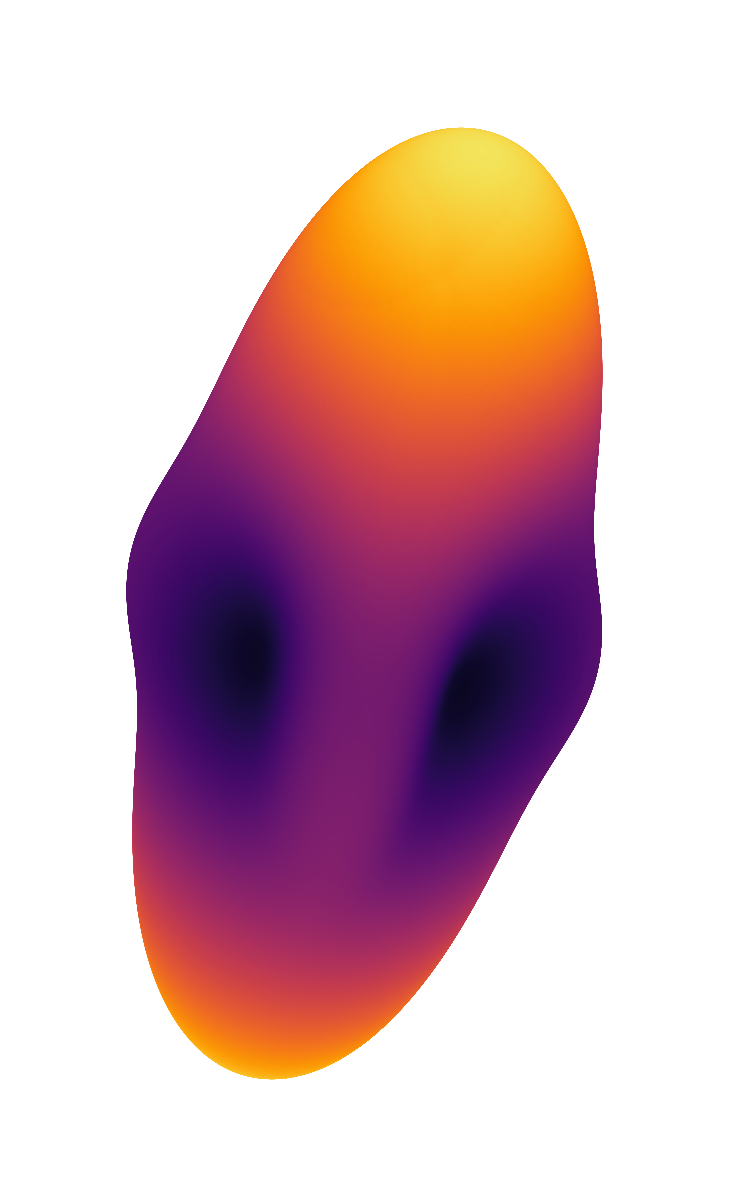}\end{tabular}\begin{tabular}{c}\includegraphics[height=55pt]{figures/DS16_legend2.pdf}\end{tabular} & $\SI{1.41}{\giga\pascal}$ & $-0.70$\\
                     25& \begin{tabular}{c}\includegraphics[height=69.4pt]{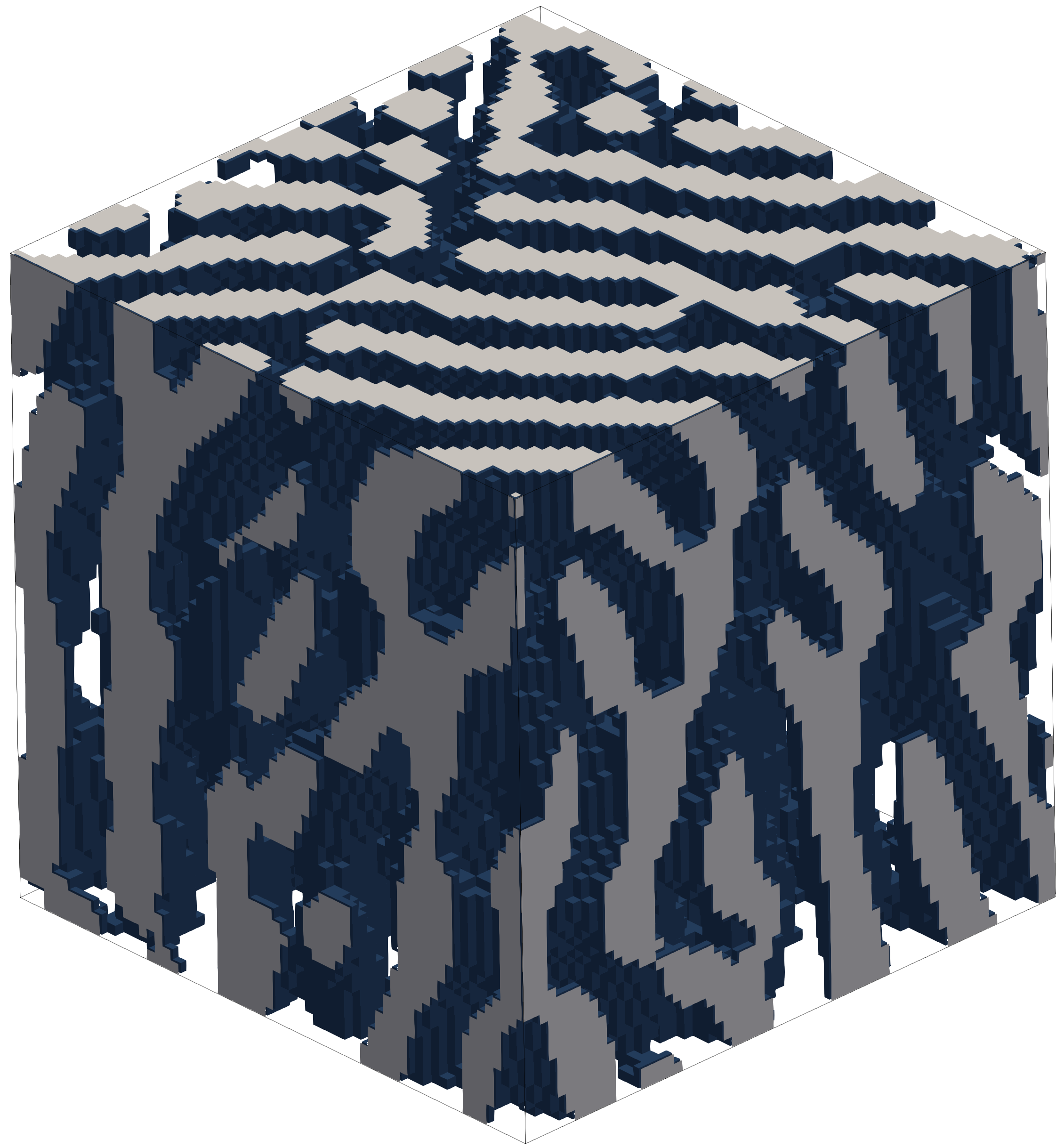} \includegraphics[height=30pt]{figures/coordinate_system.pdf}\end{tabular} & $\begin{pmatrix} \SI{0.0}{\degree} \\ \SI{15.0}{\degree} \\ \SI{32.5}{\degree} \end{pmatrix}$ & 0.55 & $\begin{pmatrix} \SI{207.0}{\degree} \\ \SI{98.6}{\degree} \\ \SI{335.1}{\degree} \end{pmatrix}$ & \begin{tabular}{c}\includegraphics[height=30pt]{figures/coordinate_system.pdf}\includegraphics[height=69.4pt]{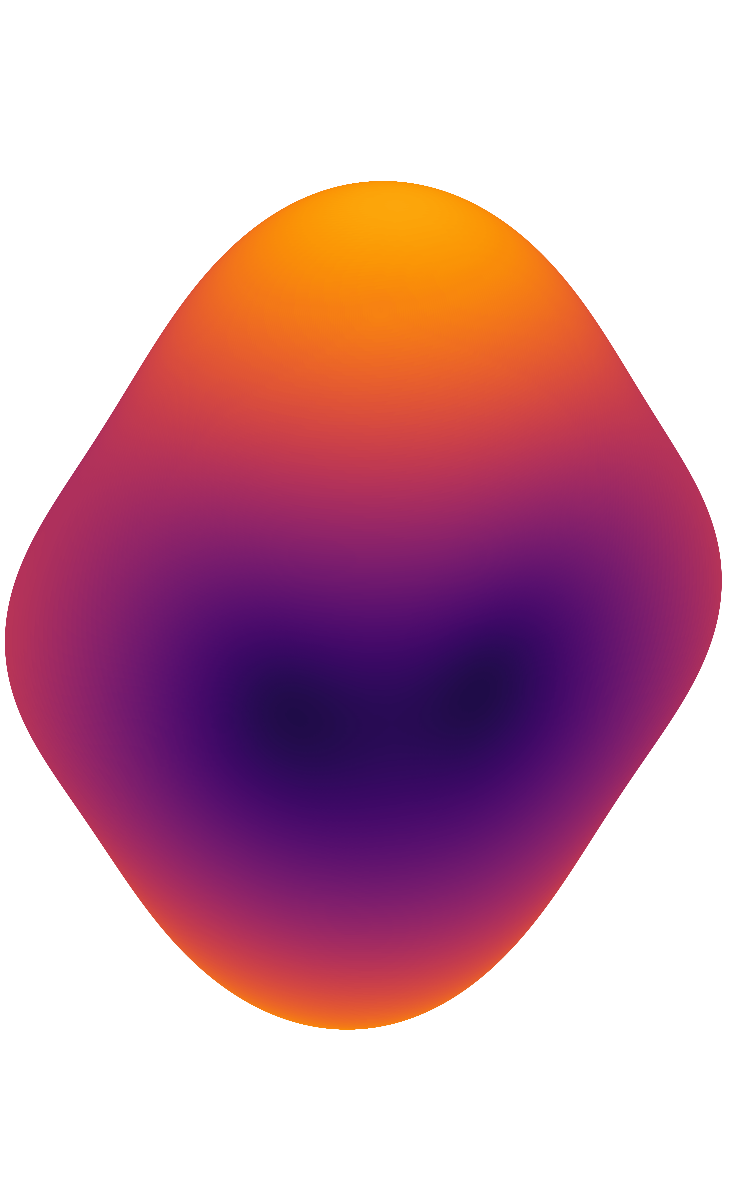}\end{tabular}\begin{tabular}{c}\includegraphics[height=55pt]{figures/DS16_legend2.pdf}\end{tabular} & $\SI{1.32}{\giga\pascal}$ & $-0.66$\\
                     30& \begin{tabular}{c}\includegraphics[height=69.4pt]{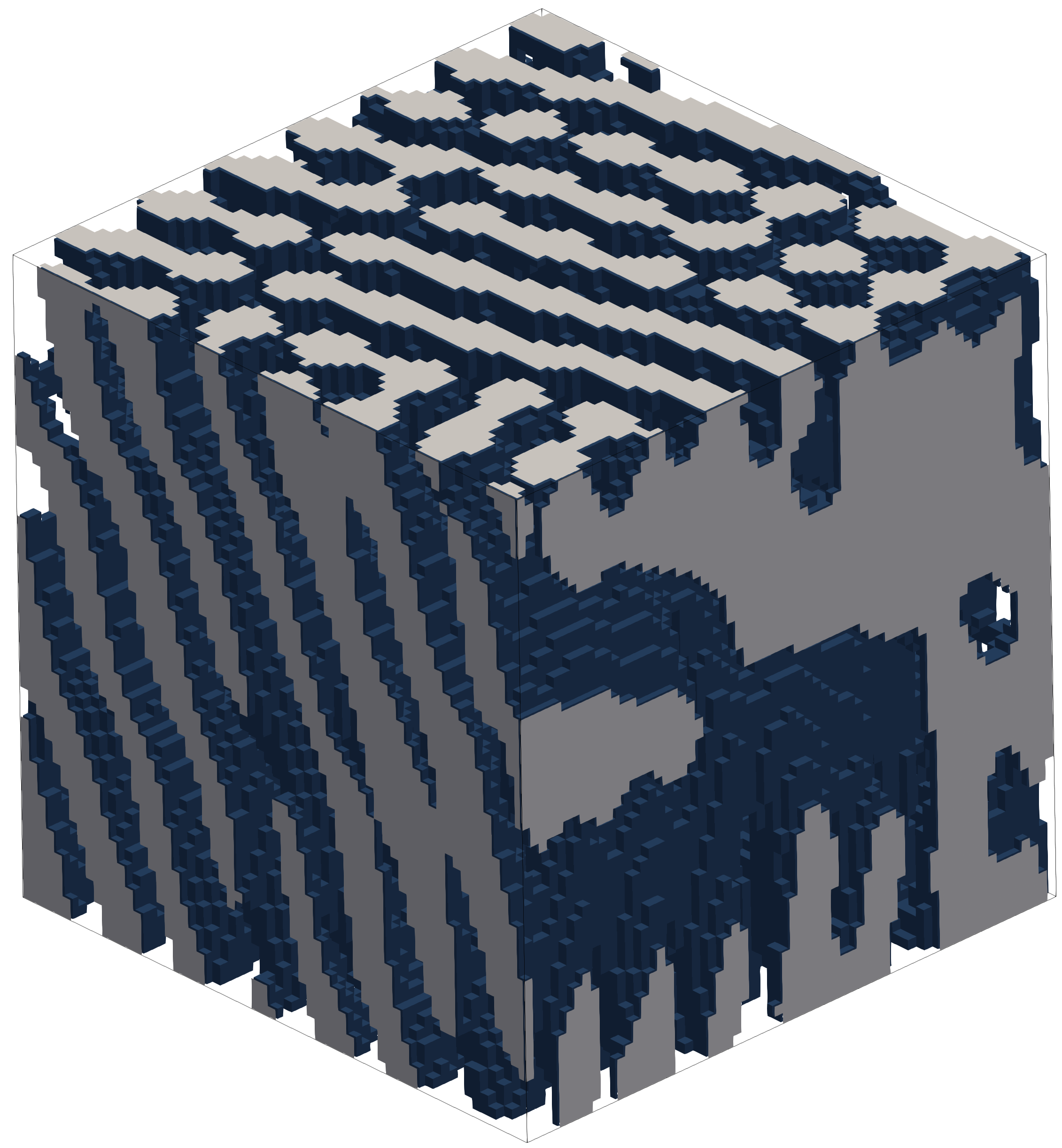} \includegraphics[height=30pt]{figures/coordinate_system.pdf}\end{tabular} & $\begin{pmatrix} \SI{0.0}{\degree} \\ \SI{15.0}{\degree} \\ \SI{22.3}{\degree} \end{pmatrix}$ & 0.55 & $\begin{pmatrix} \SI{345.1}{\degree} \\ \SI{99.0}{\degree} \\ \SI{176.2}{\degree} \end{pmatrix}$ & \begin{tabular}{c}\includegraphics[height=30pt]{figures/coordinate_system.pdf}\includegraphics[height=69.4pt]{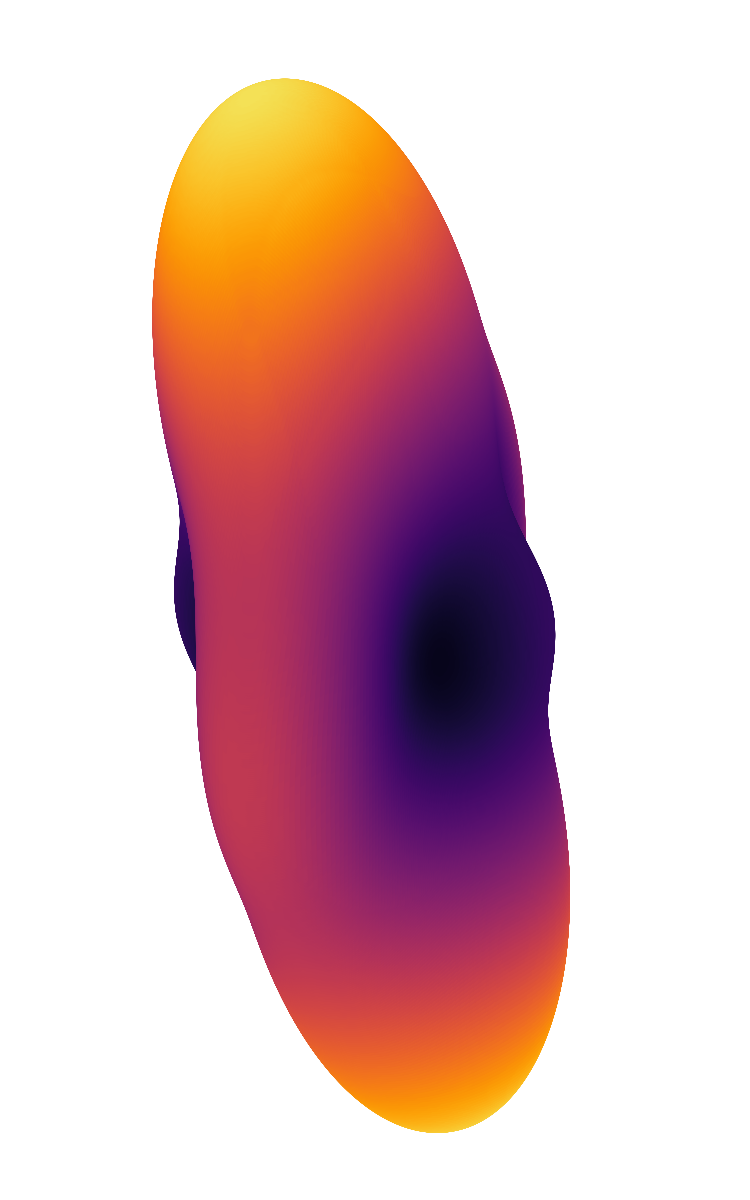}\end{tabular}\begin{tabular}{c}\includegraphics[height=55pt]{figures/DS16_legend2.pdf}\end{tabular} & $\SI{1.42}{\giga\pascal}$ & $-0.71$\\
                     \midrule
                     33& \begin{tabular}{c}\includegraphics[height=69.4pt]{figures/DS16_short16_32-3.png} \includegraphics[height=30pt]{figures/coordinate_system.pdf}\end{tabular} & $\begin{pmatrix} \SI{0.0}{\degree} \\ \SI{17.0}{\degree} \\ \SI{16.1}{\degree} \end{pmatrix}$ & $0.55$ & $\begin{pmatrix} \SI{358.5}{\degree} \\ \SI{83.6}{\degree} \\ \SI{289.0}{\degree} \end{pmatrix}$ & \begin{tabular}{c}\includegraphics[height=30pt]{figures/coordinate_system.pdf}\includegraphics[height=69.4pt]{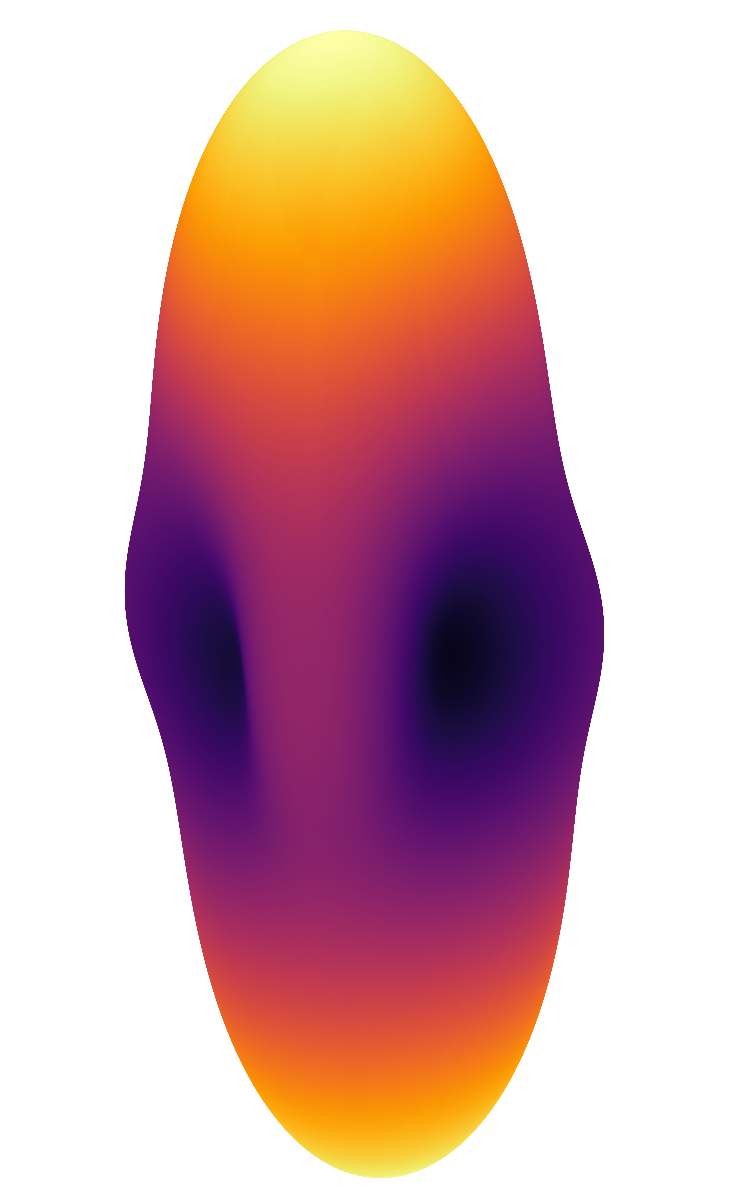}\end{tabular}\begin{tabular}{c}\includegraphics[height=55pt]{figures/DS16_legend2.pdf}\end{tabular} & $\SI{1.61}{\giga\pascal}$ & $-0.80$ \\
                     \bottomrule
                \end{tabular}
            \end{table*}
            For multiple iterations $i$, the structure, the descriptor, and the properties are given. Elastic surfaces, i.e., the visualization of the effective Young's modulus in all directions, support in understanding the preferable directions of the material. The same scaling and color-coding is used.
            It can be seen that the best structure of the initial data set ($i=0$) is of particularly high cost due to the low Young's modulus as result of the low volume fraction and no strongly pronounced anisotropy.
            In the intermediate results, the formation of anisotropic features in directions not too far away from the $z$-axis can be observed. Nonetheless, they also comprise structures with isotropic elastic responses as can be seen for iteration 10.
            From iteration 20 on, the best candidates have pronounced columnar or lamellar features. While the structure from iteration 20 is already close to the optimal choice of the morphological angles $\bm{\uptheta}$, the rotation angles $\bm{\upphi}$ are not optimal as the stiffest direction does not point in $z$-direction.
            The structure of the lowest cost at iteration 33 is characterized by the descriptor
            \begin{multline}\label{eq:6D_Tbest}
                \T^\text{6D*} := \left(\uptheta_2^* \; \uptheta_3^* \; v^*_\text{f}\; \upphi_1^*\; \upphi_2^*\; \upphi_3^* \right) \\
                \text{with} \quad
                \uptheta_2^* = \SI{17.0}{\degree},\, \uptheta_3^* = \SI{16.1}{\degree}, v^*_\text{f} = 0.55, \\\upphi_1^*=\SI{358.5}{\degree},\, \upphi_2^*=\SI{83.6}{\degree},\, \upphi_3^*=\SI{289.0}{\degree}
            \end{multline}
            and is shown in~\autoref{fig:ResCompSVE6D}, too. It is apparent that a structure \emph{close} but not identical to the perfect structure is found. The rotation vector is close to the expected value from~\autoref{eq:DS6_rotation} in terms of inducing a rotation around the $y$-axis of approximately $\SI{90}{\degree}$. This is due to the insignificance of $\upphi_1$ and $\upphi_3$ for the alignment of the columns in $z$-direction as these angles effectively cause a rotation around this axis only.

\begin{figure*}[t]
				\centering
				\includegraphics[width=\textwidth]{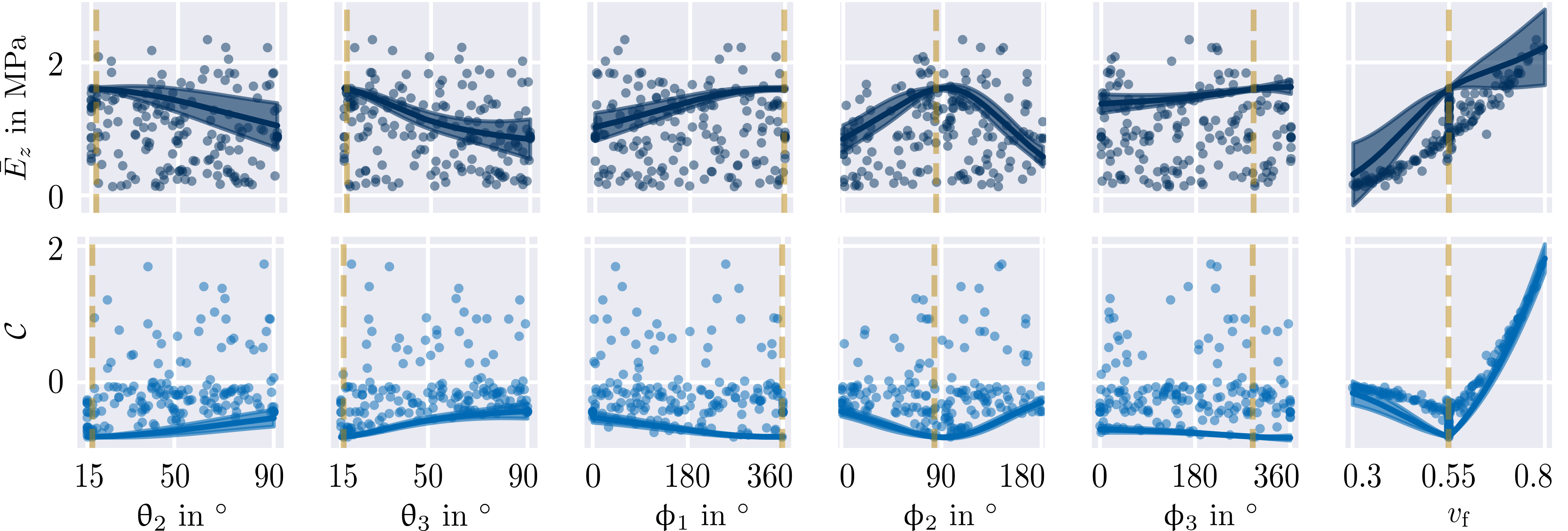}
				\caption{Effective young's modulus $\bar{E}_z$ and cost $\C$ for all structures in the augmented data set III (\autoref{tab:DataSets}) after 33 iterations. The dimensions of the 6D design space are depicted for the optimal spinodoid found, indicated by the vertical dashed line. The single plots show only 2D cross-sections. The continuous line and band indicate the Gaussian process regression's predicted mean and variance. The points are projected onto the 2D cross-section plane.}\label{fig:6D_GP}
			\end{figure*}

            This ambiguity in the rotations has to be inferred from the data in a 6D space, posing a challenge to the regression model, especially when considering the statistical deviations in the structures derived from the exact same descriptor, see~\autoref{app:Convergence}. 
            \autoref{fig:6D_GP} illustrates the Gaussian process regression model for the effective Young's modulus $\bar{E}_z$ and the distribution of the cost $\C$ for different descriptor values.
            To allow for an interpretable visualization of the 6D space, the distributions are shown as 1D cross section through the hypersurface spanned by the structure-property model. This highly limited view on the model is given for visual impression rather than for drawing solid conclusions. The descriptor values which are not variable, i.e., not the descriptor on the $x$-axis, are set to the optimal descriptor $\T^\text{6D*}$ found in the inverse design and indicated by the golden dashed line. The circular markers indicate all data points within the augmented data set after 33 iterations.

            For the specific choice of $\T=\T^\text{6D*}$ (\autoref{eq:6D_Tbest}) of the respective other descriptors, the following observations can be made:
            \begin{enumerate}[label=(\textit{\roman*})]
                \item Smaller morphological angles $\uptheta_i$ correspond to lower costs.
                \item At $\upphi_2=\SI{90}{\degree}$ is the expected minimum of cost.
                \item There is a strong correlation between Young's modulus and volume fraction for all data points.
                \item Volume fractions larger than the threshold value ${v_\text{f}^\text{thresh}}$ are effectively penalized, indicated by the increasing cost.
                \item The insignificance of $\upphi_3$ on both cost and Young's modulus is hinted by an only slightly varying model.
                \item Though known to have no influence, a large value of $\upphi_1$ seems to correlate with lower costs.
            \end{enumerate}
            Especially the last point might be caused by having data points with good other descriptors rather randomly distributed towards larger values of $\upphi_1$. However, as the specific choice of this angle is not important, it does not deteriorate the result at all.
            
            Several conclusions can be drawn from this example:
            \begin{enumerate}[label=(\textit{\roman*})]
                \item The design procedure can be applied to higher dimensional spaces (6D) even with sparse data.
                \item The ambiguities in the rotation are no insurmountable obstacle.
                \item More dimensions and few data necessitate more iterations of the design loop.
                \item Only a solution close to the perfect solution is found within 40 iterations.
            \end{enumerate}
  
	\section{Discussion}\label{sec:Discussion}
        The proposed framework for approaching the inverse design of spinodoid structures in an autonomous iterative scheme through Bayesian optimization is implemented and successfully applied to three demonstration problems of increasing dimension and complexity.
        The hypotheses that this allows for an inverse design even in the small-data regime is proven.

        From the results of the three numerical experiments, the following conclusions can be drawn:
        \begin{enumerate}[label=(\textit{\roman*})]
            \item The design procedure yields plausible results. The three demonstration problems are selected such that the optimal results are known. They are found by the present implementation, proving the concept.
            
            \item The design for multiple targeted properties is possible as shown in the second example in~\autoref{subsec:4D}. Thanks to the flexibility in formulating the cost function different design goals can be pursued. Although limited to two or four properties here, e.g., whole stiffness tensors can be achieved in principle.
            
            \item The design procedure can be applied to higher dimensional spaces (6D) even with sparse data as shown in the third example in~\autoref{subsec:6D}. Only 16 initial descriptor-property pairs are necessary to conduct the design in a 6D space.
            Compared to most machine learning-based approaches the necessary amount of initial data is extremely small. For instance, Kumar~et~al.~\cite{kumar2020} utilized roughly 20\,000 data points for optimizing the whole stiffness tensor of spinodoid structures.
            
            \item Ambiguities in the optimization problem are no insurmountable obstacle. This is demonstrated by the third example in~\autoref{subsec:6D}, where rotational angles are to be designed.
            
            \item More dimensions and few data necessitate more iterations of the design loop as can be seen from the third example. An almost perfect result but not the perfect result itself is found within limited iterations.
            
            \item Employing an efficient implementation of the Bayesian optimization, that harnesses the potential of GPUs, and the possibility to explore multiple candidates in each iteration thanks to the batched formulation of the acquisition function allows to perform the Bayesian optimization in perfectly acceptable wall-clock time, see~\autoref{tab:Resources}, on an HPC cluster, even in higher-dimensional spaces.
        \end{enumerate}

        Considering these conclusions, it can be stated that the presented inverse design approach allows to efficiently tailor the internal structure of architected materials based on few data. The latter point is of particular interest as data, especially if experimentally obtained, is expensive.

        While the illustrative examples underline the large potential of this approach, they are currently limited to one or three elastic properties. For harnessing the full potential of spinodoids, the design space of the morphological angles ought to include the zero angle $\{0\}$. In the present examples this is excluded for the sake of simplicity in the formulation of the boundary conditions of the optimization problem.
        Additionally, this augmentation approach is limited to properties for which a simulation model is available.
        However, if such a model is known, multi-fidelity Bayesian optimization could be employed for data fusion applications, where expensive experimental data is augmented by in silico computations. The utilization of statistical descriptors that allow for characterization \emph{and} reconstruction of volume elements seems promising in this context. They can form the link between real and in silico generated structures through a uniform description.

        Building up on the efficiency and remarkably low need for data, future research might apply the framework to design spinodoid structures with targeted inelastic or fatigue related properties. Due to the extensively higher cost compared to elastic properties, the generation of large data sets that, e.g., comprise whole yield surfaces, is nearly unfeasible but would be required for most machine learning-based design approaches.
        An extension from the class of spinodoid structures to general structures ought to be investigated in future research.
        Owing to the flexible choice of the acquisition function, the augmentation loop could be employed to iteratively explore large design spaces rather than targeting a specific design, resulting in possibly improved structure-property linkages of reduced variance. This feature could be employed for an adaptive sampling strategy, e.g., for more informative model-aware training set generation, when facing expensive functions.

        In conclusion, an indirect inverse design approach for spinodoid structures is presented that overcomes the obstacle of data availability by operating in the small-data regime. Examples demonstrate its potential for paving the way for advance in the inverse design of material structures.

	\section*{Author contribution statement}
    	\textbf{Conceptualization}: A.\ Ra{\ss}loff, P.\ Seibert, K.A.\ Kalina, M.\ K\"astner; \textbf{Methodology}: A.\ Ra{\ss}loff, P.\ Seibert;
        \textbf{Formal analysis and investigation}: A.\ Ra{\ss}loff, P.\ Seibert;
        \textbf{Software}: A.\ Ra{\ss}loff;
        \textbf{Visualization}: A.\ Ra{\ss}loff;
        \textbf{Writing - original draft}: A.\ Ra{\ss}loff;
        \textbf{Writing - review and editing}: A.\ Ra{\ss}loff, P.\ Seibert, K.A.\ Kalina, M.\ K\"astner;
        \textbf{Funding acquisition and resources}: M.\ K\"astner;
        \textbf{Supervision}: K.A.\ Kalina, M.\ K\"astner;
	
	\section*{Statements and Declarations}
	   The authors have no competing interests to declare that are relevant to the content of this article.
    
    \section*{Acknowledgements}
        The authors gratefully acknowledge the GWK support for funding this project by providing computing time through the Center for Information Services and HPC (ZIH) at TU Dresden.
        This work was supported by the German Research Foundation (DFG) within the Research Training Group GRK 2868 D${}^3$ -- project number 493401063.
        The group of M.\ K\"astner thanks the DFG which further supported this work under Grant number KA 3309/18-1.
	
    \bibliographystyle{unsrtnat} 
	\bibliography{Rassloff2024_InverseDesign}	
	
	\appendix
        \section{Prediction with Gaussian process regression}\label{app:GP}
            Gaussian processes can be applied to regression problems as described, e.g., in~\cite{quadrianto2010} in detail.
            Starting with $n_\text{k}$ known data points
            \begin{equation*}
                \Tilde{\mathcal{D}}=\left\{\left(\ind{i}\Tilde{\T}, \ind{i}\Tilde{\text{P}}\right)\right\}_{i=1}^{n_\text{k}} \:,
            \end{equation*}
            the goal of the regression problem is to find a function $f_\mathcal{GP}$, such that
            \begin{equation*}
                \ind{i}\Tp = f_\mathcal{GP} \left( \ind{i}\Tilde{\T} \right) + \ind{i}\varepsilon \:,
            \end{equation*}
            holds, where $\varepsilon \sim \mathcal{N}(0,\sigma^2)$ and $\Tp$ denotes the distribution of the property $\TP$ at $\TT$.
            Assuming a zero-mean prior Gaussian distribution over functions
            \begin{equation*}
                f_\mathcal{GP}(\bullet) \sim \mathcal{GP} \left(0, k(\bullet,\bullet)\right) \:,
            \end{equation*}
            any set of functions drawn from it must have joint multivariate Gaussian normal distribution, so for the points of the data set $\Tilde{\mathcal{D}}$ and for $n^\star$ unknown points 
            \begin{equation*}
                \mathcal{\Tilde{D}}^\star=\left\{\left(\ind{j}\Tilde{\T}^\star, \ind{j}\Tilde{\text{P}}^\star\right)\right\}_{j=1}^{n^\star} \:,
            \end{equation*}
            i.e.,
            \begin{equation*}
                \begin{pmatrix}
                    \ve{F} \\
                    \ve{F}^\star
                \end{pmatrix}
                \bigg|\; \ve{A}_{\TT}, \ve{A}_{\TT}^\star \sim \mathcal{N}\left(\bm{0}, \begin{bmatrix}
                    \bm{\upkappa}\left(\ve{A}_{\TT}, \ve{A}_{\TT}\right) & \bm{\upkappa}\left(\ve{A}_{\TT},\ve{A}_{\TT}^\star\right) \\
                    \bm{\upkappa}\left(\ve{A}_{\TT}^\star, \ve{A}_{\TT}\right) & \bm{\upkappa}\left(\ve{A}_{\TT}^\star, \ve{A}_{\TT}^\star\right)
                \end{bmatrix}
                \right) \:.
            \end{equation*}
            The same applies to the noise distribution
            \begin{equation*}
                \begin{pmatrix}
                    \ve{A}_\varepsilon \\
                    \ve{A}_{\varepsilon}^\star
                \end{pmatrix}
                    \sim \mathcal{N}\left(\bm{0}, \begin{bmatrix}
                    \sigma^2\ve{I} & \bm{0} \\
                    \bm{0} & \sigma^2\ve{I}
                \end{bmatrix}
                \right) \:.
            \end{equation*}

            For the sake of simplicity, to allow for the collection of all elements of an ordered set of a generic quantity $\bullet$ in an array, the definitions
            \begin{equation*}
                \ve{A}\left( \left\{ \ind{i}\bullet \right\}_{i=1}^{n} \right) := \begin{pmatrix}
                    \ind{i}\bullet \\
                    \ind{i+1}\bullet \\
                    \vdots \\
                    \ind{n}\bullet
                \end{pmatrix} \quad \text{and} 
            \end{equation*}
            \begin{equation*}
            \ve{F}_\mathcal{GP}\left( \left\{\ind{i}\bullet\right\}_{i=1}^n\right) := \begin{pmatrix}
                    f_\mathcal{GP}(\ind{i}\bullet)\\
                    f_\mathcal{GP}(\ind{{i+1}}\bullet)\\
                    \vdots\\
                    f_\mathcal{GP}(\ind{n}\bullet)
                \end{pmatrix}\:,
            \end{equation*}
            are introduced.
            Using them, the abbreviations
            \begin{gather*}
                \ve{A}_{\TT} := \ve{A}\left( \left\{ \ind{i}\TT \right\}_{i=1}^{n_\text{k}} \right), \:
                \ve{A}_{\TT}^\star := \ve{A}\left( \left\{ \ind{j}\TT^\star \right\}_{j=1}^{n^\star} \right) \\
                \ve{A}_{\TP} := \ve{A}\left( \left\{ \ind{j}\TP \right\}_{j=1}^{n_\text{k}} \right), \:
                \ve{A}_{\Tp}^\star := \ve{A}\left( \left\{ \ind{j}\Tp^\star \right\}_{j=1}^{n^\star} \right) \\
                \ve{F} := \ve{F}_\mathcal{GP}\left(\ve{A}_{\TT}\right), \:
                \ve{F}^\star := \ve{F}_\mathcal{GP}\left(\ve{A}_{\TT}^\star\right) \\
                \upkappa_{ij}\left( \ve{A}(\bullet), \ve{A}(\bullet') \right) := k\left( \ind{i}\bullet, \ind{j}\bullet' \right)
            \end{gather*}
            are defined.

            As the sums of independent Gaussian random variables are also Gaussian,
            \begin{multline*}
                \begin{pmatrix}
                    \ve{A}_{\TP} \\
                    \ve{A}_{\Tp}^\star
                \end{pmatrix}
                \bigg|\; \ve{A}_{\TT}, \ve{A}_{\TT}^\star = 
                \begin{pmatrix}
                    \ve{F} \\
                    \ve{F}_*
                \end{pmatrix} + 
                \begin{pmatrix}
                    \ve{A}_{\varepsilon} \\
                    \ve{A}_{\varepsilon}^\star
                \end{pmatrix} \\
                \sim \mathcal{N}\left(\bm{0}, \begin{bmatrix}
                    \bm{\upkappa}\left(\ve{A}_{\TT}, \ve{A}_{\TT}\right) + \sigma^2\ve{I} & \bm{\upkappa}\left(\ve{A}_{\TT},\ve{A}_{\TT}^\star\right) \\
                    \bm{\upkappa}\left(\ve{A}_{\TT}^\star, \ve{A}_{\TT}\right) & \bm{\upkappa}\left(\ve{A}_{\TT}^\star, \ve{A}_{\TT}^\star\right) + \sigma^2\ve{I})
                \end{bmatrix}
                \right)
            \end{multline*}
            holds.
            Using the rules for conditioning Gaussians, it can be rewritten as
            \begin{equation*}
                \ve{A}_{\Tp}^\star \mid \Tilde{\mathcal{D}}, \ve{A}_{\TT}^\star \sim \mathcal{N}\left( \ve{M}^\star, \ve{K}^\star \right) \;,
            \end{equation*}
            where
            \begin{equation}\label{eq:GP_pred_mean}
                \ve{M}^\star := \bm{\upkappa}\left(\ve{A}_{\TT}^\star, \ve{A}_{\TT}\right) \left( \bm{\upkappa}\left(\ve{A}_{\TT}, \ve{A}_{\TT}\right) + \sigma^2\ve{I} \right)^{-1} \ve{A}_{\TP} \quad \text{and}
            \end{equation}
            \vspace*{-24pt}
            \begin{multline}\label{eq:GP_pred_cov}
                \ve{K}^\star := \bm{\upkappa}\left( \ve{A}_{\TT},\ve{A}_{\TT}^\star \right) + \sigma^2\ve{I} \\ 
                - \bm{\upkappa}\left(\ve{A}_{\TT}^\star, \ve{A}_{\TT}\right) \left( \bm{\upkappa}\left(\ve{A}_{\TT}, \ve{A}_{\TT}\right) + \sigma^2\ve{I} \right)^{-1} \bm{\upkappa}\left(\ve{A}_{\TT},\ve{A}_{\TT}^\star\right) \;.
            \end{multline}

            In this way the distribution of a property $\Tp$ can be evaluated at each location $\TT$ in the descriptor space.
            
        \section{Numerical simulation}\label{app:DAMASK}
			The numerical simulations of the mechanical behavior are conducted using the simulation toolkit DAMASK~\cite{roters2019}. Its efficient spectral solver allows to perform the simulations directly on the voxelized spinodoids structures. However, this necessitates to model both solid and void phase.
   
            For the present study only the elastic properties are of interest, more specifically the effective directional Young's moduli $\bar{E}$ and the stiffness tensor. While a purely elastic material model would suffice, an inelastic model is employed as well. This does not reduce the performance as it is not activated for the considered load cases here. However, it allows to run possible future simulations, including inelasticity, using the same model.

            More details on the constitutive modeling can be found in~\cite{roters2019}. Hereafter, the key equations are briefly listed.
            A generalized Hooke's law
            \begin{equation*}
                \ve{S} := \boldsymbol{\mathbb{C}} : \ve{E}^\text{e} \: ,
            \end{equation*}
            where $\ve{S}$ denotes the second Piola-Kirchhoff stress tensor, $\ve E^\text{e}$ denotes the elastic Green-Lagrange strain tensor, and $\boldsymbol{\mathbb{C}}$ denotes the fourth-order stiffness tensor, is used to model the elastic response.

            A phenomenological power law
            \begin{equation*}
                \dot{\gamma}_\text{p} := \dot{\gamma}_0 \left( \sqrt{\frac{3}{2}} \frac{\|\ve S^\text{dev}\|_\text{F}}{3 \xi} \right)^n \: , 
            \end{equation*}
            where $\dot \gamma_\text{p}$ denotes the plastic strain rate, $\dot{\gamma}_0$ denotes the initial strain rate, $n$ is the stress exponent, $\|(\cdot)\|_\text{F}$ denotes the Frobenius norm of the deviatoric part $\ve S^\text{dev} := \ve{S} - \dfrac{1}{3} \text{tr}(\ve S) \ve{I}$ of the second Piola-Kirchhoff stress, and $\xi$ denotes the material resistance, is selected for the inelastic behavior.

            The evolution of the material resistance
            \begin{equation*}
                \dot{\xi} := \dot{\gamma}h_0 \left| 1 - \frac{\xi}{\xi_\infty} \right|^a \; \text{sign} \left( 1 - \dfrac{\xi}{\xi_\infty} \right) \;
            \end{equation*}
            depends on the initial hardening $h_0$, the initial~$\xi_0$ and final resistance~$\xi_\infty$, and a fitting parameter~$a$.

            The parameters of the constitutive equations are selected to model polyvinyl chloride (PVC) as the solid part of the spinodoid. The void voxels are assigned 100 times smaller stiffnesses to approximate the behavior of air and to avoid numerical issues. The parameters are listed in \autoref{tab:material_parameters}.

            For the determination the effective Young's modulus in a certain direction a small deformation ${\Delta \text{F}=5\cdot 10^{-6}}$ is applied and prescribed in DAMASK in terms of the deformation gradient $\ve F$ and first Piola-Kirchhoff stress $\ve P$, e.g., for the $x$-direction as
            \begin{equation*}
                \ve{F} = \begin{pmatrix}
                    1 + \Delta\text{F} & 0 & 0 \\
                    0 & - & 0 \\
                    0 & 0 & -
                \end{pmatrix}
                \quad \text{and} \quad
                \ve{P} = \begin{pmatrix}
                    - & - & - \\
                    - & 0 & - \\
                    - & - & 0
                \end{pmatrix}
                \:,
            \end{equation*}
            where '-' means that this coordinate is not prescribed. This desired state is to be reached in a single increment.
            The resulting fields of $\ve{F}$ and $\ve{P}$ are volume averaged per increment, translated into the Green-Lagrange strain tensor and second Piola-Kirchhof stress tensor, and the Young's modulus is computed. For more details on the homogenization, the reader is kindly referred to~\cite{seibert2023}.
            
            \begin{table}[t]
            	\centering
            	\caption{Material parameters for the numerical simulations with DAMASK.}
            	\label{tab:material_parameters}
            	\begin{tabular}{c | c  c }
                \toprule
                Parameter & PVC & Void\\ 
                \midrule
                $\mathbb{C}_{1111}$ in GPa & 7.5 & 0.075 \\ 
                $\mathbb{C}_{1122}$ in GPa & 5.0 & 0.05 \\  
                $\mathbb{C}_{1123}$ in GPa & 1.25 & 0.0125 \\ 
                $\xi_0$ in MPa & 15.0 & -\\  
                $\xi_\infty$ in MPa & 150.0 & -\\  
                $h_0$ in MPa & 10 & -\\ 
                $\dot{\gamma}_0$ & 0.001 & -\\
                n & 20 & -\\
                a & 2 & -\\ 
                \bottomrule
            	\end{tabular}
            \end{table}
            
		\section{Study on volume element size}\label{app:Convergence}
            The size of the volume element to be simulated has to be representative to a certain extent to allow for meaningful results from a single simulation. This is particularly relevant for the statistical generation of spinodoid, where the structure $\ve{S}$ is a random instance derived from a descriptor $\T$. Consequently, two spinodoids generated from the same descriptor are statistically similar to a certain extent but the voxel-valued representation in terms of there structure $\ve{S}$ is most probably different. However, the more characteristic features can be incorporated in the volume element, i.e., the larger the volume element, the more similar should be the mechanical behavior.
            In terms of the spinodoid structures this size can be easily set by the volume element length $l_\text{VE}$ as introduced in~\autoref{subsec:Spino}.

\begin{figure}[t]
    			\centering
    			\includegraphics[width=0.9\columnwidth]{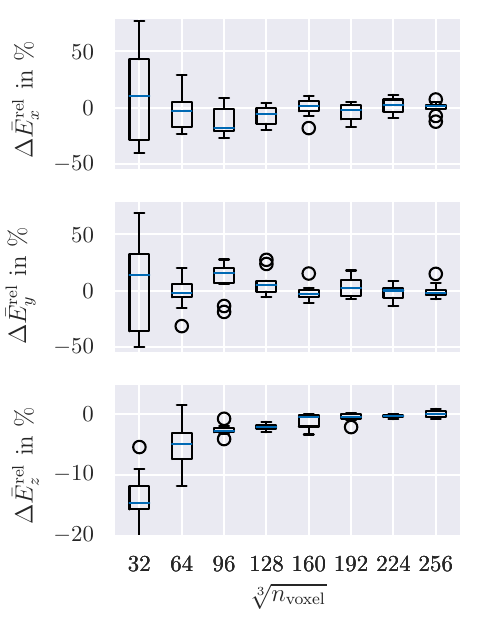}
    			\caption{Relative deviation of the effective directional Young's modulus for 10 structures of the same descriptor $\T$ consisting of the arrays $\bm{\uptheta} = (\SI{15}{\degree}, \SI{15}{\degree}, \SI{0}{\degree}),\: \bm{\upphi}=(\SI{0}{\degree}, \SI{0}{\degree}, \SI{0}{\degree})$, and $ v_\text{f}=0.5$ but different random seed for differently sized volume elements. The volume scales linearly with the number of voxels $n_\text{voxel}$ used for the spatial discretization, see \autoref{fig:AppStructures} for examples. The relative modulus is referred to the mean modulus of the structures with a resolution $256^3$. The boxes range from the first quartile (Q1) to the third quartile (Q3) and indicate the median by a line within the box. From the box, the whiskers stretch to the furthest point that is still within 1.5 times the inter-quartile range (Q3-Q1) from the box. Outliers are marked as points.}\label{fig:AppStudy}
    		\end{figure}

\begin{figure}[t]
    			\centering
    			\begin{subfigure}[b]{0.9\columnwidth}
    				\centering
    				\includegraphics[width=\textwidth]{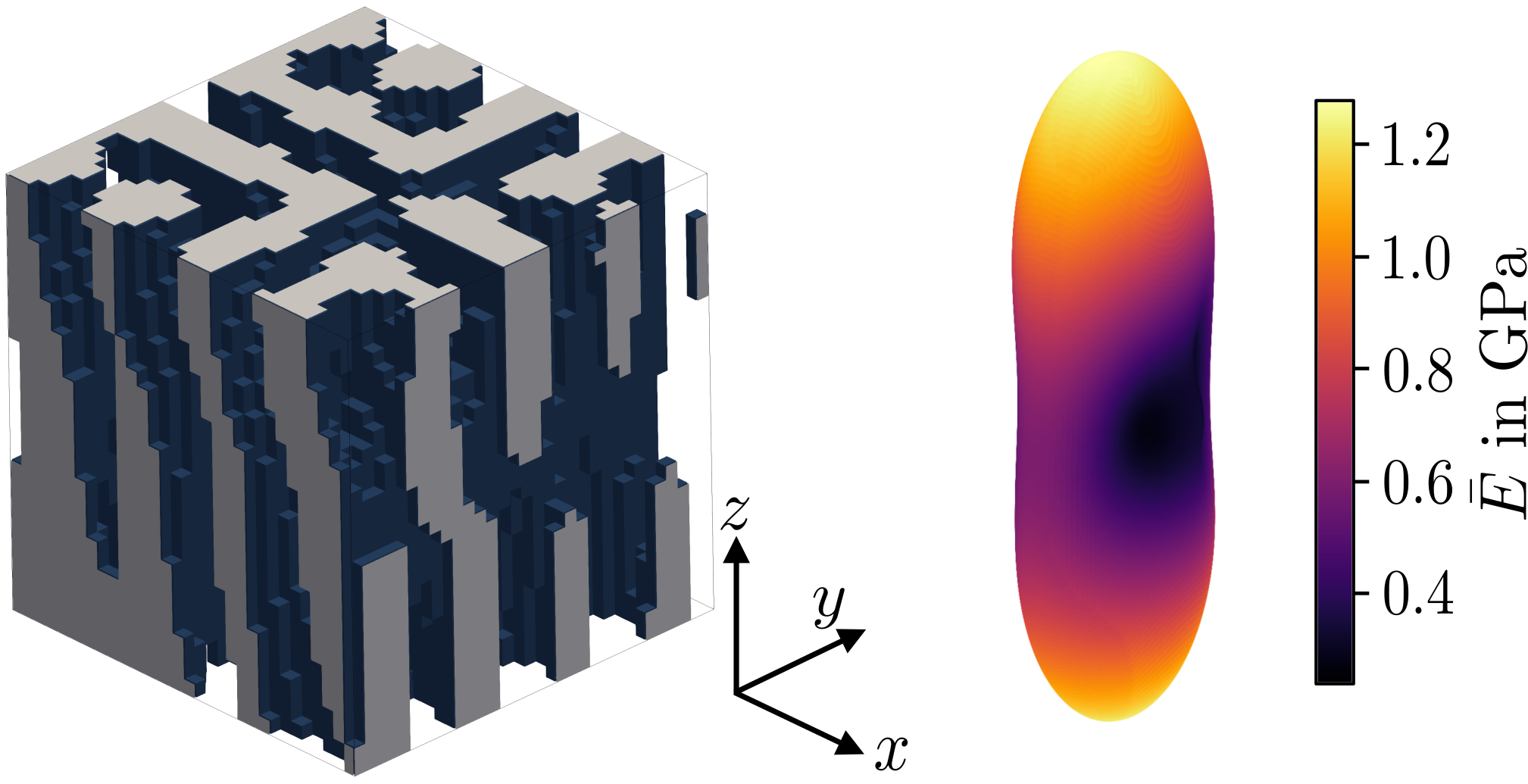}
    				\caption{$n_\text{voxel}=32^3$}
    			\end{subfigure}
    			\begin{subfigure}[b]{0.9\columnwidth}
    				\centering
    				\includegraphics[width=\textwidth]{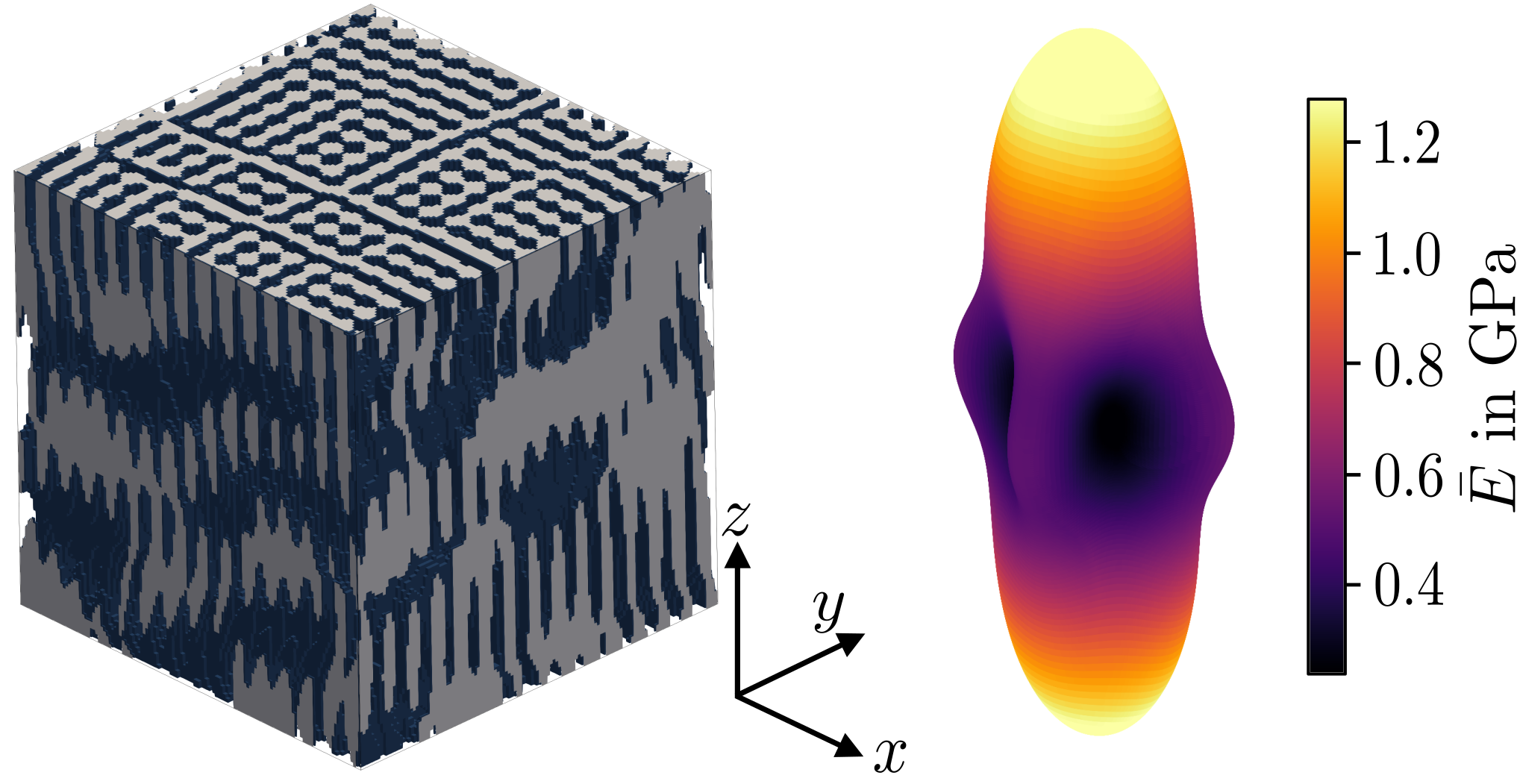}
    				\caption{$n_\text{voxel}=128^3$}
    			\end{subfigure}
    			\begin{subfigure}[b]{0.9\columnwidth}
    				\centering
    				\includegraphics[width=\textwidth]{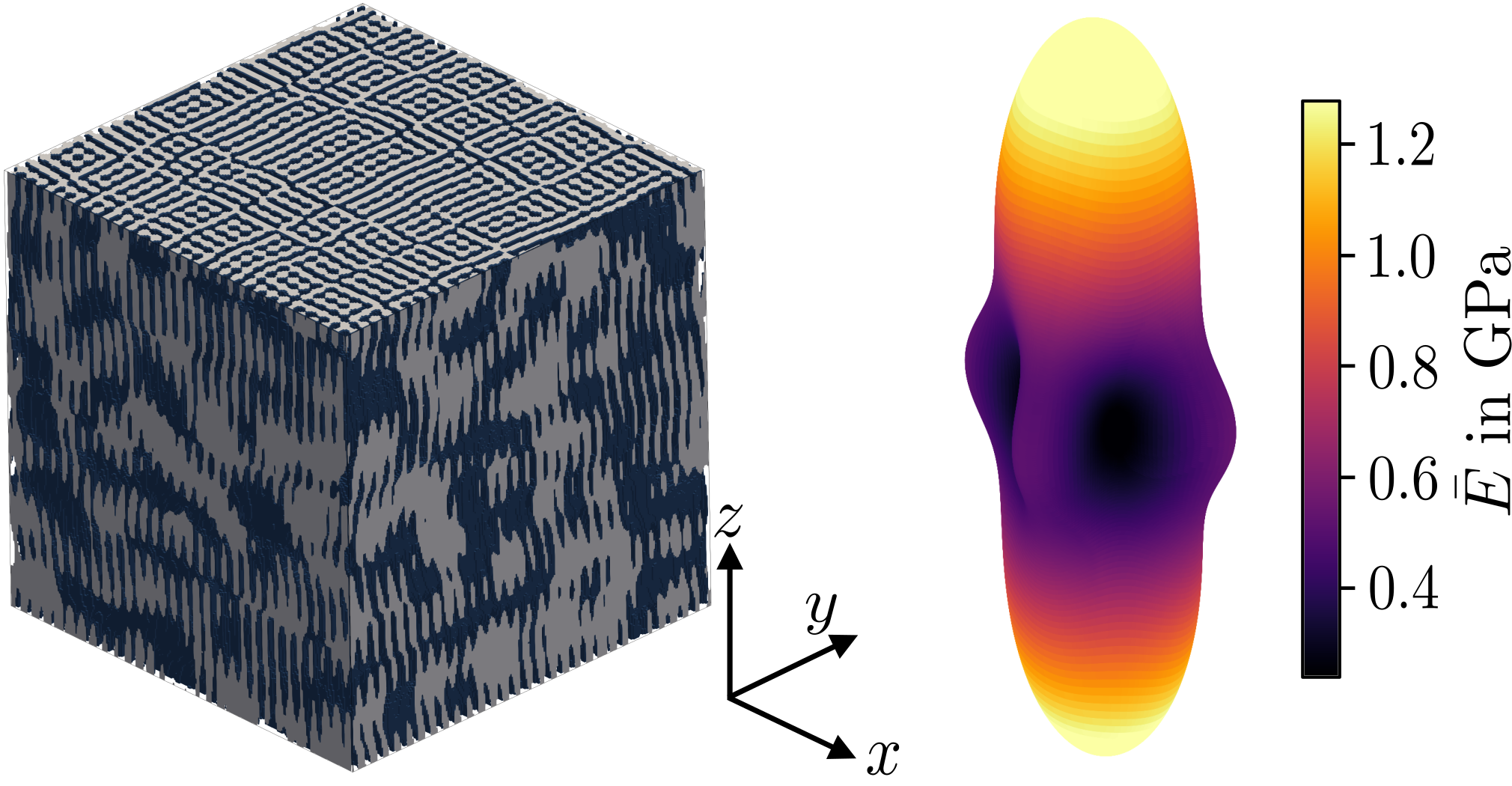}
    				\caption{$n_\text{voxel}=256^3$}
    			\end{subfigure}
    		\caption{Exemplary volume element and corresponding elastic surfaces for differently sized volume elements. The volume scales linearly with the number of voxels $n_\text{voxel}$ used for the spatial discretization.}\label{fig:AppStructures}
    		\end{figure}

            \autoref{fig:AppStudy} shows the result of a study on the representativness of volume elements of certain size. For this study, columnar spinodoid structures of different sizes but the same descriptor are generated.
            The discretization in $n_\text{voxel}$ is scaled linearly with the varying element length $l_\text{VE}$ to secure comparability. \autoref{fig:AppStructures} shows three exemplary structures of different sizes alongside their elastic surface.
            For each size 10 structures are generated from different random seeds.

            The relative deviation of the effective Young's modulus from the mean modulus of the set of the largest size ($n_\text{voxel}=256^3$) is plotted in~\autoref{app:Convergence} for the $x$-, $y$-, and $z$-direction.
            It can be seen, that the median deviation of all sets for the $x$- and $y$-direction, i.e., the directions perpendicular to the columnar features, is comparatively small. However, the smallest size ($n_\text{voxel}=32^3$) has a significantly larger variance.
            Concentrating on the relevant $z$-direction, it can be seen that both the median deviation and the variance within a set drastically decrease for larger sizes.

            As a good trade-off between higher demand for computational resources with increasing size and representativness, the discretization in $n_\text{voxel}=64^3$ is chosen for the present article.

\end{document}